\documentclass[10pt,openany,a4paper]{article}
\usepackage{geometry,color,framed}
\usepackage{amsmath,amssymb,mathtools,slashed,xcolor,mdframed}
\usepackage[vcentermath,enableskew]{youngtab}
\usepackage{cancel, pifont,multirow}
\usepackage{tikz}
\usetikzlibrary{positioning}
\usetikzlibrary{decorations.text}
\usepackage{amstext}
\usepackage{graphicx}
\usepackage[section]{placeins}
\usepackage{verbatim}
\usepackage{bm}
\usepackage{caption, subcaption, float}
\usepackage[colorlinks=true, 
            linkcolor=blue, citecolor=blue,
            menucolor = red,
            urlcolor=blue,linktoc=all]{hyperref}
\makeatletter

\newcommand{\beq}{\begin{eqnarray}}
\newcommand{\eeq}{\end{eqnarray}}
\newcommand{\non}{\nonumber\\}

\newcommand{\ben}{\begin{enumerate}}
\newcommand{\bei}{\begin{itemize}}
\newcommand{\eni}{\end{itemize}}
\newcommand{\enn}{\end{enumerate}}
\newcommand{\ra}{\rightarrow}

\newcommand{\xdownarrow}[1]{%
  {\left\downarrow\vbox to #1{}\right.\kern-\nulldelimiterspace}
}

\def\a{{\alpha}}
\def\b{{\beta}}
\def\g{{\gamma}}

\def\d{{\delta}}

\usepackage{tikz}
\usetikzlibrary{positioning}
\makeatother
\title{Deep Learning Calabi-Yau four folds with hybrid and recurrent neural network architectures}
\author{H. L. Dao\footnote{espoirdujour1162@gmail.com}}
\date{\today}

\begin{document} 
\maketitle
\begin{abstract}
In this work, we report the results of applying deep learning based on hybrid convolutional-recurrent and purely recurrent  neural network architectures to the dataset of almost one million complete intersection Calabi-Yau four-folds (CICY4) to machine-learn their four Hodge numbers $h^{1,1}$, $h^{2,1}$, $h^{3,1}$, $h^{2,2}$. In particular, we explored and experimented with twelve different neural network models, nine of which are convolutional-recurrent (CNN-RNN) hybrids with the RNN unit being either GRU (Gated Recurrent Unit) or Long Short Term Memory (LSTM). The remaining four models are purely recurrent neural networks based on LSTM. In terms of the $h^{1,1}$, $h^{2,1}$, $h^{3,1}$ and $h^{2,2}$ prediction accuracies, at 72\% training ratio,  our best performing individual model is CNN-LSTM-400, a hybrid CNN-LSTM with the LSTM hidden size of 400, which obtained 99.74\%, 98.07\%, 95.19\%, 81.01\%,  our second best performing individual model is LSTM-448, an LSTM-based model with the hidden size of 448, which obtained  99.74\%, 97.51\%, 94.24\%, and 78.63\%. These results were improved by forming ensembles of the top two, three or even four models. Our best ensemble, consisting of the top four models, achieved the accuracies of 99.84\%, 98.71\%, 96.26\%, 85.03\%. At 80\% training ratio, the top two performing models LSTM-448 and LSTM-424 are both LSTM-based with the hidden sizes of 448 and 424. Compared with the 72\% training ratio, there is a significant improvement of accuracies, which reached 99.85\%, 98.66\%, 96.26\%, 84.77\% for the best individual model and  99.90\%, 99.03\%, 97.97\%, 87.34\% for the best ensemble. By nature a proof of concept, the results of this work conclusively established the utility of RNN-based architectures and demonstrated their effective performances compared to the well-explored purely CNN-based architectures in the problem of deep learning Calabi Yau manifolds. 
\end{abstract}

\tableofcontents
\section{Introduction}
Since the pioneering works of \cite{He-17}, \cite{ruehle-17}, \cite{ks-17} that introduced artificial intelligence and machine learning (AI/ML) into string theory, a lot of progress has been made in successfully utilizing AI/ML techniques to explore and solve a wide variability of problems in string-theory-based settings. A detailed list of these problems  can be found in the recent exhaustive review \cite{ruehle-19} and references therein. In this work, we are interested in the particular string-theory setting involving Calabi-Yau (CY) manifolds, which are of paramount imporatance in the string theory compactification process that reduces the 10D/11D original theory to some phenomenologically semi-realistic four-dimensional theory with the correct spectrum of minimally supersymmetric standard model of particles \cite{he-18}, \cite{ibanez}, \cite{anderson-tasi}. More specifically, the physics of the four-dimensional spacetime resulting from this compactification is largely determined by the geometry and topology of the CY inner space. A notable instance of this is the fact the Hodge numbers of the CY three-folds used for compactifying the 10D string theory correspond to the number of generations of fundamental fermions in the 4D theory \cite{he-big-data}. 
\\\\
In the context of machine learning/deep learning CY manifolds, important examples of the AI/ML methods employed include supervised learning using arfificial neural networks (ANNs) or more conventional ML techniques (such as linear regression, decision tree/random forest, etc.) to learn the topological properties (the Hodge numbers) of the CY manifolds \cite{ef-20}, reinforcement learning to explore the vast landscape of CY compactification that can lead to viable and (semi-)realistic four-dimensional theories \cite{sm-20}. In the aforementioned problems, the CY manifolds are complete intersection CY (CICY) \cite{he-18} - a particular construction of CY as the vanishing loci of a number of homogeneous polynomial equations in an ambient space that is the product of all the individual spaces in which each polynomial equation is defined. 
\\\\
Moreover, the fact that many aspects of CY manifolds are amenable to AI/ML techniques has opened up many different avenues for new research in this direction. In particular, the recent work \cite{demirtas-20} studies the fine, regular, star triangulations  (FRSTs) of two reflexive polytopes 
from the Kreuzer Skarke dataset \cite{KS} that contains around 473 million four-dimensional reflexive polytopes whose triangulations correspond to the ambient toric varieties in which CY three-folds can be defined as hypersurfaces by employing a neural network made of ReZero layers \cite{rezero} to learn several geometric quantities. The work \cite{cy-weighted} applied both unsupervised and supervised machine learning methods to the database of weighted $\mathbb{P}^4$ admitting CY 3-fold hypersurfaces to study the CY topological data.  
The work \cite{cymetric} (see also \cite{cymetric-lectures}) used neural networks to compute the numerical Ricci-flat metrics for CICY and Kreuzer-Skarke CY manifolds. The work \cite{cy-genetic} uses genetic algorithms generates new reflexive polytopes in various dimensions used to construct the toric varieties hosting the  CY manifolds as hypersurfaces. The work \cite{cy-wp} uses ANNs to machine-learn the Hodge numbers of CY five-folds constructed as hypersurfaces in weighted projective spaces. The work \cite{cy-sas} uses ANNs and symbolic regression to machine learn the topological quantities related to the Sasakian and $G_2$-geometries of contact Calabi-Yau 7-manifolds. The reviews \cite{he-2204}, \cite{he-2303} cover an extensive list of earlier works on the topic of machine learning CY manifolds in various contexts.
\\\\
Regarding the problem of machine learning Hodge numbers of CICY manifolds, the dominant technique that has yielded impressive results comes from the field of computer vision, in which the configuration matrices of dimensions $M\times N$ are treated as gray-scale images of size $M\times N$ that can subsequently be taken as inputs to a convolutional neural network (CNN) \cite{he-big-data}. This is the case for both the well-studied CICY3 dataset \cite{cc3-1}, \cite{cc3-2}, \cite{cc3-3} and the larger, less-studied CICY4 dataset \cite{gray-13}, \cite{gray-14}. For the smaller CICY3 dataset which comprises 7890 data points of $12\times 15$ configuration matrices, there are only two Hodge numbers to learn, $h^{(1,1)}$ and $h^{(2,1)}$. The best-performing neural networks, reported in \cite{inception-cicy3}, \cite{inception}, are based on the Inception module (a specialized type of CNNs) and achieved 100\% accuracy for $h^{1,1}$ and above 50\% for $h^{2,1}$. Earlier works dealing only with the machine learning of $h^{1,1}$ are \cite{He-17}, \cite{bull}. As noted in \cite{inception-sum}, the task of machine learning $h^{2,1}$ to a high degree of accuracy for CICY3 remains an open problem. 
\\\\
For the CICY4 dataset which comprises almost one million (921,497) data points of $16\times 20$ configuration matrices with four Hodge numbers $h^{(1,1)}, h^{(2,1)}, h^{(3,1)},h^{(2,2)}$, the best-performing CNN model for this problem, CICYMiner, is again based on the Inception architecture with around $10^7$ trainable parameters, which was presented in the work of \cite{inception} (see also \cite{inception-sum} for a summary of both \cite{inception} and \cite{inception-cicy3}). The authors obtained 100\% accuracy for both $h^{(1,1)}$ and $h^{(2,1)}$, 96\% for $h^{(3,1)},$ 83\% for $h^{(2,2)}$ at a training ratio of 80 \%. At a training ratio of 30\%, the accuracies obtained dropped to 100\%, 97\%, 81\%, 49\% for $h^{(1,1)}, h^{(2,1)}, h^{(3,1)}$ and $h^{(2,2)}$, respectively. An earlier work that uses dense neural networks to study only $h^{(1,1)}$ and $h^{(3,1)}$  was reported in \cite{cicy4-he}.
\\\\
In this work, we move away from the aforementioned established approach of using purely convolutional architecture for deep learning the CICY Hodge numbers. Instead, our approach relies mainly on variants of recurrent neural networks (RNNs) such as Gated Recurrent Unit (GRU) \cite{cho-gru-1}, \cite{cho-gru-2}, \cite{gru-3} and Long-Short Term Memory (LSTM) \cite{lstm-97}, \cite{lstm-00}, \cite{lstm-rs}, \cite{lstm-v} in two different settings. The first setting involves  hybrid architectures based on CNN and RNN (both GRU and LSTM). The second setting involves purely recurrent architecture based on LSTM. By experimenting with different variations in each of the two architectures above, we obtain a number of promising results that are comparable to or even outperforming those obtained in \cite{inception}. Ultimately, with these results, we hope to demonstrate and establish the effectiveness of recurrent neural networks in this particular problem which has been primarily and exclusively explored with convolutional architectures. 
\\\\
The organization of this paper is as follows: In Section \ref{dataset}, we summarize the main characteristics of the dataset and our data preparation method. In section \ref{rnn-basics}, we recall the basic mathematical facts about recurrent neural network architecture before moving to section \ref{arch} in which  we describe in detail the different types of neural networks architectures used in this work. In section \ref{results}, we report the training results obtained for all 12 models at 72\% data split (see sections \ref{cnn-gru-models-res}, \ref{cnn-lstm-models-res}, \ref{lstm-models-res}). In section \ref{ensembles-}, we form four ensembles of the best performing models and check their performances. Next, the accuracies of all models (evaluated on the test set), including the ensembles, are presented in section \ref{acc}. In section \ref{cv5f}, we perform a 5-fold cross validation training for a single model and use the results (section \ref{cv5f-sep}) to form new ensembles (section \ref{cv5f-ens}). In section \ref{results-80}, we report the training results of the top three individual models which were retrained on the enlarged dataset at 80\% data split. In section \ref{ensemble-80}, we again form several new ensembles and check the performances of the retrained models and and their associated ensembles (see section \ref{acc-80}). 
A summary in section \ref{concl} is followed by the Appendix \ref{app} containing the sections \ref{metrics}, \ref{metrics-cv}, \ref{metrics-80} on the additional metrics for evaluating the performance of the neural networks, and the section \ref{curves} that includes the training curves of all models. The Python codes in the form of \texttt{Kaggle} notebooks \footnote{We make use of the following standard libraries: \texttt{numpy} (\href{https://numpy.org/}{https://numpy.org/}) \cite{numpy}, \texttt{pandas} (\href{https://pandas.pydata.org/}{https://pandas.pydata.org/}) \cite{pandas}, \texttt{scikit-learn} (\href{https://scikit-learn.org}{https://scikit-learn.org}) \cite{sklearn}, \texttt{matplotlib} (\href{https://matplotlib.org/}{https://matplotlib.org/}) \cite{matplotlib}, \texttt{seaborn} (\href{https://seaborn.pydata.org/}{https://seaborn.pydata.org/}) \cite{seaborn} for data handling, manipulation, visualization, and \texttt{Pytorch} (\href{https://pytorch.org/}{https://pytorch.org/}) \cite{pytorch} for building and training the neural networks.} for this work can be found at the following \texttt{GitHub} link: \href{https://github.com/lorrespz/CICY4-Deep-learning-hybrid-recurrent-NNs-main}{https://github.com/lorrespz/CICY4-Deep-learning-hybrid-recurrent-NNs-main}.
\section{Dataset and data preparation} \label{dataset}
\subsection{Overview of the CICY4 dataset}
The original CICY4 dataset containing 921,497 data points each consisting of a $r\times K$ configuration matrix and its associated topological properties (including Euler number and a set of four Hodge numbers $h^{(1,1)},h^{(2,1)},h^{(3,1)},h^{(2,2)}$) was compiled and introduced in the work of \cite{gray-14} and \cite{gray-13}. Each CICY4 manifold is fully characterized by its configuration matrix of the form
\beq
\mathcal{M} = 
\left(\begin{array}{c|ccc}
n_1 & p_1^0 & \ldots & p_K^0\\
\vdots & \vdots & \ddots & \vdots\\
n_r & p_1^r & \ldots & p_K^r \end{array}\right)\,, \label{cm}
\eeq
where $i = 1, \ldots, r$ labels the projective ambient space factors $\mathbb{P}^{n_i}$, $j = 1, \ldots, K$ labels the polynomials $p_j$. The integer $p^i_j$ is the degree of the  $j$-th polynomial in the homogeneous coordinates of the $i$-th complex projective space with dimension $n_i$. More specifically, the first column of the matrix (\ref{cm}) denotes the dimensions of the projective spaces whose product $\mathbb{P}^{n_1}\times \ldots \times \mathbb{P}^{n_r}$ forms the ambient space in which the CICY4 is embedded. Each of the subsequent columns $p_j = (p_j^i)_{i=1,\ldots,r}$ denotes the degrees of a polynomial in the ambient projective coordinates. 
\\\\
As a concrete example, consider the case mentioned in \cite{gray-13} where a configuration matrix is of the form
\beq
\left[\begin{array}{c|cc} 1 & 1 & 1 \\ 2 & 1 &2 \\ 3 & 0 & 4\end{array}\right] \,.\label{cm-1}
\eeq
In this case, the ambient space is the product space $\mathbb{P}^1 \times \mathbb{P}^2 \times \mathbb{P}^3$, as denoted by the first column of the matrix (\ref{cm-1}) above. The remaining two columns denote the multi-degrees (1,1,0) and (1,2,4) of the two polynomials in the ambient space. Each of the three entries in the multi-degree tuples corresponding to the polynomial degree in one projective space factor. Let the $\mathbb{P}^1$ coordinates be $x^k$, ($k = 0,1$), the $\mathbb{P}^2$ coordinates be $y^a$ ($a = 0,1,2)$, and the $\mathbb{P}^3$ coordinates be $z^\a$ ($\a = 0,\ldots, 3$) then the two defining polynomials are
\beq
p_1 = \sum_{i,a} c_{ia}x^i y^a, \hspace{10mm} p_2 = \sum_{i,\ldots, \delta} d_{i,ab,\alpha\beta\gamma\delta} x^i y^ay^b\, z^\a z^\b z^\g z^\d,
\eeq
where $c_{ia}$ and $d_{i,ab,\a\b\g\d}$ are complex coefficients. The CICY4 defined by (\ref{cm-1}) is the common zero locus of these two polynomials.
\\\\
Returning to (\ref{cm}), the Calabi Yau defining property is ascertained by the following condition on each row $i$ of the configuration matrix
\beq
\sum^K_{j=1} p^i_j = n_i + 1\,.
\eeq
The Hodge numbers for all CICY4 in the dataset were computed in \cite{gray-14}. In terms of the four non-trivial Hodge numbers $h^{(1,1)}$,$h^{(2,1)}$,  $h^{(3,1)}$, $h^{(2,2)}$ of the CICY4, the Euler number can be computed from the linear relationship \cite{gray-13}
\beq
\chi = 4 + 2h^{(1,1)} -4 h^{(2,1)} + 2h^{(3,1)} + h^{(2,2)} \,. \label{cons-1}
\eeq
Furthermore, the Hodge numbers also satisfy the constraint \cite{he-21}
\beq
h^{2,2} = 2(22 + 2h^{1,1} + 2h^{3,1} - h^{2,1}). \label{cons-2}
\eeq
The analysis of \cite{gray-14} and \cite{inception}  excluded 15,813 CICY4 manifolds on the account of them being product manifolds, so the dataset actually contains 905,684 CICY4 manifolds and their corresponding Hodge numbers. The mean, maximal  and minimal values for the Hodge numbers are reported in \cite{gray-14}
\beq
\begin{array}{cc}
\langle h^{1,1}\rangle = 10.1^{24}_1, & \langle h^{2,1}\rangle = 0.817^{33}_0,
\\ \\
\langle h^{3,1}\rangle = 39.6^{426}_{20}, & \langle h^{2,2}\rangle = 241^{1752}_{204}\,,
\end{array}
\eeq
where the superscripts and subcripts correspond to the maximal and minimal values. 
The range of values of $h^{2,2}$ is an order of magnitude larger than that of $h^{3,1}$, which is itself an order of magnitude larger than those of $h^{1,1}$ and $h^{2,1}$. 

\subsection{Data preparation}
The originally compiled dataset by the authors of \cite{gray-13} can be downloaded in either text format or as a Mathematica file at \cite{cicy4-original}. A more convenient format suitable for deep learning using Python code is 
 \texttt{Numpy} array \texttt{npy}, which was created by the authors of \cite{inception}, \cite{cicy4-npy}. The \texttt{npy} files can be obtained by running the script \cite{cicy4-npy} on the original text file downloaded from \cite{cicy4-original}. The final dataset format used for the deep learning task thus consists of 905,684 data points\footnote{same as that used in the work \cite{inception}, \cite{gray-14}} of $16\times 20$ configuration matrices (whose entries are rescaled to be within the range $[0,1]$) and their associated Hodge number tuples ($h^{(1,1)}$,$h^{(2,1)}$,  $h^{(3,1)}$, $h^{(2,2)}$). As noted in \cite{inception}, there is a heavy imbalance in the dataset for $h^{2,1}$, as well as for $h^{3,1}$ and $h^{2,2}$: A large portion of the dataset for these Hodge numbers (on the order of $10^5$ data points) having values in the lower ranges. Furthermore, $h^{2,2}$  and $h^{3,1}$ (to a lesser extent, $h^{2,1}$) contain many outliers in the tail end of the data. These characteristics make the regression task for $h^{3,1}$, $h^{2,2}$ much harder than that for $h^{2,1}$ and $h^{1,1}$. Known results from the previous works \cite{cicy4-he}, \cite{inception} have demonstrated the fact that although perfect accuracies could be obtained for $h^{1,1}$ and $h^{2,1}$, the accuracies for $h^{3,1}$ and $h^{2,2}$ are usually much lower.  
 \\\\
In this work, we used two sets of data for two rounds of training plus a 5-fold cross validation (CV) round of training. In the first training round, we created a data set at 72\% data split and used this to train all twelve neural network models. In the second round of training, we created a new dataset at 80\% data split and used it to train only the top three models obtained from the first round of training. In the 5-fold CV training, we use the same data sets as the first round of training. 
\bei
\item
For the 72\% data split, we divided the data into the following three subsets:
\bei
  \item The training set, containing 652,092 data points and accounting for 72\% of the full dataset, is used to train all twelve models.
  \item The validation set, containing 72,455 data points and accounting for 8\% of the full dataset, is used for validation in the training process.
  \item The test set, containing 181,137 data points and accounting for 20\% of the full dataset, is not seen by the models during the training phase but only during the inference phase where the trained models are reloaded to perform inference.
\eni
\item
For the 5-fold CV training, the same datasets as the 72\% data split were used, but with the following crucial differences:
\bei
\item The training set (72\% of the data) is merged with the validation set (8\% of the data), to form a total training set accounting for 80\% of the data. This total training set is converted into 5 training datasets corresponding to the 5 CV training folds. 
Each `fold' (where 'fold' now is used synonymously with the training set for that fold) contains five partitions. In each of the training fold, a model is trained on the data of the 4 partitions (accounting for 64\% of the full data) and validated on the remaning partition (accounting for 16\% of the full data). For a 5-fold CV, there are 5 models. Although these models all have the same architecture, they should be treated as independent models, since they are trained and validated on different subsets of the training data. 

\begin{figure}[H]
\centering
\includegraphics[width = 0.6\textwidth]{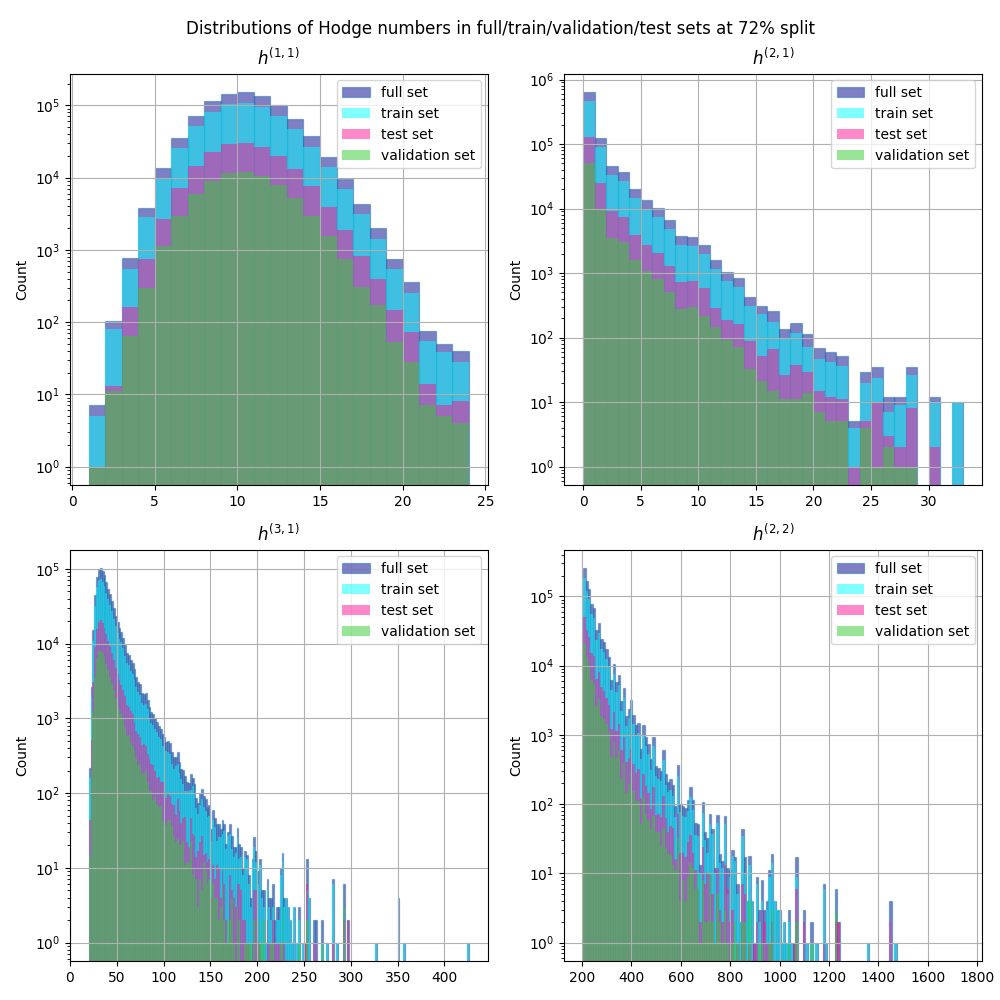}
\caption{Histograms of the four Hodge numbers $h^{1,1}$ (top left), $h^{2,1}$ (top right), $h^{3,1}$ (bottom left), and $h^{2,2}$ (bottom right) for the full, training, validation and test datasets. Figure adapted from Figure 1 of \cite{inception}.}\label{hodge_dist}
\end{figure}

\begin{figure}[H]
\centering
\includegraphics[width = 0.6\textwidth]{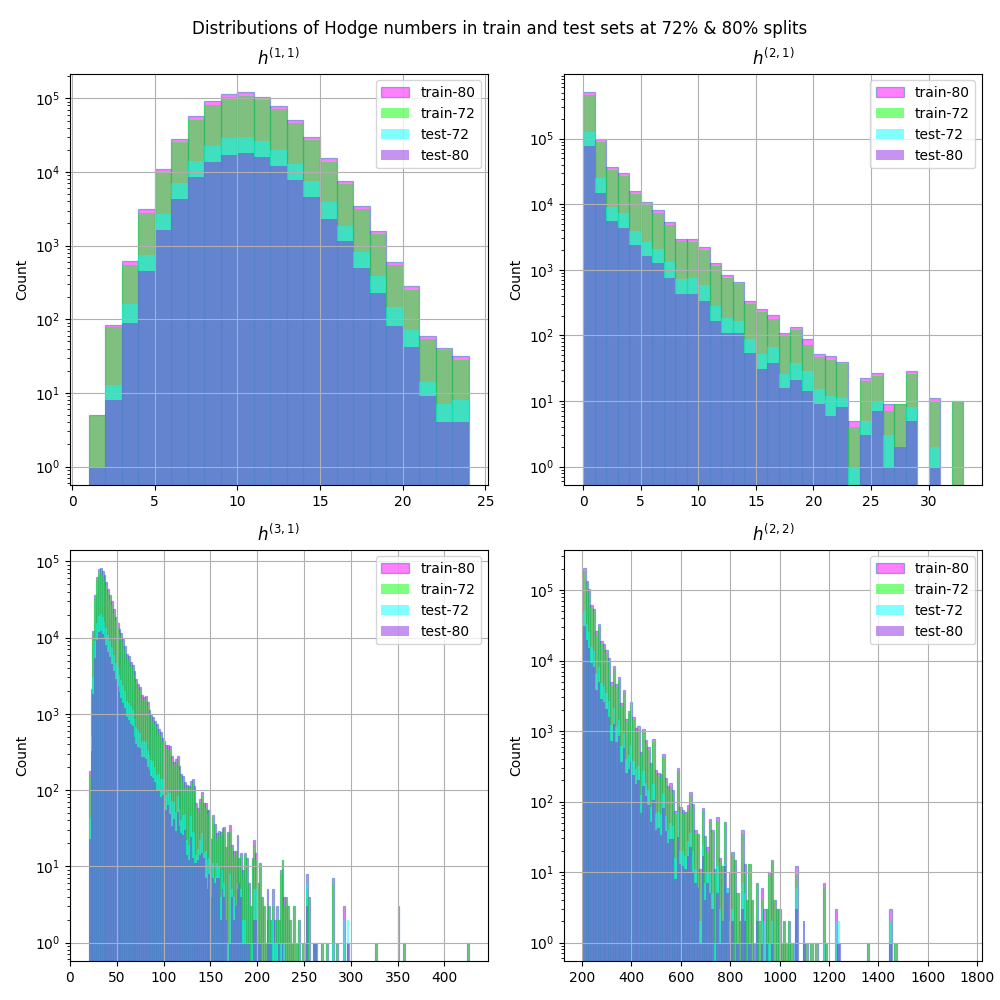}
\caption{Histograms of the four Hodge numbers $h^{1,1}$ (top left), $h^{2,1}$ (top right), $h^{3,1}$ (bottom left), and $h^{2,2}$ (bottom right) for the  training and test datasets at 72\% and 80\% data splits.}\label{old-v-new}
\end{figure}

\item The test set is the same as the one used in the 72\% data split, and accounts for 20\% of the full data set. 
\item The distributions of the four Hodge numbers in each of the training folds are shown in Figs. \ref{hodge_fold_1}, \ref{hodge_fold_2}, \ref{hodge_fold_3}, \ref{hodge_fold_4}, \ref{hodge_fold_5} in Appendix \ref{5fold_cv_hodge_dist}. It can be seen from the figures there that the distributions in all five CV folds resemble the full dataset to the same degree, with none of the folds appearing significantly different from another. 
\eni
\item
For the 80\% data split, 
\bei
\item the new training set, containing 724,547 data points and accounting for 80\% of the full dataset, is created by merging 40\% of the old test set with the old training set. This training set is used to train only the top three models from the first round. 
\item The validation set stays the same as before
\item The new test set, containing 108,682 data points and accounting for 12\% of the full dataset, is created by using the remaining 60\% of the old test set (after the 40\% was taken away). 
\eni
 \eni
From this point onwards, we will refer to these datasets as the 72\% dataset and the 80\% dataset. 
The distributions of the Hodge numbers in the full dataset plus the train, validation and test subsets at the 72\% data split are shown in Fig. \ref{hodge_dist}. For comparison, the distributions of the train and test sets at both the 72\% data split and the 80\% data split are plotted together in Fig.\ref{old-v-new}. 

\section{Basics of recurrent neural networks} \label{rnn-basics}
In this section, we briefly recall some pertinent mathematical facts about general recurrent neural networks (RNNs) and their more sophisticated GRU/LSTM variants for the benefits of the physicist readers who might not be familiar with deep learning. 
While our work makes use of both convolutional and recurrent architectures, we will not touch on the basics of convolutional neural networks, since many excellent texts already exist on CNN on the topic of deep learning CY manifolds as CNNs are the main tool for deep learning in these works (see, for example, the works \cite{ruehle-19}, \cite{he-18}, \cite{he-21}, \cite{he-big-data} which also discuss the basics of deep learning using a simple feed forward/dense articial neural network in great detail). 
\\\\
Due to their recurrent connection, RNNs excel at handling sequential data which involves long term dependency of present data on past data (see Chapter 10 of the text book \cite{deep-learning}). Concretely, let the input data be of the form $N\times T\times D$ where $N$ is the number of data samples (in a batch, typically), $T$ is the number of time steps in the data sequence, and $D$ is the number of features. When $D=1$, the input is a vector representing a sequence, while for $D>1$, the input is a two-dimensional matrix. The practice of using specialized RNNs such as  LSTM and GRU for handling image data is not new - there exists a large volume of work reporting the effectiveness of LSTM/GRU-based architectures (both in the pure or hybrid forms with CNNs) in the computer vision literature, e.g. \cite{taki-22},\cite{convlstm}.
\begin{itemize}
  \item \textbf{Recurrent Neural Network (RNN)}: 

Given an input $\mathbf{x} = (x_1, x_2, \ldots, x_T)$ (representing a sequence) of size $T\times D$ (here $D=1$), the RNN with the hidden dimension of size $M$ outputs the hidden state $h_t$ of size $M$ at time step $t$ as a function of $x_t$ and the previous hidden state $h_{t-1}$
\beq
h_t = \sigma_h\left( W_i x_t + W_h h_{t-1} + b_h\right)\,,
\eeq
where $\sigma_h$ is a nonlinear activation function such as sigmoid, ReLU or a $\tanh$ function,  $W_i$ of size $D\times M$ is the input layer to hidden layer weight matrix and $W_h$ of size $M\times M$ is the hidden layer to hidden layer  weight matrix, and $b_h$ of size $M$ is the bias vectors. 
At time step $t = 0$, the first hidden state $h_0$ is typically initialized to zero.
The output $y_t$ is calculated from $h_t$ as follows
\beq
y_t = f\left( W_o h_t + b_o\right)\,,
\eeq
where $f$ is a nonlinear activation function,  $W_o$ of size $M\times M$ is the output weight matrix, $b_o$ of size $M$ is the output bias. $W_i, W_h, W_o, b_h, b_o$ are learnable parameters to be determined during training phase of the network.
Schematically, an RNN can be represented by the following diagram in Fig. \ref{rnn-0}. 
\begin{figure}[H]
\centering
\begin{tikzpicture}[node distance = 2cm, thick]%
        \node[circle, draw] (1) {$y(t)$};
        \node[circle, draw](2) [below of = 1] {$h(t)$};
        \node (3)[circle, draw] [below of = 2] {$x(t)$};
        \draw[->] (2) -- node [right] {$W_o$} (1);
        \path[->, thick, decoration={text along path, text={$W_h$}}] 
        (2) edge [out=315, in=35, loop] (2);
        \draw[->] (3) -- node [right]{$W_i$} (2);

    \end{tikzpicture}
    \caption{A computational graph/diagram for RNN.}
    \label{rnn-0}
    \end{figure}
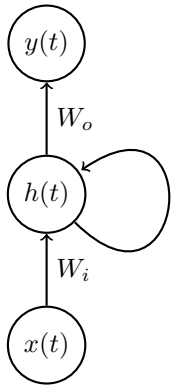
 When  the time $t$ is `unfolded' horizontally, we obtain the following diagram in Fig.\ref{rnn-1}
\begin{figure}[H]
\centering
\begin{tikzpicture}[node distance = 2cm, thick]%
        \node[circle, draw] (1) {$y_2$};
        \node[circle, draw] (1a)[left of =1] {$y_1$};
        \node[circle, draw] (1b)[right of =1] {$y_3$};
        \node[circle, draw](2) [below of = 1] {$h_2$};
        \node[circle, draw](2a) [left of = 2] {$h_1$};
        \node[circle, draw](2o) [left of = 2a] {$h_0$};
        \node[circle, draw](2b) [right of = 2] {$h_3$};
        \node[] (2c)[right of = 2b] {$\ldots$};
        \node[circle, draw] (2d)[right of = 2c] {$h_T$};
        \node[circle, draw] (1d)[above of = 2d] {$y_T$};
        \node[circle, draw] (3d)[below of = 2d] {$x_T$};
        \node (3)[circle, draw] [below of = 2] {$x_2$};
        \node (3a)[circle, draw] [left of = 3] {$x_1$};
        \node (3b)[circle, draw] [right of = 3] {$x_3$};
        \draw[->] (2) -- node [right] {$W_o$} (1);
        \draw[->] (2a) -- node [right] {$W_o$} (1a);
        \draw[->] (2b) -- node [right] {$W_o$} (1b);
        \draw[->] (2o) -- node [above] {$W_h$} (2a); 
        \draw[->] (2a) -- node [above] {$W_h$} (2); 
        \draw[->] (2) -- node [above] {$W_h$} (2b);
        \draw[->] (2b) -- node [above] {$W_h$} (2c);
        \draw[->] (2c) -- node [above] {$W_h$} (2d);
        \draw[->] (3) -- node [right]{$W_i$} (2);
        \draw[->] (3a) -- node [right]{$W_i$} (2a);
        \draw[->] (3b) -- node [right]{$W_i$} (2b);
        \draw[->] (3d) -- node [right]{$W_i$} (2d);
        \draw[->] (2d) -- node [right]{$W_o$} (1d);

    \end{tikzpicture}
    \caption{An unfolded computational graph/ diagram of RNN.} \label{rnn-1}
    \end{figure}
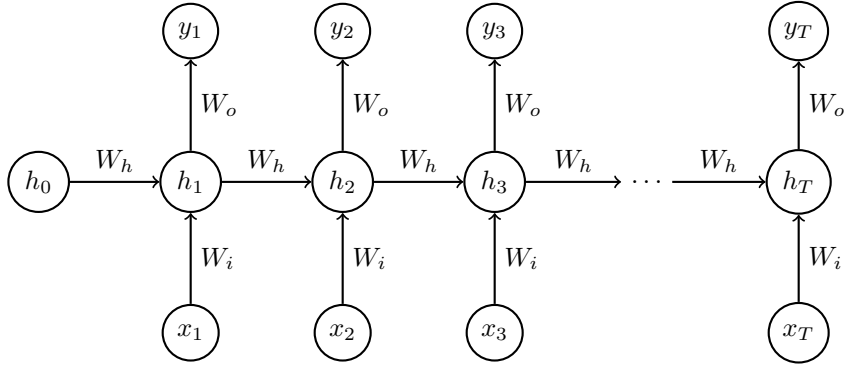
\item \textbf{Long Short-term Memory (LSTM)}:
The LSTM network was introduced in 1997 the paper \cite{lstm-97} and refined structurally in 2000 in \cite{lstm-00}. The most commonly used variant of LSTM in the literature is \cite{lstm-v}. A comprehensive study of the computational functionality of each LSTM variants was done in \cite{lstm-rs}.  Compared to the general RNN discussed above, LSTM is much more sophisticated with four gates controlling how information is directed and retained through its network. The four gates are input gate $i$, cell gate $c$, forget gate $f$, and output gate $o$.
The hidden state $h$ is now a composition function of all these gates. 
\\\\
Given an input $\mathbf{x} = (x_1, x_2, \ldots, x_T)$ (representing a sequence) of size $T\times D$ (here $D=1$), the LSTM with the hidden dimension of size $M$ calculates the  input gate $i_t$, forget gate $f_t$, cell gate $c_t$, and output gate $o_t$ (all of size $M$) as functions of the input $x_t$ at time step $t$, the hidden state $h_{t-1}$ and cell state $c_{t-1}$ at time step $(t-1)$
\beq
i_t &=& \sigma\left( W_i x_t + W_{hi} h_{t-1} + b_h\right),
\non
f_t &=& \sigma\left(W_f x_t + W_{hf} h_{t-1} + b_f \right),
\non
o_t &=& \sigma\left(W_o x_t + W_{ho}h_{t-1} +b_o\right),\label{lstm-i-f-o}
\eeq
so that the output states are the cell state $c_t$ and $h_t$, which are
\beq
c_t &=& f_t \odot c_{t-1} + i_t \odot \tanh\left( W_g x_t + W_{hg} h_{t-1} + b_g\right),\non
h_t &=& o_t \odot \tanh(c_t) \label{lstm-out}\,.
\eeq
As before, $\sigma$ is the sigmoid function, $\odot$ is the Hadamard product, all the $W$'s are weight matrices, and all the $b$'s are the bias vectors, which are the learnable parameters during the training process. Since $i_t$, $f_t$, $o_t$ all have the sigmoid activation function, their values are constrained to be within the range $[0,1]$. The output cell state $c_t$ is a sum of the previous cell state $c_{t-1}$ multiplied by the forget gate $f_t$  and the cell gate $g_t$ multiplied by the input gate $i_t$. Specifically,
\bei
\item $i_t$ determines how much new information ($\tanh\left( W_g x_t + W_{hg} h_{t-1} + b_g\right)$) is kept in $c_t$,
\item $f_t$ determines how much the previous cell state $c_{t-1}$ is retained in $c_t$. 
\eni
The hidden state $h_t$ is a squashed version of the cell state $c_t$, with the amount of squashness determined by $o_t$. 
\\\\
 From the equations (\ref{lstm-i-f-o}, \ref{lstm-out}) above, one can see that there are two types of recurrence in LSTM: an outer recurrence involving the previous state $h_{t-1}$, and an inner recurrence involving the previous cell state $c_{t-1}$. 
\item \textbf{Gated Recurrent Unit (GRU)}:
The GRU network was proposed in 2014 by \cite{cho-gru-1}, \cite{cho-gru-2}. Structurally, GRU is simpler than LSTM with three gates: reset gate $r$, update gate $z$, and new gate $n$.  
Given an input $\mathbf{x} = (x_1, x_2, \ldots, x_T)$ (representing a sequence) of size $T\times D$ (here $D=1$), the GRU with the hidden dimension of size $M$ calculates the reset gate $r_t$, the update gate $u_t$, and the new gate $n_t$ (all of size $M$) as functions of the input $x_t$ at time step $t$ and the hidden state $h_{t-1}$ at time step $(t-1)$
\beq
r_t &=& \sigma(W_r x_t + W_{hr} h_{t-1} + b_r),
\non
u_t &=& \sigma(W_u x_t + W_{hu} h_{t-1} + b_u),
\non
n_t &=& \tanh(W_n x_t + r_t \odot (W_{hn} h_{t-1} + b_n)),
\eeq
where $\sigma$ is the sigmoid activation function (which scales the output to be within the range $[0,1]$), $W_{r/z/n}$, $W_{hr/u/n}$ are the weight matrices and $b_{r/u/n}$ are the bias vectors (all learnable parameters).
The goal is to output the hidden state $h_t$ of size $M$ at time step $t$ as a function of $z_t$, $n_t$ and $h_{t-1}$
\beq
h_t = (1-u_t)\odot n_t + u_t \odot h_{t-1},
\eeq
where $\odot$ is the Hadamard (element-wise) product. The main simplification of GRU compared with its LSTM counterpart  is the fact that a single gate, the update gate $z$ performs the job of the LSTM forget gate $f$ and input gate $i$. Depending on the update gate function $u_t$, $h_t$ is the sum of two terms, one involving the previous state $h_{t-1}$ and the other involving a new state $n_t$.  Specifically,
\bei
\item
mostly new information is present in $h_t$ ($h_t \approx n_t$) when $u_t<<1$,
\item mostly old/previous information is retained in $h_t$ ($h_t \approx h_{t-1}$) when $(1-u_t) << 1$
\item a mixture of old and new information is present for any intermediate $u_t$ value in $[0,1]$. 
\eni
\end{itemize}

\section{Neural network architectures} \label{arch}
In this section, we describe in detail the neural networks used for deep learning the four Hodge numbers. All neural networks comprise of a feature extractor part and a feed forward part as in Fig. \ref{gen-a} below.
\begin{figure}[h]
\centering
\begin{tikzpicture}[node distance = 1cm, thick]%
        \node (1) {\boxed{\begin{matrix}16\times 20 \\\text{matrix}\end{matrix}}};
        \node (2) [right=of 1] {\boxed{\begin{array}{c}\text{Feature}\\ \text{extractor}\\ \end{array}}};
        \node (3) [right=of 2] {\boxed{\begin{array}{c}\text{Feed forward network}\\\boxed{\begin{matrix}\text{Dense} \\ (V_F,1024)\end{matrix}}
\\\downarrow\\
\boxed{\begin{matrix}\text{Dense}\\ (1024,4)\end{matrix}}\end{array}}};
        \node (4) [right=of 3] {\boxed{\begin{matrix} h^{(1,1)} \\ h^{(2,1)}\\h^{(3,1)} \\ h^{(2,2)}\end{matrix}}};
        \draw[->] (1) -- node [midway] {} (2);
        \draw[->] (2) -- node [midway] {} (3);
        \draw[->] (3) -- node [midway]{} (4);

    \end{tikzpicture}
    \caption{A schematic diagram of the overall architecture of all neural networks.} \label{gen-a}
    \label{gen-arch}
    \end{figure}
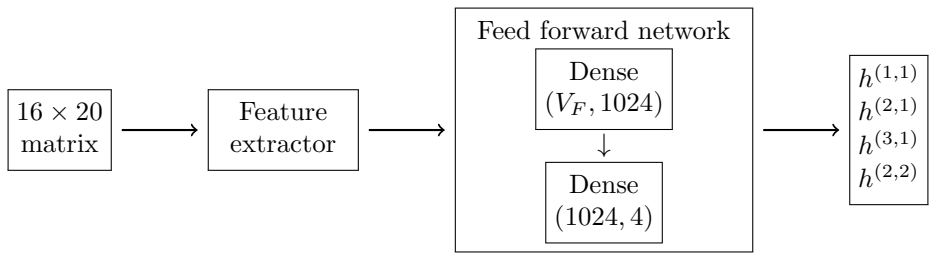
 For CNN-RNN hybrids, the feature extractor is a combination of both the CNN and the RNN (either GRU or LSTM). For LSTM-based models, the feature extractor is just the LSTM part. The output of the feature extractor is a vector of size $V_F$. Since our main goal in this work is to explore the effectiveness of RNN-based architecture in deep learning the Hodge numbers of CICY4, the same CNN block is used for all CNN-RNN models, with the only exception being the ResNet-RNN models, where we experimented with a ResNet-inspired CNN block.
The feed forward part is largely the same for all networks, and contains two fully connected (dense) layers, the first one of size $(V_F, 1024)$ and the second one of size $(1024, 4)$. In the following sections, we will describe in detail the exact composition of the feature extractor for each variant of the network considered.
\subsection{CNN-RNN hybrid neural networks} \label{arch-cnn-rnn}
There are three types of CNN-RNN hybrid models considered in this work: CNN-GRU, CNN-LSTM and ResNet-GRU/LSTM. In the following sections, we will go through the details of each type.
\subsubsection{CNN-GRU hybrid}
\begin{figure}[H]
\centering
\begin{tikzpicture}[node distance = 1cm, thick]%
        \node (1) {\boxed{\begin{matrix}16\times 20 \\\text{matrix}\end{matrix}}};
        \node (2) [right=of 1] {\boxed{\begin{array}{c}
\text{Feature extractor}\\
\boxed{\begin{matrix}\text{CNN block} \\ \text{feature vector size 384}\end{matrix}} \\\\
\boxed{\begin{matrix}\text{GRU block} \\ \text{hidden size $M$}\end{matrix}}
\end{array}}};
        \node (3) [right=of 2] {\boxed{\begin{array}{c}\text{Feed forward network}\\\boxed{\begin{matrix}\text{Dense} \\ (M+384,1024)\end{matrix}}
\\\downarrow\\
\boxed{\begin{matrix}\text{Dense}\\ (1024,4)\end{matrix}}\end{array}}};
        \node (4) [right=of 3] {\boxed{\begin{matrix} h^{(1,1)} \\ h^{(2,1)}\\h^{(3,1)} \\ h^{(2,2)}\end{matrix}}};
        \draw[->] (1) -- node [midway] {} (2);
        \draw[->] (2) -- node [midway] {} (3);
        \draw[->] (3) -- node [midway]{} (4);
    \end{tikzpicture}
    \caption{A schematic diagram of the architecture of CNN-GRU hybrid neural networks.}
    \label{cnn-gru}
    \end{figure}
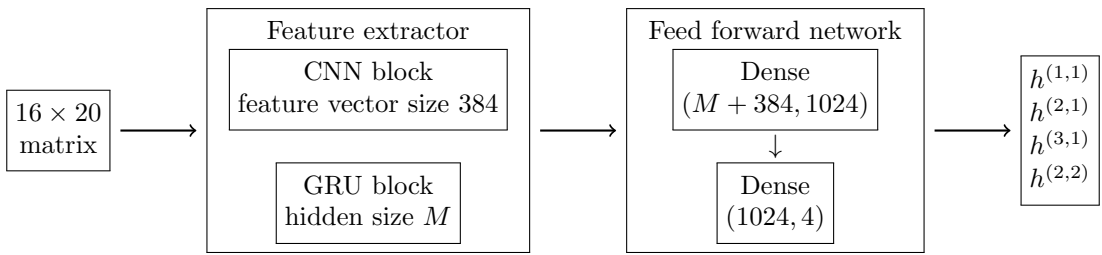

The first type of model in the family of CNN-RNN hybrid models is CNN-GRU. We will describe the building blocks of this model in detail, since the rest of the CNN-RNN hybrid models are very similar with only some minor differences.
The schematic of this type of neural network is shown in Fig.\ref{cnn-gru}.
The feature extractor stage comprises a CNN block and a GRU block in parallel.
\begin{itemize}
  \item \textit{CNN block}: 
The input to the CNN block is a tensor representing a batch of images of size $(N, H, W, C)$ where $N$ is the batch size, $H$ is the height, $W$ is the width $C$ is the number of color
channels. A colored image has three color channels so $C = 3$, while a gray scale image has only one color channel $C = 1$. A gray scale image is practically a matrix with $H$
being the number of rows and $W$ being the number of columns. For our case, $H = 16, W = 20, C = 1$.  
Structurally, the CNN block has 2 convolutional layers. Each convolutional layer, as implemented in \texttt{PyTorch} \footnote{\href{https://pytorch.org/docs/stable/generated/torch.nn.Conv2d.html}{https://pytorch.org/docs/stable/generated/torch.nn.Conv2d.html}} is identified by the following four parameters (with
  the rest of the parameters set to default values):
  \beq
  \texttt{(in\_channels, out\_channels, kernel\_size, stride)}\,.
  \eeq
For the first CNN layer,
\begin{itemize}
  \item the \texttt{in\_channels} parameter is 1, denoting the one color channel of gray scale image representing the $16\times 20$ matrix,
  \item  the \texttt{out\_channels} parameter is 128, denoting the number of convolutional filters,
  \item the \texttt{kernel\_size} parameter is 4, meaning that each filter is $4\times 4$ in size
  \item the \texttt{stride} parameter is set to 1 (the default parameter).
\eni
For the second CNN layer,
\begin{itemize}
  \item the \texttt{in\_channels} parameter is 128, which is the number of filters from the first CNN layer,
  \item  the \texttt{out\_channels} parameter is 64, denoting the number of convolutional filters in this layer,
  \item the \texttt{kernel\_size} parameter is 3, meaning that each filter is $3\times 3$ in size
  \item the \texttt{stride} parameter is set to 1 (the default parameter).
\eni
Each convolutional layer is followed by a max pooling layer that downsamples the output from the convolutional layer by a factor of 2.
After the second pooling layer, a \texttt{ReLU} activation and a \texttt{Flatten} layer are applied to the output to obtain a feature vector of size
384. The schematic of this CNN block is shown in Fig.\ref{cnn-block}.
\begin{figure}[h]
\centering
\begin{tikzpicture}[node distance = 1cm, thick]%
        \node (1) {\boxed{
\begin{array}{c}
\text{CNN block}\\\\
\boxed{\text{Conv2d}: (1, 128,4,1)}\\
\downarrow\\
\boxed{\text{MaxPool2d: (2,2)}}\\
\downarrow\\
\boxed{\text{Conv2d}: (128, 64, 3,1)}\\
\downarrow\\
\boxed{\text{MaxPool2d}: (2,2)}\\
\downarrow\\
\boxed{\text{ReLu activation}}\\
\downarrow\\
\boxed{\text{Flatten}}
\end{array}}};
        \node (2) [right=of 1] {\boxed{
\begin{matrix}\text{Feature vector 1}\\ \text{Size 384}\end{matrix}}};
        \draw[->] (1) -- node [midway] {} (2);
    \end{tikzpicture}
    \caption{Details of the CNN block used in all neural networks.}
    \label{cnn-block}
    \end{figure}
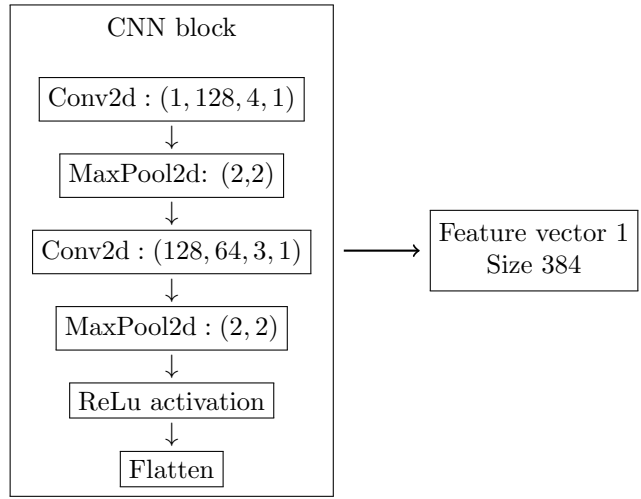
\item \textit{GRU block}: The GRU block is identitfied by the following parameters, as implemented in \texttt{PyTorch}\footnote{\href{https://pytorch.org/docs/stable/generated/torch.nn.GRU.html}{https://pytorch.org/docs/stable/generated/torch.nn.GRU.html}}
\beq
\texttt{(input\_size, hidden\_size, num\_layers)}. \label{gru-params}
\eeq

Given a batch of input sequences of dimensions $(N, T, D)$ where $N$ is the batch size, $T$ is the sequence length, $D$ is the number of features,
the \texttt{input\_size} parameter of the GRU is taken to be $D$, the \texttt{hidden\_size} parameter (typically denoted as $M$) and \texttt{num\_layers} (typically denoted as $L$) are parameters that can be chosen to optimize the obtained results. The output feature vector size\footnote{Taking into account the batch size $N$, the hidden state $h$ has dimension $(L, N, M)$, while the output $y$ has the dimension $(N, T, M)$.  The feature vector, which is the output of interest to us,  is the $y$ value at the last time step $t=T$, so the output has dimension $(N,M)$.} is always the same as $M$. In this case, with the input being $16\times 20$ matrices, we have $T = 16$, $D = 20$, and $M$ is varied and $L$ is chosen to be 2. The schematic of this block is shown in Fig.\ref{gru-block}.
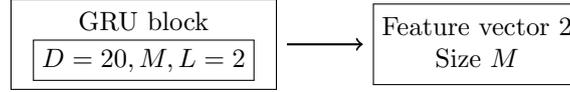
\begin{figure}[h]
\centering
\begin{tikzpicture}[node distance = 1cm, thick]%
        \node (1) {\boxed{
\begin{array}{c}
\text{GRU block}\\
\boxed{D=20, M, L = 2}
\end{array}}};
        \node (2) [right=of 1] {\boxed{
\begin{matrix}\text{Feature vector 2}\\ \text{Size $M$}\end{matrix}}};
        \draw[->] (1) -- node [midway] {} (2);
    \end{tikzpicture}
    \caption{Details of the GRU block used in all CNN-GRU neural networks.}
 \label{gru-block}
    \end{figure}
\item \textit{After the feature extractor}: we concatenate the two feature vectors - `Feature vector 1' from the CNN block of size 384 and `Feature vector 2' from the GRU block of size $M$- into a single feature vector of size ($M+384)$. 
The new feature vector is passed through a feed forward network consisting of two dense layers of sizes $(M+384, 1024)$ and (1024, 4) with no activation. The final  output is  a vector of size 4 representing the four Hodge numbers $h^{(1,1)}, h^{(2,1)}, h^{(3,1)},h^{(2,2)}$.
\end{itemize}
In this work, we experimented with various $M$ values to observe its impacts on the results while maintaining the moderately small size of the neural networks for fast training time on a GPU. First, the GRU hidden size $M$ is chosen to be the same as the CNN block output feature vector, $M = 384$. This first CNN-GRU model is named CNN-GRU-384. One more model was considered for $M = 416$, leading to the model CNN-GRU-416.

\subsubsection{CNN-LSTM hybrid}
The second type of model in the family of CNN-RNN hybrid models is CNN-LSTM.
The architecture of this network is identical to the CNN-GRU network (Fig.\ref{cnn-gru}) described in the previous section, with the only exception being the replacement of the GRU block by the LSTM block. Mathematically, LSTM is more complex than GRU and LSTM has been shown to outperform GRU in the majority of cases due to its complexity (as detailed in the previous section).  The schematic for the network architecture (Fig.\ref{cnn-lstm}) remains essentially the same with the only difference being the LSTM block replacing the GRU block.
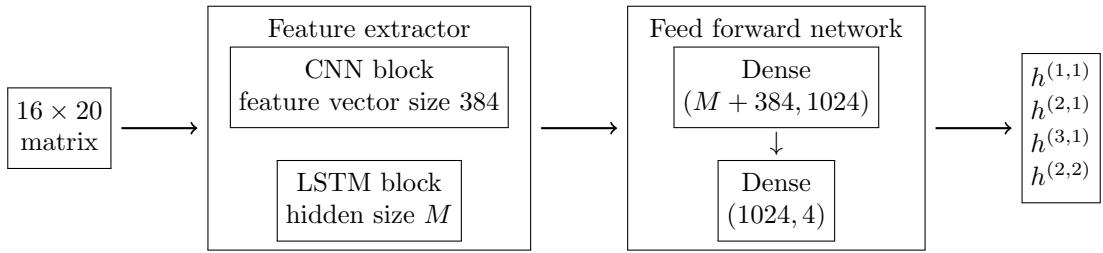
\begin{figure}[H]
\centering
\begin{tikzpicture}[node distance = 1cm, thick]%
        \node (1) {\boxed{\begin{matrix}16\times 20 \\\text{matrix}\end{matrix}}};
        \node (2) [right=of 1] {\boxed{\begin{array}{c}
\text{Feature extractor}\\
\boxed{\begin{matrix}\text{CNN block} \\ \text{feature vector size 384}\end{matrix}} \\\\
\boxed{\begin{matrix}\text{LSTM block} \\ \text{hidden size $M$}\end{matrix}}
\end{array}}};
        \node (3) [right=of 2] {\boxed{\begin{array}{c}\text{Feed forward network}\\\boxed{\begin{matrix}\text{Dense} \\ (M+384,1024)\end{matrix}}
\\\downarrow\\
\boxed{\begin{matrix}\text{Dense}\\ (1024,4)\end{matrix}}\end{array}}};
        \node (4) [right=of 3] {\boxed{\begin{matrix} h^{(1,1)} \\ h^{(2,1)}\\h^{(3,1)} \\ h^{(2,2)}\end{matrix}}};
        \draw[->] (1) -- node [midway] {} (2);
        \draw[->] (2) -- node [midway] {} (3);
        \draw[->] (3) -- node [midway]{} (4);
    \end{tikzpicture}
    \caption{A schematic of the architecture of CNN-LSTM hybrid neural networks.}
    \label{cnn-lstm}
    \end{figure}
Since the only difference from the previously described network involves the new LSTM block with all other parts remain the same, we will only describe this new component.
The LSTM layer\footnote{\href{https://pytorch.org/docs/stable/generated/torch.nn.LSTM.html}{https://pytorch.org/docs/stable/generated/torch.nn.LSTM.html}} is implemented in the same way as the GRU layer in \texttt{PyTorch}, hence it has the same parameters as explained in (\ref{gru-params}).
\begin{figure}[H]
\centering
\begin{tikzpicture}[node distance = 1cm, thick]%
        \node (1) {\boxed{
\begin{array}{c}
\text{LSTM block}\\
\boxed{D=20, M, L = 2}
\end{array}}};
        \node (2) [right=of 1] {\boxed{
\begin{matrix}\text{Feature vector 2}\\ \text{Size $M$}\end{matrix}}};
        \draw[->] (1) -- node [midway] {} (2);
    \end{tikzpicture}
    \caption{Details of the GRU block used in all CNN-LSTM neural networks.}
 \label{lstm-block}
    \end{figure}
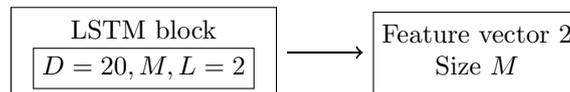
Just as the CNN-GRU case above, we consider various values for the hidden size $M$ of the LSTM block of the CNN-LSTM networks. Our first choice of $M$ is again the same as the CNN block output feature vector, $M = 384$. This first CNN-LSTM model is named CNN-LSTM-384. Three more models were considered for $M=256$, $M = 400$ and $M = 416$, leading to the models CNN-LSTM-256, CNN-LSTM-400, CNN-LSTM-416. All models have moderately small size not exceeding $3.5\times 10^6$ parameters.
\subsubsection{ResNet-RNN hybrid}
The third type of model in the CNN-RNN hybrid models that we will consider is called ResNet-RNN (where the RNN block can be either GRU or LSTM). This model is an experimentation to gauge the performance of those CNN variants with a residual connection (inspired by the ResNet model of \cite{resnet-15}) in extracting features usable for this problem at hand .
The architecture of this model is the same as that depicted in Fig.\ref{cnn-gru} or Fig.\ref{cnn-lstm}, with the only difference being the CNN block that is now replaced by a ResNet-inspired block. The overall architecture is shown in Fig.\ref{resnet-gru}.
\begin{figure}[H]
\centering
\begin{tikzpicture}[node distance = 1cm, thick]%
        \node (1) {\boxed{\begin{matrix}16\times 20 \\\text{matrix}\end{matrix}}};
        \node (2) [right=of 1] {\boxed{\begin{array}{c}
\text{Feature extractor}\\
\boxed{\begin{matrix}\text{ResNet block} \\ \text{feature vector size 256}\end{matrix}} \\\\
\boxed{\begin{matrix}\text{GRU/LSTM block} \\ \text{hidden size $M$}\end{matrix}}
\end{array}}};
        \node (3) [right=of 2] {\boxed{\begin{array}{c}\text{Feed forward network}\\\boxed{\begin{matrix}\text{Dense} \\ (M+256,1024)\end{matrix}}
\\\downarrow\\
\boxed{\begin{matrix}\text{Dense}\\ (1024,4)\end{matrix}}\end{array}}};
        \node (4) [right=of 3] {\boxed{\begin{matrix} h^{(1,1)} \\ h^{(2,1)}\\h^{(3,1)} \\ h^{(2,2)}\end{matrix}}};
        \draw[->] (1) -- node [midway] {} (2);
        \draw[->] (2) -- node [midway] {} (3);
        \draw[->] (3) -- node [midway]{} (4);
    \end{tikzpicture}
    \caption{A schematic diagram of the ResNet variant of the CNN-RNN hybrid neural networks.}
\label{resnet-gru}
    \end{figure}
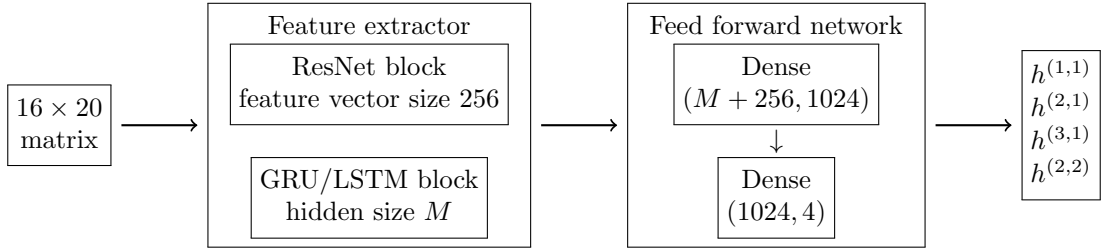

The ResNet block has two branches, a short-cut branch and a convolution branch whose outputs are added together to form a single output (Fig. \ref{resnet}).
\begin{figure}[H]
\centering
\begin{tikzpicture}[node distance = 2cm, thick]%
        \node (1) {\boxed{\begin{matrix}16\times 20 \\\text{matrix}\end{matrix}}};
        \node (2) [right=of 1] {\boxed{\begin{matrix}\text{Convolution} \\ \text{branch}\end{matrix}}};
        \node (4) [below=1cm of 2] {\boxed{\begin{matrix}\text{Short-cut} \\\text{branch}\end{matrix}}};
        \node (3) [right=of 2] {\boxed{\begin{matrix}\text{Feature}\\\text{vector $A$}\\ \text{(size 256)}\end{matrix}}};
        \node (5) [right=of 4] {\boxed{\begin{matrix}\text{Feature}\\\text{vector $B$} \\ \text{(size 256)}\end{matrix}}};
        \node (6) [right=of 3] {};
        \node (7) [below=0.5cm of 6] {\boxed{\begin{matrix}\text{ResNet feature vector}\\ \text{(size 256)}\end{matrix}}};
        \draw[->] (1) -- node [midway,above] {} (2);
        \draw[->] (1) to [bend right] node [midway,below]{} (4);
        \draw[->] (2) -- node [midway,above] {} (3);
        \draw[->] (4) -- node [midway,above]{} (5);
        \draw[->] (5) to [bend right] node [midway,below]{} (7);
        \draw[->] (3) to [bend left] node [midway,above]{} (7);
    \end{tikzpicture}
    \caption{A schematic diagram of the convolution branch in the ResNet variation of the CNN block.}
    \label{resnet}
    \end{figure}
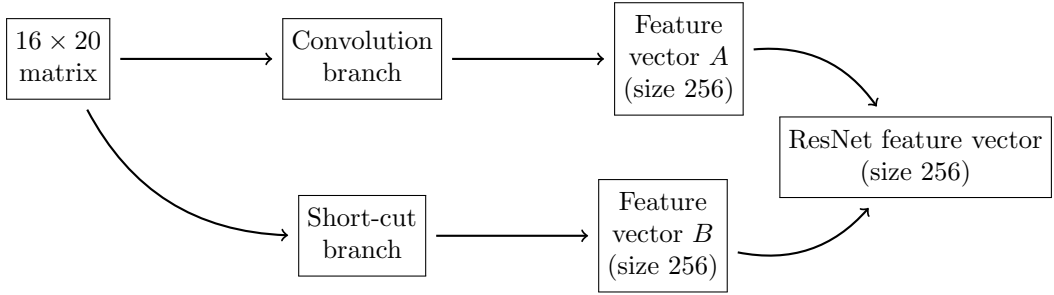
\begin{itemize}
\item The convolution branch (see Fig.\ref{convo-branch}) has three layers of convolutions, with the first two followed by a max pooling layer and a \texttt{ReLU} activation. The last convolution layer is followed by a \texttt{Flatten} layer. The output of the convolution branch is a feature vector (called `feature vector A') of size 256.
\begin{figure}[H]
\centering
\begin{tikzpicture}[node distance = 1cm, thick]%
        \node (1) {\boxed{
\begin{array}{c}
\text{Convolution branch}\\\\
\boxed{\text{Conv2d}: (1, 128,4,1)}\\
\downarrow\\
\boxed{\text{MaxPool2d: (2,2)}}\\
\downarrow\\
\boxed{\text{ReLU activation}}\\
\downarrow\\
\boxed{\text{Conv2d}: (128, 64, 3,1)}\\
\downarrow\\
\boxed{\text{MaxPool2d}: (2,2)}\\
\downarrow\\
\boxed{\text{ReLU activation}}\\
\downarrow\\
\boxed{\text{Conv2d}: (64, 128, 2,1)}\\
\downarrow\\
\boxed{\text{Flatten}}
\end{array}}};
        \node (2) [right=of 1] {\boxed{
\begin{matrix}\text{Feature vector A}\\ \text{Size 256}\end{matrix}}};
        \draw[->] (1) -- node [midway] {} (2);
    \end{tikzpicture}
    \caption{Details of the convolution branch of the ResNet block used in ResNet-RNN hybrid neural networks.}
 \label{convo-branch}
    \end{figure}
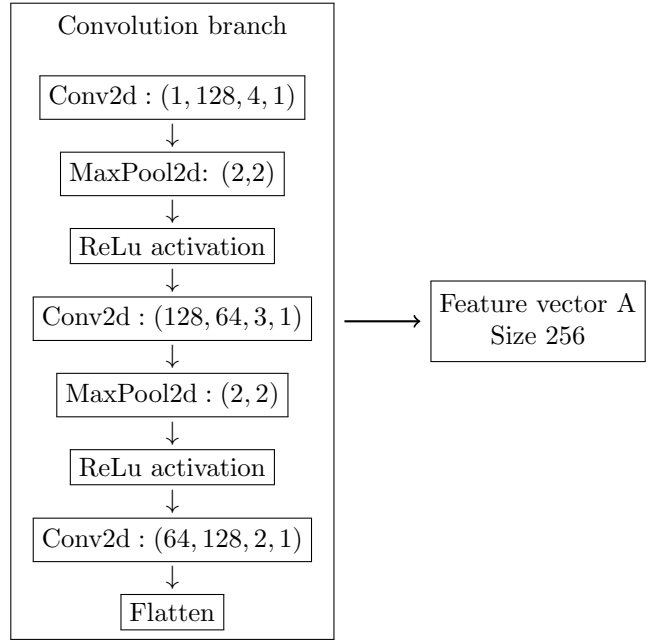
\item The short-cut branch (see Fig.\ref{short-cut}) consists of a \texttt{Flatten} layer followed by a \texttt{Dense} layer with size (320, 256) (Fig \ref{short-cut}). The output of the shortcut branch is a feature vector (called `feature vector B') of size 256.
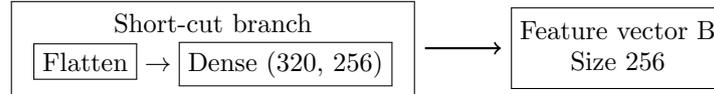
\begin{figure}[H]
\centering
\begin{tikzpicture}[node distance = 1cm, thick]%
        \node (1) {\boxed{
\begin{array}{c}
\text{Short-cut branch}\\
\boxed{\text{Flatten}}\rightarrow \boxed{\text{Dense (320, 256)}}
\end{array}}};
        \node (2) [right=of 1] {\boxed{
\begin{matrix}\text{Feature vector B}\\ \text{Size 256}\end{matrix}}};
        \draw[->] (1) -- node [midway] {} (2);
    \end{tikzpicture}
    \caption{Details of the short-cut branch  used in the ResNet block of the ResNet-RNN neural networks.}
\label{short-cut}
    \end{figure}
\item The output of the convolution branch, `feature vector $A$ 'of size 256, is added to the output of the short-cut branch - `feature vector $B$'' of size 256, to form a new feature vector that will eventually be concatenated with the feature vector coming from the RNN block to form a final feature vector of size $256+M$, as shown in Fig.\ref{resnet}.
\item For this work, we consider the GRU or LSTM  hidden size $M$ to be 256 (which is the same size as the feature vector of the ResNet part) and 400. These choices of $M$ lead to the models ResNet-GRU/LSTM-256 and ResNet-GRU/LSTM-400.
\end{itemize}
\subsection{LSTM-based neural networks}
We now move on to the second category of neural networks used in this study. These neural networks are purely RNN-based (using LSTM) with no CNN involved. The overall architecture of this network still consists of a feature extractor and a feed forward network as shown in Fig.\ref{lstm-based}, but the feature vector now comes entirely from the recurrent operation.
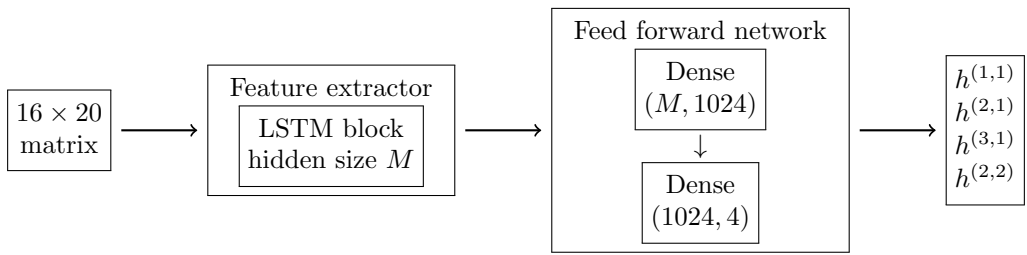
\begin{figure}[H]
\centering
\begin{tikzpicture}[node distance = 1cm, thick]%
        \node (1) {\boxed{\begin{matrix}16\times 20 \\\text{matrix}\end{matrix}}};
        \node (2) [right=of 1] {\boxed{\begin{array}{c}
\text{Feature extractor}\\
\boxed{\begin{matrix}\text{LSTM block} \\ \text{hidden size $M$}\end{matrix}}
\end{array}}};
        \node (3) [right=of 2] {\boxed{\begin{array}{c}\text{Feed forward network}\\\boxed{\begin{matrix}\text{Dense} \\ (M,1024)\end{matrix}}
\\\downarrow\\
\boxed{\begin{matrix}\text{Dense}\\ (1024,4)\end{matrix}}\end{array}}};
        \node (4) [right=of 3] {\boxed{\begin{matrix} h^{(1,1)} \\ h^{(2,1)}\\h^{(3,1)} \\ h^{(2,2)}\end{matrix}}};
        \draw[->] (1) -- node [midway] {} (2);
        \draw[->] (2) -- node [midway] {} (3);
        \draw[->] (3) -- node [midway]{} (4);
    \end{tikzpicture}
    \caption{A schematic diagram of the architecture of the LSTM-based neural networks.}
\label{lstm-based}
    \end{figure}
As before, we experimented with different choices of the hidden size $M$ of the LSTM-based networks. In choosing the suitable $M$, one of our considerations is that the size of the final network should be moderately small for a reasonable training time. Another consideration is that $M$ and the total network parameters should roughly be in the same range as those used in the hybrid case for the ease of performance comparison among different models. As such, our choices of $M=400, 424, 448, 456$ lead to the models LSTM-400, LSTM-424, LSTM-448, LSTM-456.
\section{Training results (using 72\% dataset)} \label{results}
In this section, we report the results obtained after the first round of training in which all twelve models was trained on the 72\%  dataset. Our training is done entirely on \texttt{Kaggle} (an AI/ML competition hosting platform) - \href{https://www.kaggle.com}{https://www.kaggle.com} - using the P100 GPU unit, which comes with a time usage limit\footnote{of 12 hours per notebook runtime session, (see \href{https://www.kaggle.com/docs/notebooks}{https://www.kaggle.com/docs/notebooks})
as well as a maximum of 30 hours per week, renewable at the start of each week. See \href{https://www.kaggle.com/docs/efficient-gpu-usage}{https://www.kaggle.com/docs/efficient-gpu-usage}}.
All networks were trained using \texttt{HuberLoss}\footnote{\href{https://pytorch.org/docs/stable/generated/torch.nn.HuberLoss.html}{https://pytorch.org/docs/stable/generated/torch.nn.HuberLoss.html}}, \texttt{AdamW} optimizer\footnote{\href{https://pytorch.org/docs/stable/generated/torch.optim.AdamW.html}{https://pytorch.org/docs/stable/generated/torch.optim.AdamW.html}} and a batch size of 128.  The use of \texttt{HuberLoss} for training ensures a robust regression performance in the presence of outliers since \texttt{HuberLoss} combines the advantages of both the Mean Squared Error loss and the Mean Absolute Error (MAE) loss by using a $\delta$ parameter as follows:
\beq
L = \begin{dcases} \frac{1}{2} \left( y_\text{pred} - y_\text{target}\right)^2, & |y_\text{pred} - y_\text{target}| < \delta
\\
\delta\left(|y_\text{pred} - y_\text{target}| - \frac{1}{2}\,\delta \right), & |y_\text{pred} - y_\text{target}| > \delta
\end{dcases} \label{hloss}
\eeq
where $\delta = 1$ (the default parameter),  $y_\text{pred}$ is the model's output (a vector of size four), $y_\text{target}$ is the actual target variables to be predicted (a vector of size four containing the four Hodge numbers)\footnote{Since our training task is a multi-regression, the actual loss per data sample during the training is the mean value of the four individual losses for the four Hodge numbers.}.
\\\\
Furthermore, we used the learning rate scheduler \texttt{ReduceLROnPlateau}\footnote{\href{https://pytorch.org/docs/stable/generated/torch.optim.lr_scheduler.ReduceLROnPlateau.html}{https://pytorch.org/docs/stable/generated/torch.optim.lr\_scheduler.ReduceLROnPlateau.html}} which automatically reduces the learning rate if the training stage hits a plateau for more than 10 epochs. The starting learning rate is fixed to 0.01, with a decay factor of 0.1. During the training process, we observed that most of the networks had a `preferred' learning rate of 0.001 that was kept for the longest duration (in the order of hundreds of epochs). Once the learning rate dropped to $10^{-4}$, the losses only decreased for a few epochs before the learning rate was adjusted to $10^{-5}$, at which point the network had practically converged.
We used a maximum of 550 epochs for training, but the actual number of training epochs varied for different models due to their different sizes. With an early stopping mechanism built in to stop the training if the validation loss does not improve for 20 epochs (when the model has achieved convergence), training was automatically stopped and the network parameters were saved at the best possible validation loss value. All models, with the exception of CNN-LSTM-416, achieved convergence well before 550 epochs.  
\\\\
The total number of parameters and total training time in hours for all networks are recorded in Table \ref{params}.
\begin{table}[H]
\centering
\begin{tabular}{lcc}
\hline\hline
& \text{Parameters ($\times 10^6$)} & \text{Total training time (hrs)}
\\ \hline\hline
\text{CNN-GRU-384} &  2.222  & 4.16 \\
\text{CNN-GRU-416} &    2.488   &  4.66 \\
\text{ResNet-GRU-256} &  1.329  &  3.42\\
\text{ResNet-GRU-400} & 2.337 & 5.86 \\
\text{CNN-LSTM-256}  &  1.547  & 4.00\\
\text{CNN-LSTM-384} &  2.674   & 4.77\\
\text{CNN-LSTM-400}& 2.842  & 6.37\\
\text{CNN-LSTM-416} & 3.017  &  8.82\\
\text{LSTM-400} &  2.373 &  3.77\\
\text{LSTM-424} &   2.637 & 4.17 \\
\text{LSTM-448} &   2.915& 4.83\\
\text{LSTM-456} & 3.011 & 4.45 \\
\hline\hline
\end{tabular}
\caption{Parameters and total training time of all models used in this work. Two models, ResNet-LSTM-256 and ResNet-LSTM-400 that were initially part of the models planned for training, are not included in this Table since their training was terminated early due to the very slow decrease of their training/validation losses compared to the rest of the models.}
\label{params}
\end{table}
It is important to note that the number of parameters for the CNN block in the CNN-RNN hybrid models is much smaller than that for the RNN block. More specifically, in the models utilizing the conventional CNN block, the total number of CNN parameters is roughly $7.6 \times 10^4$ (75,968), while for those models utilizing the ResNet-style CNN block, the total number of CNN parameters is roughly $1.9 \times 10^5$ (191,168). In the majority of the cases, these parameters account for less than 10\% of the total network parameters. Hybrid models using the ResNet block takes longer to train compared to those using the conventional CNN block. In particular, the training of both ResNet-LSTM-256 and ResNet-LSTM-400 was terminated early due to the extremely slow decrease in both the training and validation losses of these models compared to the rest, which is indicative of a much longer overall training time with a possibly high loss at the point of convergence. As such, these models are not counted in the total number of models explored in this work. 
\subsection{CNN-GRU hybrid neural networks} \label{cnn-gru-models-res}
In this section, we report the training results for the four CNN-GRU models: CNN-GRU-384, CNN-GRU-416, ResNet-GRU-256, and ResNet-GRU-400. 
\begin{itemize}
\item The obtained test accuracies for the 4 Hodge numbers for the four CNN-GRU hybrid networks are given in Table \ref{cnn-gru-acc}.
\begin{table}[h]
\begin{center}
\begin{tabular}{|c|c|c|c|c|}
\hline
\,\,\,\, & $h^{(1,1)}$ & $h^{(2,1)}$ & $h^{(3,1)}$ & $h^{(2,2)}$
\\ \hline
\text{CNN-GRU-384} &  98.24& 84.85& 69.29& 31.73\\
\text{CNN-GRU-416} & 98.25& 88.64& 77.69& 45.37\\
\text{ResNet-GRU-256} & 97.87&  84.05 &  69.54 & 37.14 \\
\text{ResNet-GRU-400}& $\mathbf{99.22}$ &  $\mathbf{92.46}$ &  $\mathbf{86.62}$ & $\mathbf{59.44}$\\
\hline
\end{tabular}
\caption{Obtained accuracies of CNN-GRU models during inference on the test dataset}\label{cnn-gru-acc}
\end{center}
\end{table}
\item The best performing models among the four variants of CNN-GRU type is ResNet-GRU-400 (results noted in bold), while the worst performing model is CNN-GRU-384. A noteworthy trend of the results in the Table \ref{cnn-gru-acc} is that all CNN-GRU models do relatively well in predicting $h^{1,1}$ (97.8\% to 99\% accuracy) and $h^{2,1}$ (84\% to 92 \%), they do worse when it comes to predict $h^{3,1}$ (69.2 \% to 86.6\%), and much worse for the case of $h^{2,2}$ (31.7\% to 59.4\%).

\item Increasing the size of the hidden dimension $M$ of the GRU unit led to a significant improvment in the results, as the obtained accuracies increase when $M$ goes from 384 to 416 for CNN-GRU and from 256 to 400 for ResNet-GRU. Furthermore, as the convolutional part goes, the ResNet unit used seems to outperform the CNN unit, judging from the better performance of ResNet-GRU-256 (with $1.33\times 10^6$ parameters) compared to CNN-GRU-384 (with $2.22\times 10^6$ parameters), and of ResNet-GRU-400 (with $2.34\times 10^6$ parameters) compared to CNN-GRU-416 (with $2.49\times 10^6$ paramters), especially when ResNet-based hybrid networks have fewer parameters than their CNN-based counterparts.

\item For comparison among all CNN-GRU models, see Fig.\ref{cnn-gru-all} which shows the validation accuracies obtained during training for all models in logarithmic scale. The training curves showing both training and validation losses for individual CNN-GRU models can be found in Fig.\ref{cnn-gru-results}.
\begin{figure}[H]
\centering
\includegraphics[width = 0.65\textwidth]{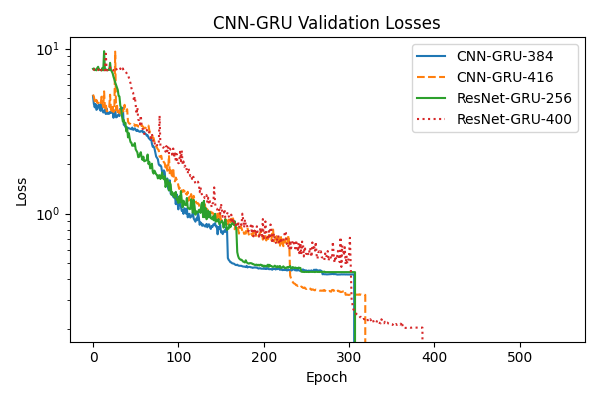}
\caption{Validation losses of all CNN-GRU models (in log scale).}\label{cnn-gru-all}
\end{figure}
\end{itemize}
\subsection{CNN-LSTM hybrid neural networks} \label{cnn-lstm-models-res}
In this section, we report the training results for the four CNN-LSTM models: CNN-LSTM-256, CNN-LSTM-384, CNN-LSTM-400, and CNN-LSTM-416. While the first three CNN-LSTM models achieved convergence during the training phase, the last model, CNN-LSTM-416 did not converge, even after 550 epochs. The results obtained by CNN-LSTM-416 are still included for reference in this section, but these would not be used in later sections for comparison with other models.  
\begin{itemize}
\item The obtained test accuracies for the 4 Hodge numbers for the four CNN-LSTM hybrid networks are given in Table \ref{cnn-lstm-acc}.
\begin{table}[h]
\begin{center}
\begin{tabular}{|c|c|c|c|c|}
\hline
\,\,\,\, & $h^{(1,1)}$ & $h^{(2,1)}$ & $h^{(3,1)}$ & $h^{(2,2)}$
\\ \hline
\text{CNN-LSTM-256} &  99.31 & 94.37 & 88.61&62.33\\
\text{CNN-LSTM-384}  &  99.13 & 94.77& 87.85 &  62.32\\
\text{CNN-LSTM-400}  &  $\mathbf{99.74}$ & $\mathbf{98.07}$ & $\mathbf{95.19}$ & $\mathbf{81.01}$\\
\text{CNN-LSTM-416}  &  92.15 & 77.59 & 51.62 & 16.37\\
\hline
\end{tabular}
\caption{Obtained accuracies of different CNN-LSTM models during inference on the test dataset.}\label{cnn-lstm-acc}
\end{center}
\end{table}
\item CNN-LSTM hybrid models provide a big improvement in the obtained accuracies for all 4 Hodge numbers compared to their CNN-GRU counterparts, with the exception of CNN-LSTM-416. This is evident from the $h^{1,1}$ accuracy being in around 99\%, the $h^{2,2}$ accuracy being in the range $94\% - 98\%$, $h^{3.1}$ accuracy being in the range $88\% - 95\%$, and the $h^{2,2}$ accuracy being in the range $62-81\%$.

\item The best performing CNN-LSTM hybrid model is CNN-LSTM-400 (with the train/test results noted in bold font). This is also the best model out of the 12 models considered in this work. A notable trend of these models' results is the relatively stable performance observed at $M = 256$ and $M=284$. Peak performance is attained at $M=400$. When increasing $M$ further to 416, the performance drops dramatically due to very slow convergence. In fact, the worst performing model of all models considered is CNN-LSTM-416, which also has the largest number of parameters ($3.017\times 10^6$). While all models achieved convergence after an average of around 300 epochs, CNN-LSTM-416's training still progressed on at epoch 550 (the cutoff). If training was allowed to continue, the model would presumably continue to improve, but due to the time constraint with Kaggle's GPU usage, it was not possible to extend the training time. 

\item The obviously superior performance of LSTM-based hybrid networks over GRU-based ones can be seen when comparing the performances of CNN-LSTM-256 and CNN-LSTM-384 versus those of CNN-GRU-384 and CNN-GRU-416 (Fig.\ref{lstm-vs-gru}). In particular, it is noted that the lighter CNN-LSTM-256 (at $1.55\times 10^6$ parameters) outperforms both the heavier CNN-GRU-384 ($2.22\times 10^6$) and CNN-GRU-416 ($2.49\times 10^6$). This is a clear indication of the efficiency of LSTM architecture over GRU's in this task. In general, when comparing models with different underlying architectures, what matters most is the differences in the architectures rather than the number of parameters, since it is the architecture that determines how well a model can extract and learn from useful features\footnote{While it might be fair to compare the performances of models from the same family (with the same underlying architectures) using solely their number of parameters, the same cannot be said for models from different families with different underlying architectures.}. In this case, LSTMs with their more sophisticated double recurrent loops have proven better than their simpler GRU counterparts. 
\begin{figure}[H]
\centering
\includegraphics[width = 0.8\textwidth]{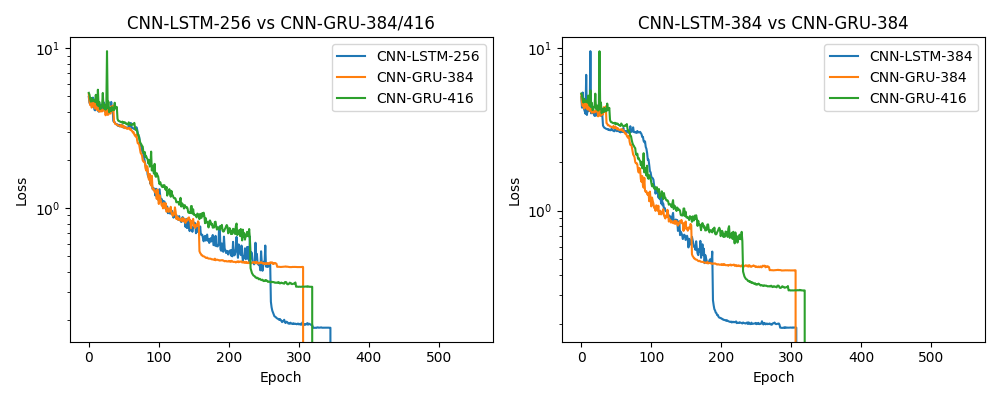}
\caption{Validation accuracies of CNN-LSTM-256 and CNN-LSTM-384 versus CNN-GRU-384/416 networks.}\label{lstm-vs-gru}
\end{figure}

\item For comparison among the six CNN-LSTM models, see Fig.\ref{cnn-lstm-all-results} which shows the validation losses (log scale) of all CNN-LSTM models.  The training curves showing training and validation losses for each individual CNN-LSTM model can be found in Fig.\ref{cnn-lstm-all} (in Appendix \ref{curves}).
\begin{figure}[H]
\centering
\includegraphics[width = 0.65\textwidth]{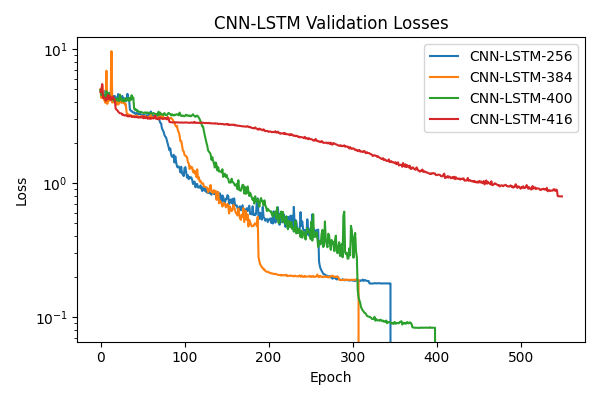}
\caption{Validation losses of all CNN-LSTM models (log scale). The lowest loss value (the green curve) corresponds to the best model CNN-LSTM-400. The red curve, corresponding to CNN-LSTM-416, did not drop to zero but ended at a very high value at the end of epoch 550 (signaling a failure to converge). }\label{cnn-lstm-all-results}
\end{figure}
\eni


\subsection{LSTM-based neural networks} \label{lstm-models-res}
In this section, we report the training results for the four LSTM models: LSTM-400, LSTM-424, LSTM-448 and LSTM-456. 
\begin{itemize}
\item The obtained test accuracies for the 4 Hodge numbers for the different LSTM-based networks are given in Table \ref{lstm-acc}. 
\begin{table}[h]
\centering
\begin{tabular}{|c|c|c|c|c|}
\hline
\,\,\,\, & $h^{(1,1)}$ & $h^{(2,1)}$ & $h^{(3,1)}$ & $h^{(2,2)}$
\\ \hline
\text{LSTM-400} & 99.38 & 94.14 & 88.01 & 65.05\\
\text{LSTM-424} & 99.56&  97.07 & 93.19& 74.47\\
\text{LSTM-448}  &  $\mathbf{99.74}$ & $\mathbf{97.51}$ &  $\mathbf{94.24}$ & $\mathbf{78.63}$ \\
\text{LSTM-456} & 99.35 & 94.01 & 87.78&  64.58\\
\hline
\end{tabular}
\caption{Obtained accuracies of differet LSTM-based models during inference on the test set}\label{lstm-acc}
\end{table}
\item Purely recurrent, LSTM-based model have very similar performance to CNN-LSTM hybrid models, as is evident from the results in Table \ref{lstm-acc} above. For the four LSTM models trained, the $h^{1,1}$ accuracy is at least 99\%, the $h^{2,2}$ accuracy is at least 94\%, the $h^{3,1}$ accuracy is in the range 88\%-94\%, while the $h^{2,2}$ accuracy is in the range 65\%-78\%. More specifically, LSTM-400 (with $2.37\times 10^6$ parameters) has almost the same performance as CNN-LSTM-256 (with $1.54\times 10^6$ parameters) and CNN-LSTM-384 (with $2.67\times 10^6$ parameters), while LSTM-448 (with $2.92\times 10^6$ parameters) performs almost as well as CNN-LSTM-400 (with $2.84\times 10^6$ parameters). This seems to suggest that, as far as feature extractor goes, recurrent unit performs at the same level as convolutional unit, at least for this specific task of extracting useful features from the CICY4 configurational matrices to predict the Hodge numbers.

\item In the category of LSTM-based model, the best performing model is LSTM-448 (results noted in bold), while the worst-performing model is LSTM-456.
LSTM-448 and LSTM-424 are the second and third best performing model out of all 12 models considered in this work. A rise in performance of the LSTM-based models is observed with an increase in the size of $M$, from $M =400$ to 424, peaking at  448,  before dropping when $M$ is increased further to 456.

\item For comparison among the four LSTM-based models, the validation losses of all four models are shown in Fig.\ref{lstm-all-results} in logarithmic scale.
\begin{figure}[H]
\centering
\includegraphics[width =0.65\textwidth]{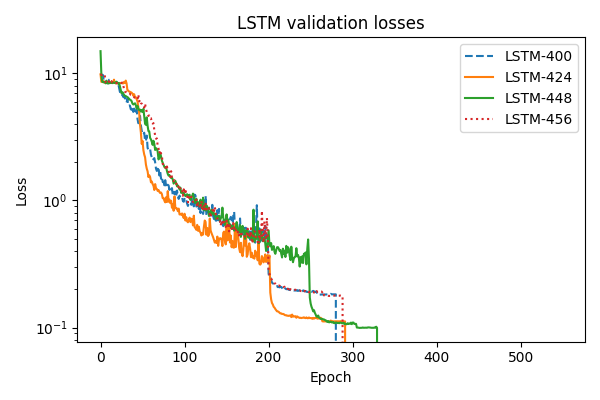}
\caption{Validation losses of all LSTM models (log scale). In the plot, the green curve with the smallest loss corresponds to LSTM-448.}\label{lstm-all-results}
\end{figure}
\item The training curves of individual LSTM-based models showing training and validation losses can be found in Fig.\ref{lstm-all-models}.
\end{itemize}
\subsection{Ensembles of best models} \label{ensembles-}
Forming ensemble is a common technique to improve the overall performance, especially when there are several trained models ready to be deployed.
Due to the fact that each model has extracted their own useful features and  thus learnt a different representation of the data, they might have different performance strengths in different parts of the data distribution. For instance, a model that performs well on a certain range  of the data might not do well on a different range of the same data. In other words, each model has its own region of competence. In this sense, the diversity of model architectures in the ensemble potentially leads to a more robust performance and better generalization on the entire data set. In an ensemble, the predictions of all the constituent models are averaged, either evenly by taking the mean - as is the case in this section, or unevenly by taking an weighted average. 
\\\\
For the task at hand, the specific ensembles that we considered are listed in Table \ref{ensemble}, together with their associated test accuracies (best results are listed in bold).
\begin{table}[h]
\centering
\begin{tabular}{|c|l|c|c|c|c|}
\hline
Ensemble & Composition &$h^{(1,1)}$ & $h^{(2,1)}$ & $h^{(3,1)}$ & $h^{(2,2)}$
\\ \hline
Ens-1 & $\begin{array}{l}\text{LSTM-448}\\\text{CNN-LSTM-400}\end{array}$
 & $\mathbf{99.80}$ & 98.32&  95.72 & 82.93\\ \hline &&&&& \\
Ens-2 &$\begin{array}{l}\text{LSTM-448}\\ \text{LSTM-424}\\\text{CNN-LSTM-400}\end{array}$
& 99.80 & $\mathbf{98.40}$ & $\mathbf{95.80}$ & $\mathbf{83.02}$\\ \hline &&&&&\\
Ens-3 &$\begin{array}{l}\text{LSTM-448}\\ \text{LSTM-424}\\\text{CNN-LSTM-384}\\ \text{CNN-LSTM-400}\end{array}$
&  99.78 & 98.22& 95.22& 80.17\\\hline &&&&&\\
Ens-4& $\begin{array}{l}\text{LSTM-448}\\ \text{LSTM-424}\end{array}$
& 99.71&  97.97&  94.97&  80.15\\
\hline
\end{tabular}
\caption{Test accuracies for various ensembles of several best performing models. While the $h^{1,1}$ accuracy of both Ens-1 and Ens-2 are rounded to 99.80\%,  the unrounded $h^{1,1}$ test accuracy of Ens-1 is 99.804\%, for that of Ens-2 is 99.797\%.} \label{ensemble}
\end{table}
\bei
\item The first ensemble, Ens-1, consists of the two best performing models, CNN-LSTM-400 and LSTM-448.
\item The second ensemble, Ens-2, consists of the three best performing models, CNN-LSTM-400, LSTM-448, and LSTM-424.
\item The third ensemble, Ens-3, consists of the three best performing models, CNN-LSTM-400, LSTM-448, LSTM-424, plus CNN-LSTM-384.
\item The fourth ensemble, Ens-4, consists of the two best performing LSTM-based models,  LSTM-448, LSTM-424.
\eni
The first two ensembles lead to the test accuracy results that are improvements of the best model, CNN-LSTM-400, while the third ensemble has better accuracy than CNN-LSTM-400 only for the first three Hodge numbers $h^{1,1}, h^{2,1}, h^{3,1}$. The fourth and last ensemble does not lead to an improvement in accuracy for any Hodge numbers compared to CNN-LSTM-400, but it does lead to an improvement in acccuracy for three out of four Hodge numbers ($h^{2,1}$, $h^{3,1}, h^{2,2}$) compared to LSTM-448 (which is the best LSTM-based model). 
Interestingly, overall, we do not have an absolute best ensemble for all 4 Hodge numbers. Instead, we have one ensemble (Ens-1) that has the best accuracy for $h^{(1,1)}$ and another ensemble (Ens-2), that has the best accuracies for the remaining three Hodge numbers.
\subsection{Comparison of accuracy} \label{acc}
The train and test accuracies for the prediction of all four Hodge numbers by each of the individual models and the three ensembles considered in this work can be visualized in Fig.\ref{accuracies-4x} and Fig.\ref{accuracies}. 
\\\\
In Fig.\ref{accuracies-4x}, the train and test accuracies for all four Hodge numbers are plotted together with no ordering of the model performance. From the train accuracies in the first subplot of Fig.\ref{accuracies-4x}, it is noteworthy to point out that only individual three models, LSTM-424, LSTM-448 and CNN-LSTM-400 (plus the the three ensembles), achieved almost 100\% accuracy for all 4 Hodge numbers, while the rest of the models do not. Examining the graphs showing the train and test accuracies of all models, the ease of regression for predicting the Hodge numbers is obvious, with $h^{1,1}$ being the easiest, followed by $h^{2,1}$, $h^{3,1}$ and lastly $h^{2,2}$. While all models achieve a consistently high accuracy of nearing 100\% for $h^{1,1}$, or at least 85\% for $h^{2,1}$, their performances vary wildly when it comes to $h^{3,1}$ and $h^{2,2}$, as is evident from the graphs (the green and red lines) in the second subplot of Fig.\ref{accuracies-4x}.
\\\\
In Fig.\ref{accuracies}, the test accuracy for each of the Hodge number is plotted separated to show the ordering of the model performance. 
 It is important to note that different models perform differently when it comes to predicting the four different Hodge numbers. A model that predicts $h^{(1,1)}$ most accurately might not do so when it comes to $h^{(2,1)}$, $h^{(3,1)}$ or $h^{(2,2)}$. Similarly, a model that predicts $h^{(3,1)}$ most accurately might not do so when it comes to the rest of Hodge numbers.
The four best performing models in terms of acurracy for each of the Hodge numbers (in descending order), not including the ensembles, are:
\begin{itemize}
  \item $h^{(1,1)}$:  CNN-LSTM-400 $\ra$ LSTM-448 $\ra$ LSTM-424 $\ra$ LSTM-400\,.
  \item $h^{(2,1)}$: CNN-LSTM-400 $\ra$ LSTM-448  $\ra$ LSTM-424 $\ra$ CNN-LSTM-384\,.
  \item $h^{(3,1)}$: CNN-LSTM-400 $\ra$ LSTM-448  $\ra$ LSTM-424 $\ra$ CNN-LSTM-256\,.
  \item $h^{(2,2)}$: CNN-LSTM-400 $\ra$ LSTM-448 $\ra$ LSTM-424 $\ra$ LSTM-400\,.
\end{itemize}
The four best performing models in terms of acurracy for each of the Hodge numbers (in descending order), including the ensembles, are:
\begin{itemize}
  \item $h^{(1,1)}$: Ens-1  $\ra$ Ens-2$\ra$ Ens-1 $\ra$ CNN-LSTM-400\,.
  \item $h^{(2,1)}$: Ens-2  $\rightarrow$ Ens-1  $\rightarrow$ Ens-3$\ra$ CNN-LSTM-400\,.
  \item $h^{(3,1)}$:  Ens-2  $\rightarrow$ Ens-1  $\rightarrow$ Ens-3 $\ra$ CNN-LSTM-400\,.
    \item $h^{(2,2)}$: Ens-2  $\rightarrow$ Ens-1  $\rightarrow$CNN-LSTM-400  $\ra$ Ens-3\,.
\end{itemize}

In section \ref{metrics}, the model performances are ranked in terms of additional metrics such as MSE (see Table \ref{mse-i-f} and Fig. \ref{mse-i-f}), MAE (see Table \ref{mae-i} and Fig. \ref{mae-i-f}),  $R^2$ (see Table \ref{r2-i} and  Fig. \ref{r2-i-f}), and collectively (see Table \ref{mse-mae-r2-all} and Fig. \ref{scores-all}).

\begin{figure}[h]
\centering
\includegraphics[width = 0.7\textwidth]{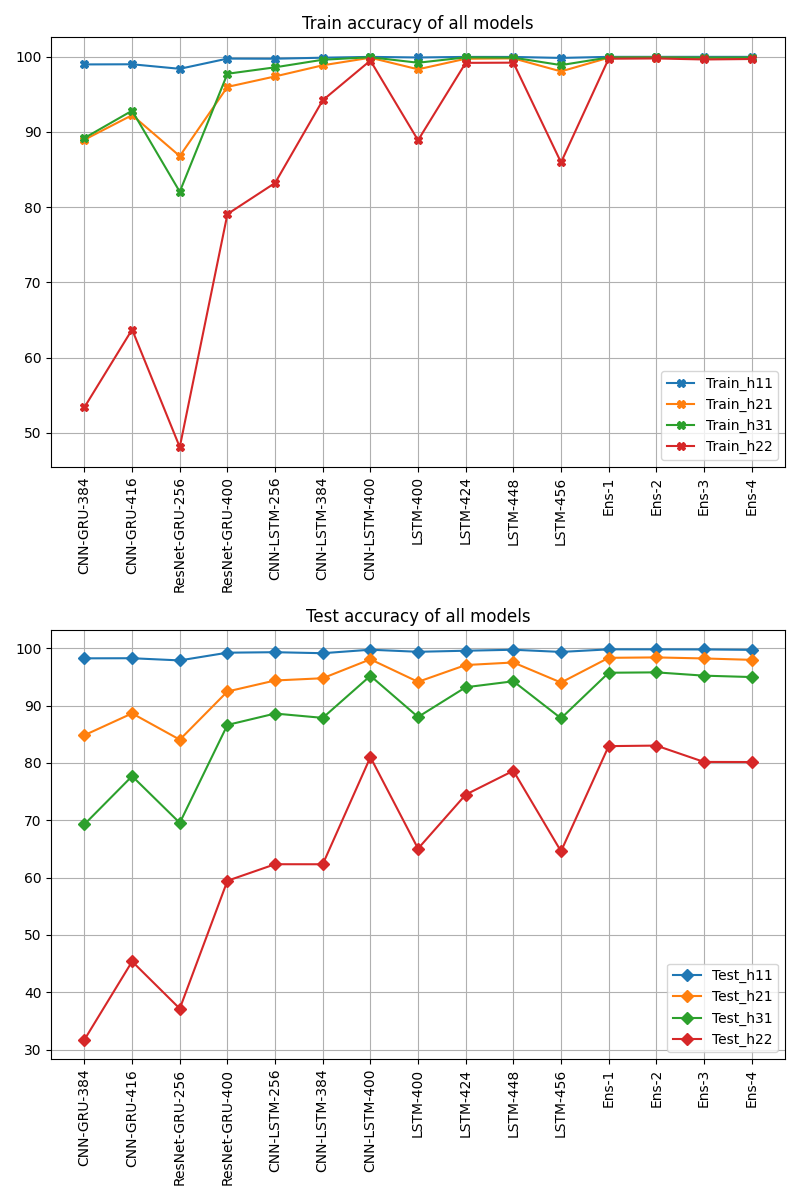}
\caption{Train and test accuracies of all models (including ensembles) evaluated on the 72\% dataset for 4 Hodge numbers.}\label{accuracies-4x}
\end{figure}

\begin{figure}[h]
\centering
\includegraphics[width = 0.75\textwidth]{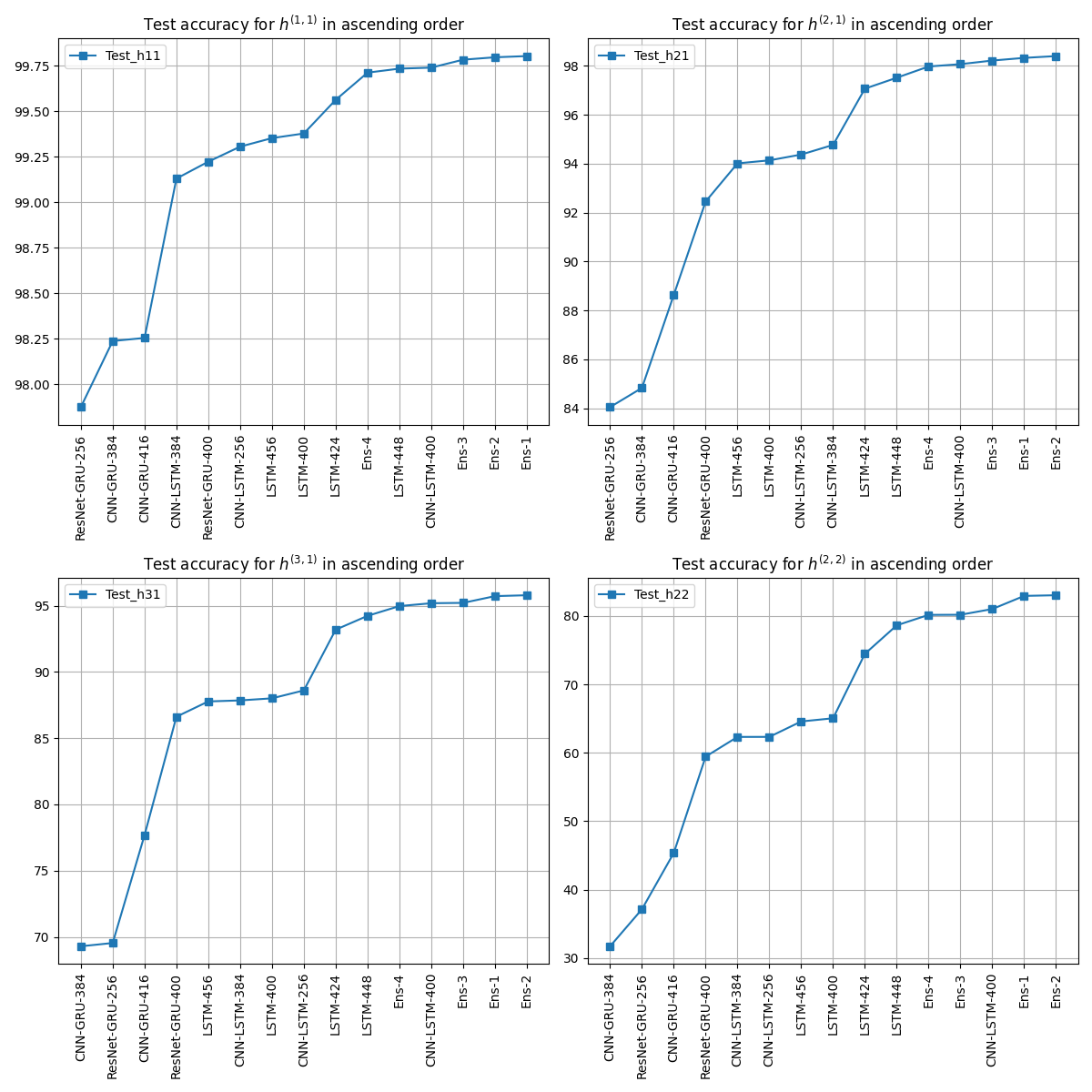}
\caption{Test accuracies of all models (including ensembles) evaluated on the test set of the 72\% dataset for each of the four Hodge numbers in ascending order (with worst to best models arranged from the left to the right in each diagram).}\label{accuracies}
\end{figure}
\newpage
\section{Training results: 5-fold cross validation}  \label{cv5f}
In this section, we report the results obtained from training a single model, LSTM-448, using 5-fold cross validation (CV) on the combined training and validation sets from the 72\% dataset. For the cross-validation experiments, we initially chose two models: CNN-LSTM-400 and LSTM-448, but CNN-LSTM-400 training was terminated early due to its slow convergence on the first fold of data. The training process is the same as that used for training the 12 models of Table \ref{params} and has been described in detail in the previous section. In each fold, a new LSTM-448 model was trained from scratch using the dataset corresponding to that fold (see the discussion on the data preparation in Section \ref{dataset}). For the distributions of the training and validation data in each fold, see Appendix \ref{5fold_cv_hodge_dist}. The final outcomes of the 5-fold CV experiment are the five different LSTM-448 models LSTM-448-f0, LSTM-448-f1, LSTM-448-f2, LSTM-448-f3, and LSTM-448-f4. 
\subsection{Individual models} \label{cv5f-sep}
The test accuracies obtained for the five LSTM-448 models are compiled in Table \ref{5f-sep}. For comparison, the validation losses obtained during the training phase of these models are plotted together in Fig. \ref{lstm-5fold-val}, while the full training curves (with both the training and validation losses) of each model are shown in Fig.\ref{lstm-5fold-models} in Appendix \ref{curves}.
\begin{table}[H]
\centering
\begin{tabular}{|c|c|c|c|c|c|}
\hline
Fold & Model & $h^{1,1}$ &$h^{2,1}$ &$h^{3,1}$ &$h^{2,2}$ \\
\hline
0 & LSTM-448-f0 & 99.70 & 96.78& 92.27& 75.61\\
1 & LSTM-448-f1 & 99.81 &  98.29& 95.35 & 81.98\\
2 & LSTM-448-f2 & 99.04&  91.43 & 82.04& 58.75\\
3& LSTM-448-f3 & 99.59 & 95.29 & 90.20 & 68.58 \\
4& LSTM-448-f4 & 99.71 &  97.55& 93.88& 78.75\\
\hline
\end{tabular}
\caption{Test accuracies for the five LSTM-448 models trained using 5-fold cross validation, evaluated on the test set of the 72\% dataset.}\label{5f-sep}
\end{table}
Recalling  the fact that the test set used in the 5-fold CV experiments is the exact same test set used in the 72\% dataset (which accounts for 20\% of the total data), and that each of the five models in Table \ref{5f-sep} was trained and validated separately on different subsets of the total training data (a combination of the training and validation sets of the 72\% dataset), these models can be evaluated individually as standalone models or in some ensembles consisting of several models together. As standalone models, the worst performing one is LSTM-448-f2 trained on fold 2, while the best performing model is LSTM-448-f1 (trained on fold 1), followed by LSTM-448-f4 (trained on fold 4) and LSTM-448-f0 (trained on fold 0) - see Fig.\ref{lstm-5fold-val}. LSTM-448-f1 actually outperforms CNN-LSTM-400 from Table \ref{cnn-lstm-acc} (with the accuracies of 99.74\%, 98.07\%, 95.19\%, 81.01\%), which is the best model trained on the 72\% dataset (see Fig.\ref{acc-CV-4x}). 

\begin{figure}[H]
\centering{}
\includegraphics[width = 0.75\textwidth]{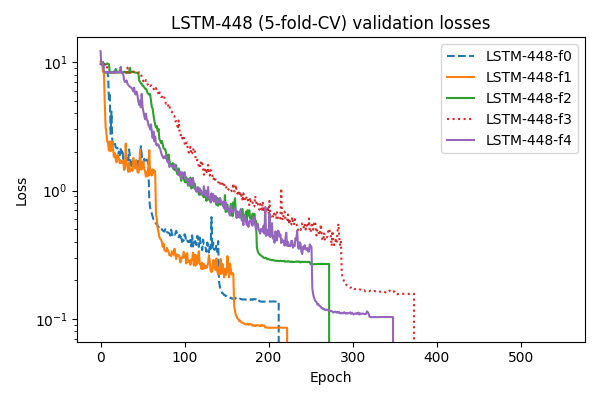}
\caption{Validation losses of the five LSTM models in 5-fold cross validation training.}\label{lstm-5fold-val}
\end{figure}

\subsection{Ensembles} \label{cv5f-ens}
Next, five ensembles were formed using the 5 models (LSTM-448-f0 to LSTM-448-f4) together with the two top performing models (CNN-LSTM-400, LSTM-448) from the first round in various combinations as listed below. It bears repeating the fact that it is a valid practice to combine the LSTM-448 models from the 5-fold CV training with those models trained on the 72\% dataset, since these models essentially use the same training data and are evaluated on the same test data. 
\ben
\item LSTM-448-5f consists of the five LSTM-448 models listed in Table \ref{5f-sep}. This ensemble has the lowest accuracies of all the ensembles in Table \ref{5f-ens}.
\item LSTM-448-f0f1f4  consists of the three top performing models in Table \ref{5f-sep}, LSTM-448-f0, LSTM-448-f1 and LSTM-448-f4. 
\item LSTM-448-f1f4 consists of the top two performing models in Table \ref{5f-sep}, LSTM-448-f1 and LSTM-448-f4.
\item Ens-f1f4-CL400  has the same composition has LSTM-448-f1f4, plus the model CNN-LSTM-400 from Table \ref{cnn-lstm-acc} trained on the 72\% dataset. 
\item Ens-0f1f4-CL400  has the same composition as Ens-f1f4-CL400, plus the model LSTM-448 from Table \ref{lstm-acc}. This ensemble has the best accuracies for all the Hodge numbers. 
\enn 
Some further comments are in order regarding the ensembles formed above:
\bei
\item While it is a common practice to include the results of all folds in a cross validation training to form an
ensemble, a well proven technique in machine learning competitions (as hosted on \texttt{Kaggle}, see for example the discussion in \cite{kaggle-scp}) is to
form weighted ensembles where the models trained on different CV folds are weighted differently
to obtain the best results when evaluated on the test dataset. 
There are various ways to obtain the optimal weights to be used in the weighted ensemble \cite{tsf-ens}. A very popular method is `hill climbing' that starts with the best model and  cycles through the list
of all other available models together with a list of weights between a certain range to pick the second
best model with a certain weight to form the first ensemble. Next, the third
best model with a certain weight is chosen by repeating the same process of cycling through
the remaining models to form a new ensemble with the models chosen from the previous iterations. The process continues until the results stop improving, i.e. added models no longer change the obtained results. 
\item In the described process above, models resulting from CV training can be treated independently\footnote{In a practical sense, a neural network model is nothing but a collection of its weights and biases so a different set of weights and biases constitutes an entirely different model. Models trained 5-fold CV are in fact independent models following this practical sense.} and added to the mix in the same way as a model not trained using CV, as long as all models are evaluated on the same test set. 
The result of this exercise might lead to some optimized ensembles with certain CV folds not included in the final mix. 
In this work, due to the time and computational resource constraints, we did not carry out the optimization of the weights to form the most optimized weighted ensembles. Instead, we heuristically form the ensembles above by selecting the top performing models from both CV and non-CV training. 
\eni
The test accuracies of the five ensembles are gathered in Table \ref{5f-ens}.
\begin{table}[h]
\centering
\begin{tabular}{|c|c|c|c|c|c|}
\hline
 Ensemble & Composition & $h^{1,1}$ &$h^{2,1}$ &$h^{3,1}$ &$h^{2,2}$ \\
 \hline
LSTM-448-5f & $\begin{array}{l} \text{LSTM-448-f0}\\ \text{LSTM-448-f1}\\\text{LSTM-448-f2}
\\\text{LSTM-448-f3}\\\text{LSTM-448-f4}\end{array}$ & 
 99.75&  97.74& 94.14& 76.81\\ \hline
LSTM-448-f0f1f4 & $\begin{array}{l} \text{LSTM-448-f0}\\ \text{LSTM-448-f1}\\ \text{LSTM-448-f4}\\\end{array}$ &   
99.80& 98.31& 95.30& 82.44\\ \hline
LSTM-448-f1f4 & $\begin{array}{l} \text{LSTM-448-f1} \\\text{LSTM-448-f4}\end{array}$ & 
99.80 &  98.46&  95.58& 83.04 \\ \hline
Ens-f1f4-CL400 & $\begin{array}{l} \text{LSTM-448-f1}\\ \text{LSTM-448-f4} \\ \text{CNN-LSTM-400}\end{array}$ & 
 99.83& 98.68& 96.11 & 84.67 \\ \hline
Ens-0f1f4-CL400 & $\begin{array}{l} \text{LSTM-448-f1}\\ \text{LSTM-448-f4}\\ \text{LSTM-448} \\ \text{CNN-LSTM-400}\end{array}$ & 
 $\mathbf{99.84}$& $\mathbf{98.71}$ & $\mathbf{96.26}$ & $\mathbf{85.03}$ \\ \hline
\end{tabular}
\caption{Test accuracies for the ensembles formed from the LSTM-448 models trained using 5-fold CV and models trained on the 72\% dataset evaluated on the test set of the 72\% dataset. Best results are noted in bold.}\label{5f-ens}
\end{table}
Compared to the ensembles of Table \ref{ensemble} formed using the models trained on the 72\% dataset, the ensembles of Table \ref{5f-ens} provide a significant improvement in the results. This can be visualized in Fig.\ref{acc-CV-4x} in which the accuracies of the models in Table \ref{5f-ens} and Table \ref{5f-sep} are ranked together with the top four performing models from the first round of training (Ens-2, Ens-3, Ens-1 from Table \ref{ensemble}, CNN-LSTM-400 from Table \ref{cnn-lstm-acc}). 
In terms of the accuracies as listed in Table \ref{5f-ens} above, the best ensemble is Ens-0f1f4-CL400, whose accuracy performance exceeds the best ensemble Ens-2 created from the models trained on the 72\% dataset. The second best model is Ens-f1f4-CL400 whose accuracy also exceeds Ens-2. The difference between Ens-0f1f4-CL400 and Ens-f1f4-CL400 is the addition of the non-CV model LSTM-448. While it might appears counterintuitive that the addition of LSTM-448, whose performance ranks below CNN-LSTM-400 on the 72\% dataset, elevates the performance of Ens-0f1f4-CL400 above Ens-f1f4-CL400, it is perfectly logical once one recalls that different models have different regions of competence that might be complimentary to one another's. This is why ensembles often outperform their constituent models due to their robustness and better capability to generalize when evaluated on the same test set. \footnote{Heuristically, it is not impossible to have an ensemble (let’s
call it Ens-AB) that is made of Model A with 55\% accuracy and Model B with 65\% accuracy
obtaining 70\% accuracy. It is again not impossible that adding another model, Model C with 60\%
accuracy, leads to an even better ensemble (let’s call it Ens-ABC) with 72\% accuracy. In this
case, Model B at 65\% accuracy clearly outperforms Model C (at 60\%), but the new ensemble
Ens-ABC including Model C can outperforms the old ensemble (Ens-AB). In AI/ML competitions, this is a scenario that happens repeatedly and winning solutions often make use of ensembles of multiple models in this exact sense (\cite{kaggle-elc}, \cite{kaggle-hc}, \cite{kaggle-hms}).}
Additional model rankings and performances can be found in Table \ref{mse-i-cv} and  Fig.\ref{mse-i-f-cv} in terms of the MSE metric,  Table \ref{mae-i-cv} and Fig. \ref{mae-i-f-cv} in terms of the MAE metric,  Table \ref{r2-i-cv} and Fig.\ref{r2-i-f-cv} in terms of the $R^2$ metric, Table \ref{scores-all-CV} and Fig.\ref{scores-all-CV-f} for all three metrics.

\begin{figure}[H]
\centering
\includegraphics[width = 0.78\textwidth]{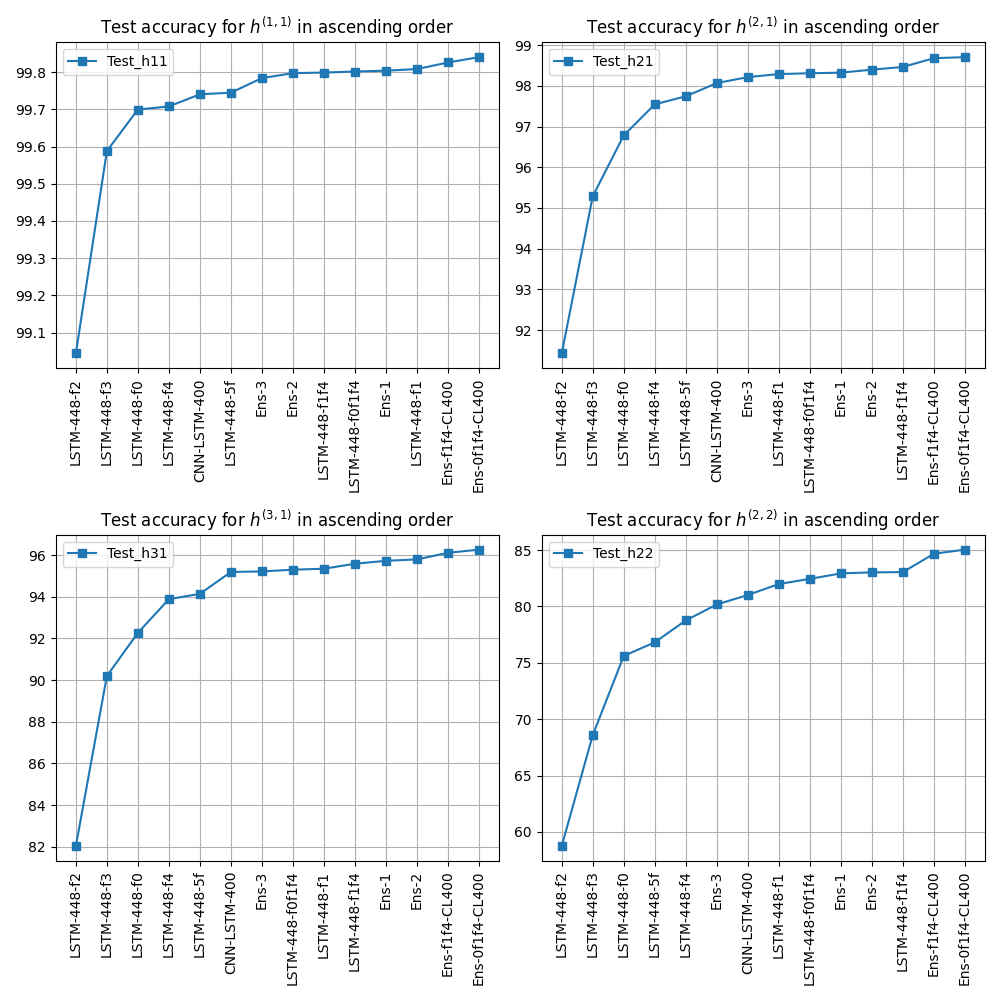}
\caption{Test accuracies of all models in Tables \ref{5f-sep} \ref{5f-ens}, as well as the top 4 models from the first round of training (Ens-2, Ens-3, Ens-1 from Table \ref{ensemble}, CNN-LSTM-400 from Table \ref{cnn-lstm-acc}), evaluated on the test set of the 72\% dataset for each of the four Hodge numbers in ascending order (with worst to best models arranged from the left to the right in each diagram).}\label{acc-CV-4x}
\end{figure}

\section{Training results (using 80\% dataset)} 
\subsection{Individual models}\label{results-80}
In this section, we compiled the results obtained in the second round of training in which we chose the top three models (CNN-LSTM-400, LSTM-448, LSTM-424) from the first round of training and retrained them on the enlarged 80\% dataset. The details of the training process (all parameters and hardware) remain the same as described in section \ref{results}. The only difference is that we used the saved model checkpoints from the previous training round as our starting point for this round. There is no risk of data leakage because our training set was enlarged by adding new data from the test set, which was completely unseen by the trained models during the training phase of the first round. The training in this round took considerably shorter than before thanks to the use of the trained models' saved weights and biases, which already have been optimized for a large part of the dataset. The obtained results are gathered in Table \ref{res-80}, and the training curves of the three models can be found in Fig.\ref{2nd-round-results} in Appendix \ref{curves}. 

\begin{table}[H]
\centering
\begin{tabular}{cccccc}
\hline\hline
& Model & $h^{1,1}$ & $h^{2,1}$ & $h^{3,1}$ & $h^{2,2}$\\
\hline \hline
& CNN-LSTM-400    & 99.67& 96.56& 91.71& 73.72 \\
& CNN-LSTM-400-d72  
  & 99.73& 98.04& 95.09& 80.86 \\
& LSTM-424        & 99.82 & 98.19& 95.79& 81.46 \\
& LSTM-448        & $\mathbf{99.85}$& $\mathbf{98.66}$ & $\mathbf{96.26}$& $\mathbf{84.77}$ \\
& LSTM-448-f1 & 99.80& 98.27& 95.27& 81.94\\
\hline
\end{tabular}
\caption{Test accuracies obtained by retraining CNN-LSTM-400, LSTM-424 and LSTM-448 in the second training round. CNN-LSTM-400-d72 is the trained model on the 72\% dataset from the first round. LSTM-448-f1 is the model trained on fold 1 of the 5-fold CV training. The best results are noted in bold.} \label{res-80}
\end{table}
The top performing model in this case is LSTM-448, followed by LSTM-424. The worst performing model is CNN-LSTM-400.  The performances of the three models can be visualized by looking at the curves of the validation losses obtained during the retraining phase (see Fig. \ref{2nd-val-all}). The full training curves (with training and validation losses) for the three models can be found in Fig.\ref{2nd-round-results} in section \ref{curves}.
\begin{figure}[H]
\centering
\includegraphics[width = 0.65\textwidth]{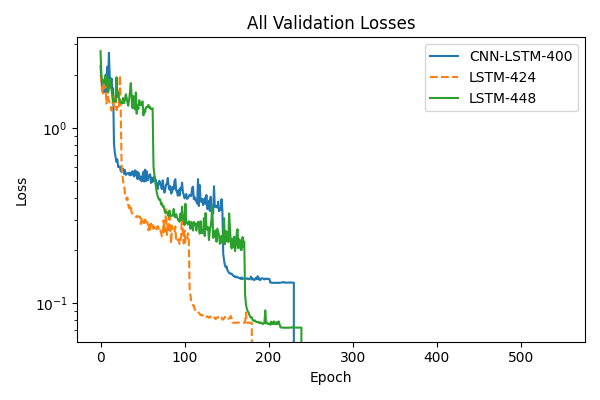}
\caption{Validation losses (in log scale) of the three models trained from scratch listed in Table 8.}\label{2nd-val-all}
\end{figure}
Interestingly, while the models LSTM-424 and LSTM-448 both obtained improved accuracies with more data, CNN-LSTM-400 - the top performing model from the first round - actually performed significantly worse on the enlarged dataset. While the reason for this observation remains not well understood, we hypothesize that the poor performance of CNN-LSTM-400 with more data has to do with the hybrid nature of this neural network, since its feature extractor is composed of both the CNN and RNN parts, which have different performances when it comes to extracting useful features. On the other hand, the LSTM-based networks rely solely on the LSTM module to extract features so there is no issue regarding mismatched performances. 
As an experiment, we used the model CNN-LSTM-400-d72 originally trained on the 72\% dataset to evaluate its performance on the smaller test set of the 80\% dataset. 
Since the results of CNN-LSTM-400d72 turned out to be much better compared to those obtained from the retraining, we will use this model (referring to it simply as CNN-LSTM-400 from now on) instead of the model trained on the enlarged 80\% dataset to form ensembles with the retrained LSTM-based models and also for any evaluation purposes that involve the 80\% dataset. Furthermore, we also used the model LSTM-448-f1 (trained on fold 1 of the 5-fold CV training) to re-evaluate its performance on the test set of the 80\% dataset. Using the trained models CNN-LSTM-400-d72  and LSTM-448-f1 to evaluate their performances on the new test set is justified by the fact that these models were trained on the smaller datasets not overlapping with the data from the test set. 
\newpage
\subsection{Ensembles} \label{ensemble-80}
Using the models in Table \ref{res-80} above, five new ensembles were formed and evaluated on the test set of the 80\% dataset. The results are collected in Table \ref{ens-res-80}.
\begin{table}[H]
\begin{center}
\begin{tabular}{|c|l|c|c|c|c|}
\hline
Ensemble & Composition &$h^{1,1}$ & $h^{2,1}$ & $h^{3,1}$ & $h^{2,2}$\\
\hline 
 Ens-80-1&$\begin{array}{l} 
 \text{LSTM-424} \\ \text{LSTM-448} \end{array}$   
 & 99.88 & 98.85& 96.86& 86.19\\ \hline 
Ens-80-2 &$\begin{array}{l}
 \text{LSTM-448} \\ \text{CNN-LSTM-400} \end{array}$   
     & 99.85 & 98.77& 96.60& 85.78\\\hline 
 Ens-80-3 &$\begin{array}{l}  \text{LSTM-424} \\
 \text{LSTM-448}\\ \text{CNN-LSTM-400} \end{array}$ &     
 99.88 & 98.91 & 96.96 & 86.78\\\hline
 Ens-80-4 &$\begin{array}{l}  \text{LSTM-424} \\
 \text{LSTM-448}\\\text{LSTM-448-f1}\\ \text{CNN-LSTM-400} \end{array}$ & 
 $\mathbf{99.90}$ &  $\mathbf{99.03}$ &  $\mathbf{97.07}$ & $\mathbf{87.34}$\\
 \hline
 Ens-80-5 &$\begin{array}{l}  \text{LSTM-424} \\
 \text{LSTM-448}\\\text{LSTM-448-f1}\end{array}$ & 
 99.89& 98.93& 96.95& 87.12\\
\hline
\end{tabular}
\caption{Test accuracies of the ensembles formed from the individual models in Table \ref{res-80}. The best results are noted in bold.} \label{ens-res-80}
\end{center}
\end{table}
The overall best ensemble is Ens-80-4 consisting of the newly retrained LSTM-448, LSTM-424, the CNN-LSTM-400 model trained on the 72\% dataset and the LSTM-448-f1 model trained on fold 1 (using only 64\% training data) of the 5-fold CV training. 
\subsection{Comparison of accuracy} \label{acc-80}
The performances of all models (including the ensembles of Table \ref{ens-res-80}) on the test set of the 80\% dataset are ranked separately for each of the $h^{1,1}, h^{2,1}, h^{3,1}, h^{2,2}$ Hodge numbers in terms of their accuracies and shown in Fig. \ref{acc-80-4x}. In section \ref{metrics-80}, the model performances are ranked in terms of additional metrics such as MSE (see Table \ref{mse-80-i-f} and Fig. \ref{mse-80-i-f}), MAE (see Table \ref{mae-80-i} and Fig. \ref{mae-80-i-f}),  $R^2$ (see Table \ref{r2-80-i} and  Fig. \ref{r2-80-i-f}), and collectively (see Table \ref{scores-80-all} and Fig. \ref{scores-all-80}). 
\begin{figure}[H]
\centering
\includegraphics[width = 0.8\textwidth]{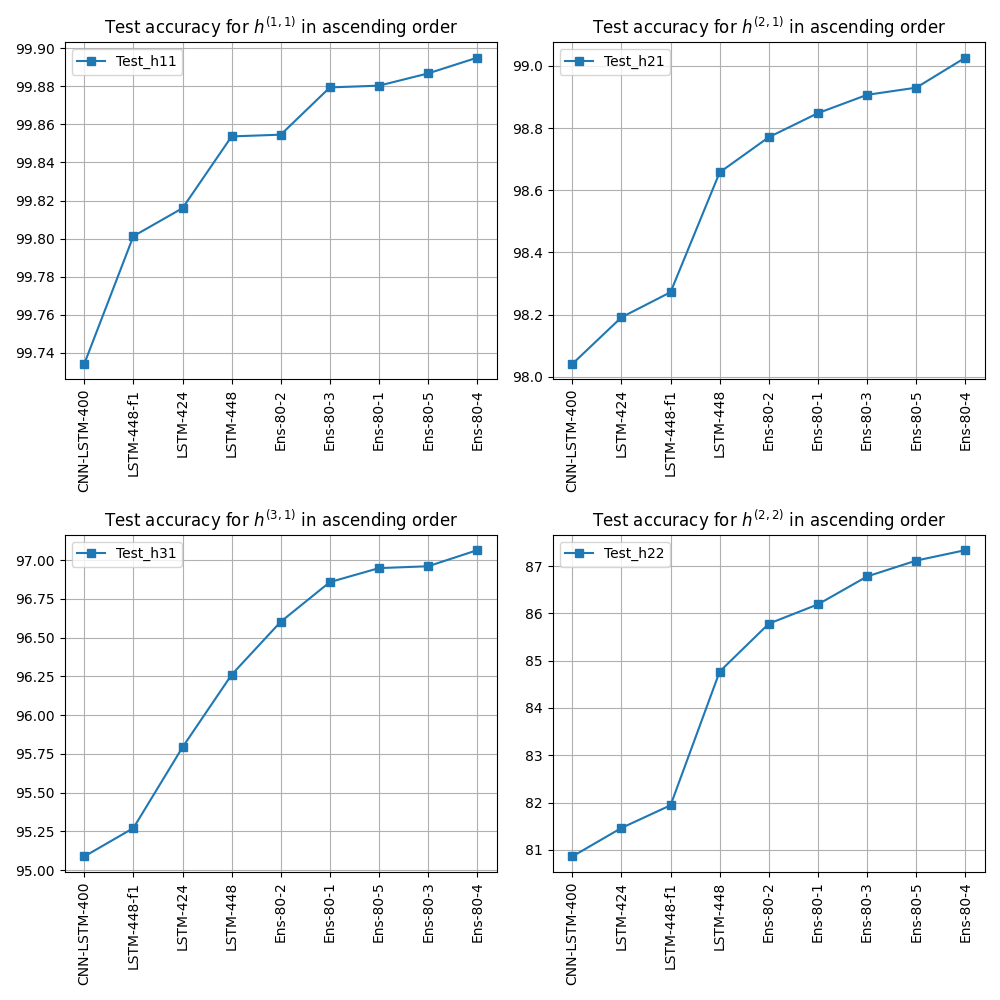}
\caption{Test accuracies of all models (including ensembles) evaluated on the test set of the 80\% dataset for each of the four Hodge numbers in ascending order (with worst to best models arranged from the left to the right in each diagram).}\label{acc-80-4x}
\end{figure}
\newpage
\section{Summary \& Outlook} \label{concl}
In this work, we explored and experimented with hybrid convolutional-recurrent and purely recurrent neural network architectures for the task of deep learning the Hodge numbers of CICY4 dataset. We emphasize that our work is by nature a proof-of-concept type, exploring and establishing the utility of RNN-based architectures for the task of regressing or learning the Hodge numbers, rather than a full-fledged characterization of which RNN architectures perform best.  In this sense, it should be noted that while we used \cite{inception} as a benchmark to compare all our RNN-based models against, there are significant differences in the way the problem was approached. Unlike \cite{inception}, we did not start with four separate neural network models to learn each Hodge number independently\footnote{as done in \cite{inception}, the goal of this was to find the best architecture and parameters on which to base the final model, `CICYMiner', capable of learning all 4 Hodge numbers simultaneously}, but chose to begin instead with a single RNN-based model (with a different architecture variant in each experiment) to learn all four Hodge numbers simultaneously. This approach was adopted based on the context that CNN-based architectures were already the established and default type of neural networks to use for this type of problem since 2017 (\cite{He-17}), and it was only a matter of finding out which CNN architecture was the best to use in subsequent works. In this work, the starting point of the investigation was more fundamental, in the sense that RNN-based architectures had never been applied to this kind of problem, so the natural question that arose was: \textit{Can we apply an RNN to this and will we obtain similar results to those obtained by CNNs? } In this context, the most computationally efficient way to carry out this task is to create one single deep-learning model to regress all four Hodge numbers, and
contrast the performance of this model with the best-performing CNN model already established. After carrying out multiple experiments involving training twelve models listed in Table \ref{params} at 72\% training ratio (including the 5-fold cross validation training of LSTM-448 in Table \ref{5f-sep}) and retraining the top three models listed in Table \ref{res-80} at 80\% training ratio, as well as forming various ensembles of the best performing models (Tables \ref{ensemble}, \ref{5f-ens}, \ref{ens-res-80}), our main findings are summarized below.
\bei
\item[\ding{85}]For the hybrid CNN-RNN models, the CNN block, containing only about 10\% of the total trainable parameters, is designed to be simple and light weight compared to the RNN block, which is intended to play the more dominant role. Two variants of CNN block were used: a custom CNN network with two convolutional layers, and a ResNet-inspired CNN network with two branches, a main branch containing three convolutional layers, and a short-cut branch with a single dense layer (with no bias). Due to the longer training time of hybrid networks utilizing the ResNet block, the majority of the CNN-RNN hybrid models use the custom CNN block. Varying the hidden size $M$ of the RNN block with the RNN being either GRU or LSTM let us observe the effect of $M$ on the model performance. The general trend observed is an increase in the obtained accuracy of the Hodge number predictions when $M$ is increased. When the peak performance is achieved at a certain $M$, increasing $M$ further leads to a decrease in performance or a failure to converge, the cause of which is likely to be overfitting. Furthermore, it is observed that CNN-LSTM models significantly outperform CNN-GRU models, a result perhaps attributable to the fact that LSTMs in general often outperform GRUs due to their more sophisticated recurrent architecture. The results of CNN-RNN hybrid models can be found in Tables \ref{cnn-gru-acc}, \ref{cnn-lstm-acc}.
\item[\ding{85}]
For the purely recurrent LSTM-based models, their observed performances are comparable to those CNN-LSTM models of roughly the same size (see Table \ref{lstm-acc}). Similar to the trend observed with CNN-RNN, increasing the size of $M$ leads to an increase in performance until a peak is reached, after which performance drops due to overfitting.  
\item[\ding{85}]
 Collectively, the results for our three best performing individual models at each training ratio are gathered in Table \ref{best-models}.
\begin{table}[H]
\centering
\begin{tabular}{ |c|c|c|c|c|c|c|c| } 
\hline
 Training ratio & Rank & Model & $h^{1,1}$ & $h^{2,1}$& $h^{3,1}$& $h^{2,2}$  \\
\hline
\multirow{3}{4em}{72\% } & \#1  & CNN-LSTM-400  & 99.74& 98.07& 95.19& 81.01\\ 
& \#2 & LSTM-448 &  99.74& 97.51& 94.24& 78.63\\ 
& \#3 & LSTM-424 & 99.56 & 97.07 &93.19 & 74.47\\ 

\hline
\multirow{3}{4em}{80\%} & \#1  & LSTM-448 & 99.85& 98.66& 96.26 &84.77\\ 
& \#2 & LSTM-424 &  99.82 & 98.19 & 95.79 & 81.46\\
& \#3 & CNN-LSTM-400  & 99.73 & 98.04 &95.09 & 80.86\\ 
\hline
\end{tabular}
\caption{Top three performing models at 72\% and 80\% training ratio, not including the models from the 5-fold CV training.}\label{best-models}
\end{table}
\item[\ding{85}]
With the 5-fold cross validation training of the model LSTM-448 that resulted in the five models LSTM-448-f0, LSTM-448-f1, LSTM-448-f2, LSTM-448-f3 and LSTM-448-f4 (Table \ref{5f-sep}) which can be evaluated on either the test set of the 72\% dataset or that of the 80\% dataset, the model ranking in Table \ref{best-models} changed. The best models are now listed in Table \ref{best-models-cv}.
\begin{table}[H]
\centering
\begin{tabular}{ |c|c|c|c|c|c|c|c| } 
\hline
 Training ratio & Rank & Model & $h^{1,1}$ & $h^{2,1}$& $h^{3,1}$& $h^{2,2}$  \\
\hline
\multirow{3}{4em}{72\% } & 
\#1 & LSTM-448-f1 & 99.81 & 98.29 & 95.35& 81.98 \\
&\#2  & CNN-LSTM-400  & 99.74& 98.07& 95.19& 81.01\\ 
& \#3 & LSTM-448 &  99.74& 97.51& 94.24& 78.63\\ 
\hline
\multirow{3}{4em}{80\%} & \#1  & LSTM-448 & 99.85& 98.66& 96.26 &84.77\\ 
& \#2 &LSTM-448-f1  & 99.80 &98.27& 95.27 &81.94\\ 
& \#3 & LSTM-424 &  99.82 & 98.19 & 95.79 & 81.46\\
\hline
\end{tabular}
\caption{Top three performing models at 72\% and 80\% training ratio, including the model LSTM-448-f1 from the 5-fold CV training.}\label{best-models-cv}
\end{table}
\item[\ding{85}]
To improve the above results beyond those obtained by the best individual model at each training ratio, ensembles of several top performing models were created. The ensemble results in each case considered always proved to be better than the results of the individual models forming the ensemble (see Tables \ref{ensemble}, \ref{5f-ens}, \ref{ens-res-80} and Figs. \ref{accuracies}, \ref{acc-CV-4x}, \ref{acc-80-4x}). This is indicative of the fact that the variations in the model architectures led to different models learning different useful features of the data and thus excelling at predicting different (perhaps complementary) ranges of the data. The best ensembles, together with their associated accuracies, evaluated on the test set of the 72\% and 80\% datasets are listed in Table \ref{best-ens}.
\begin{table}[H]
\centering
\begin{tabular}{ |c|c|c|c|c|c|c| } 
\hline
 Training ratio &Ensemble  & Composition &$h^{1,1}$ & $h^{2,1}$& $h^{3,1}$& $h^{2,2}$  \\
\hline
72\%  &Ens-0f1f4-CL400 & 
$\begin{array}{l} \text{LSTM-448-f1}\\ \text{LSTM-448-f4} \\\text{LSTM-448} \\\text{CNN-LSTM-440}\end{array}$ &99.84 & 98.71 & 96.26 & 85.03\\ \hline
80\% &Ens-80-4 &
 $\begin{array}{l} \text{LSTM-448-f1}\\\text{LSTM-448}\\\text{LSTM-424} \\\text{CNN-LSTM-440}\end{array}$ &
 99.90 & 99.03& 97.07 & 87.34\\ 
\hline
\end{tabular}
\caption{Accuracies obtained by the best ensembles at 72\% and 80\% training ratio.}\label{best-ens}
\end{table}
\item[\ding{85}]
In comparison with the work \cite{inception} in which the authors employed a specialized inception-based network with around $10^7$ parameters (whose training time took approximately 5 days on a single NVIDIA V100 GPU)  and obtained the results of 100\% for both $h^{(1,1)}$ and $h^{(2,1)}$, 96\% for $h^{(3,1)}$ and 83\% for $h^{(2,2)}$ at 80\% training ratio, our results are really promising, given the fact that our best models (LSTM-448, LSTM-448-f1, LSTM-424, CNN-LSTM-400) are all around three times smaller and requiring much shorter training time (around 7 hours fully on the \texttt{Kaggle} P100 GPU for both rounds of training). While we did not achieve 100\% accuracy for either $h^{1,1}$ or $h^{2,1}$, we did achieve better accuracies for $h^{3,1}$ and $h^{2,2}$ compared to \cite{inception} for both the case of an individual model (LSTM-448 at 80\% training ratio) and the case of  ensembles (Ens-0f1f4-CL400 at 72\% and Ens-80-4 at 80\% training ratios). This demonstrates the suitability and effectiveness of recurrent-based models (both in the pure and hybrid forms) for this specific task of regressing the Hodge numbers. 
\eni 
\indent Finally, we would like to conclude this work by noting that given the relatively simple choice of our neural network architectures, there is still a lot of room for experimentation and improvement. The reported results in this paper are by no means the best possible ones that could be obtained by these model architectures. In particular, since the CNN part of the hybrid models is light weight and simple, its structure could be modified (either by adding more convolutional layers or changing the numbers of filters in the existing convolutional layers) and tuned, while keeping the RNN structure fixed, to observe the change in the results. The same kind of modifications can be applied to the ResNet-inspired variant of the CNN-RNN hybrid models. For the RNN block, we have kept the numbers of layers $L$ fixed to be 2 throughout this work, but it might be interesting to vary this number to observe its impact. During our initial experimentations, we observed that choosing $L=2$ yieled better results than $L=1$ or $L=3$ (or higher), but this might change when a different CNN structure is considered. For both the hybrid and purely recurrent architectures, there is an option to make the RNNs bidirectional but we decided not to, because bidirectionality will double the number of parameters in the network and increase the training time significantly. Furthermore, due to the constraint of the GPU usage allowed by the online platform \texttt{Kaggle}, the systematic scanning (parameter optimization by a dedicated library such as \texttt{Optuna}\footnote{\href{https://optuna.org/}{https://optuna.org/}}) \cite{optuna} of the hidden size $M$ of all GRU/LSTM networks to find the optimum $M$ at which the highest accuracy can be obtained could not be carried out.\footnote{To be able to perform an effective scan for $M$, a sizeable portion (in the order of at least tens of thousands of samples) of the training set should be used, which leads to very long scanning time well exceeding the allowed GPU time limit.} It would be useful to perform a scanning for the RNN hidden size $M$ (and perhaps the number of layers $L$) if there were a better access to GPU resources. 
\\\\
\indent A more drastic change would involve doing away with convolutional or recurrent structures altogether in favor of experimenting with the transformer-based architecture \cite{transformer}, \cite{vit-trans} which has been rapidly replacing both CNNs in computer vision and RNNs in natural language processing as the state-of-the-art technique. An approach that we would take when using a transformer-based architecture for this task of learning the Hodge numbers involves treating the CICY4 configuration matrices as images and decomposing them into patches that can be fed into a vision transformer (ViT) neural network \cite{vit-trans}\footnote{The ViT network is a custom network written from scratch using PyTorch, not a pretrained one}. An experiment implementing this is already under way, but the initial indications are not promising - a single epoch of training using ViT takes about five minutes, about five times longer than the longest time taken by the largest model (CNN-LSTM-416)\footnote{using the same Kaggle P100 GPU}. Moreover, given the fact that due to the lack of certain inductive biases (such as translation equivariance and locality), transformer-based networks do not perform as well as CNNs when the dataset contains less than one million data points \cite{vit-trans} (which is the case with the CICY4 dataset), we might already have an early indication of the result of the experiment involving transformer-based models. 
\\\\
\noindent\textbf{Acknowledgements} 
\\
The author is an unaffiliated, independent researcher - this work was possible thanks solely to the publicly free and available GPU cloud computing resources offered by \texttt{Kaggle} (\href{https://www.kaggle.com}{https://www.kaggle.com}). 

\section{Appendices} \label{app}
\subsection{MSE, MAE, R-squared metrics (72\% dataset)} \label{metrics}
In this section, we look at additional metrics typically used to evaluate the performance of models designed for regression task. In particular, the three metrics that we use are Mean Squared Error (MSE), Mean Absolute Error (MAE), and $R^2$ score. For a single regression task, these scores are calculated as follows
\beq
MSE&=& \frac{1}{N}\sum_{i=1}^N(y_\text{pred} - y_\text{target})^2 \non
MAE &=& \frac{1}{N}\sum_{i=1}^N|y_\text{pred} - y_\text{target}|  \non
R^2 &=& 1-\frac{\sum_{i=1}^N(y_\text{pred} - y_\text{target})^2}{\sum_{i=1}^N(y_\text{pred} - \langle y_\text{target}\rangle)^2} 
\eeq
where $N$ is the size of the vectors $y_\text{pred}$ and $y_\text{target}$ (typically $N$ is the number of samples in a batch), $\langle y_\text{target}\rangle$ is the mean of the target $y_\text{target}$.
For a multi-regression task such as the one in this work, where $y_\text{pred}$ and $y_\text{target}$ can take the form $(N, P)$ with $P$ being the total number of regressions, the final MSE, MAE and $R^2$ scores are the mean of  individual score calculated from each individual regression task. 
\beq
\text{Score}&=& \frac{1}{P}\sum_{i=1}^P \text{Score}_i, \hspace{10mm} (\text{Score} = MSE, MAE, R^2).
\eeq
MAE and MSE scores take value in the range of $[0, \infty)$, $R^2$ score takes value in the range $(-\infty, 1]$. For a perfect agreement, MSE and MAE scores should be zero, while the $R^2$ score should be 1. 
In Table. \ref{mse-mae-r2-all}, we present the MSE, MAE and $R^2$ scores for the 11 models (excluding the non-convergent CNN-LSTM-416) plus the four ensembles (Table \ref{ensemble}). 
\begin{table}[H]
\centering
\begin{tabular}{clcccc}
\hline\hline
& Model & MSE & MAE & $R^2$ &\\
\hline \hline
& CNN-GRU-384     &  3.064 & 0.584  &0.976 &\\
& CNN-GRU-416    &   2.516 & 0.446&  0.982 &\\
& ResNet-GRU-256  & 3.442  &0.586 &0.973 &\\
& ResNet-GRU-400   & 1.797 & 0.291 &0.989 &\\
& CNN-LSTM-256    &  1.565 &0.251&0.992 &\\
& CNN-LSTM-384    &  1.794& 0.263 & 0.992 &\\
& CNN-LSTM-400    &  0.982 & 0.117 & 0.997 &\\
& LSTM-400        &  1.524 & 0.255 & 0.991& \\
& LSTM-424        &  1.025 & 0.156 &0.995&\\
& LSTM-448        &  0.921 &  0.137 &0.996 &\\
& LSTM-456        &  1.397 &0.253 &0.991&\\
& Ens-1     &  0.708&   0.104 &0.997&\\
& Ens-2      &  $\mathbf{0.650}$ & $\mathbf{0.100}$ &$\mathbf{0.998}$&\\
& Ens-3      &  0.725 &0.115 &0.997&\\
& Ens-4      &  0.733 &0.119 & 0.997 &\\
\hline\hline
\end{tabular}
\caption{MSE, MAE, R2 scores of 11 models plus the four ensembles. The best results are noted in bold font.} \label{mse-mae-r2-all}
\end{table}
The MSE/MAE/$R^2$ scores above are the mean values of the individual MSE/MAE/$R^2$ scores for each of the four Hodge numbers. From the results above, we see that in all metrics, the best model is Ens-2. The specific ranking of all models according to the three metrics can be found in Fig.\ref{scores-all}. 

\begin{figure}[H]
\centering
\includegraphics[width = \textwidth]{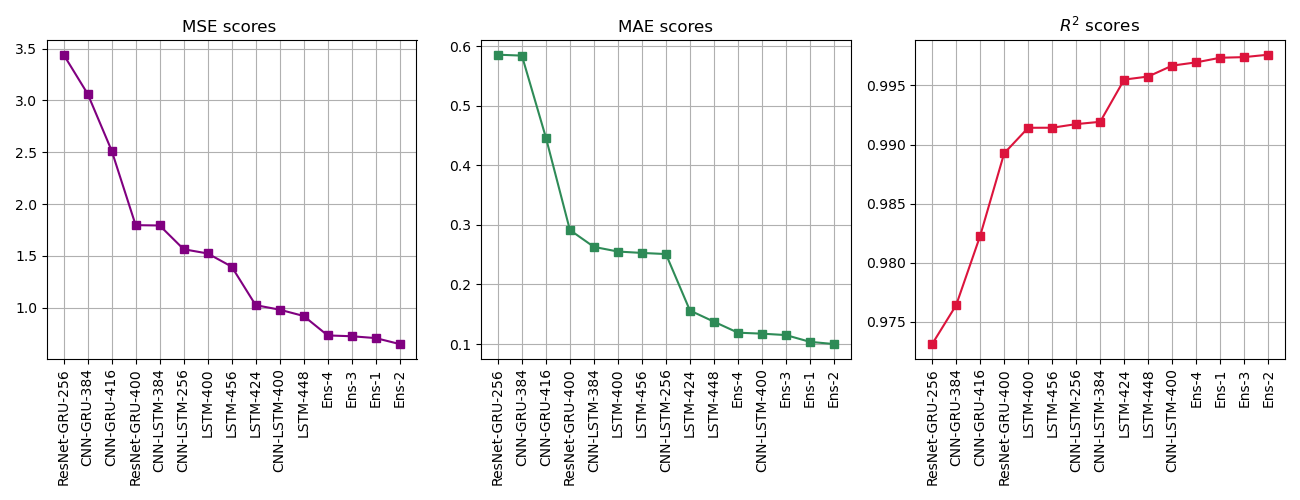}
\caption{Model performances ranked in terms of the MSE/MAE/$R^2$ scores (with worst to best models arranged from the left to the right in each diagram).}\label{scores-all}
\end{figure}
For references, we also include the individual MSE scores, MAE scores and $R^2$ scores for all the models for each of the Hodge numbers in Table \ref{mse-i}, Table \ref{mae-i} and Table \ref{r2-i}, respectively. Fig. \ref{mse-i-f}, Fig. \ref{mae-i-f} and Fig. \ref{r2-i-f} provide visualizations of the performances of all models in terms of the MSE, MAE and $R^2$ scores.
\begin{table}[H]
\centering
\begin{tabular}{clcccc}
\hline\hline
 & Model & $h^{1,1}$ & $h^{2,1}$ & $h^{3,1}$ & $h^{2,2}$ \\
\hline  \hline     
& CNN-GRU-384 & 0.026 & 0.289 & 0.718 & 11.222\\
& CNN-GRU-416 &0.022 & 0.213 & 0.593 & 9.236 \\
&ResNet-GRU-256 & 0.033 &0.328 &0.816 &12.590\\
&ResNet-GRU-400  &0.016 &0.125 &0.429 & 6.618\\
& CNN-LSTM-256  &0.013  &0.094 &0.372  &5.781\\
& CNN-LSTM-384  &0.012  &0.090 &0.422 & 6.652\\
& CNN-LSTM-400  &0.004 & 0.035 & 0.225 &3.663\\
& LSTM-400  & 0.012 &0.099 &0.364 & 5.620\\
& LSTM-424 & 0.007 &0.049 & 0.246 &3.798\\
& LSTM-448 & 0.005 &0.048 &0.214 & 3.416\\
& LSTM-456  &0.013 &0.100 &0.343 &5.132\\
& Ens-1 &  0.003 & 0.029 &0.162 & 2.637\\
& Ens-2 & $\mathbf{0.002}$ & $\mathbf{0.026}$ & $\mathbf{0.150}$ &$\mathbf{2.422}$\\
& Ens-3 & 0.003 &0.028 &0.167 &2.704\\
& Ens-4 & 0.004 &0.034 &0.171 &2.725   \\  
\hline
\end{tabular}
\caption{MSE scores of all models for each of the Hodge numbers. The best results are noted in bold font. Taking the mean of the four MSE scores for each model gives us the MSE score (in Table \ref{mse-mae-r2-all}) for the same model.}\label{mse-i}
\end{table}

\begin{figure}[H]
\centering
\includegraphics[width = 0.8\textwidth]{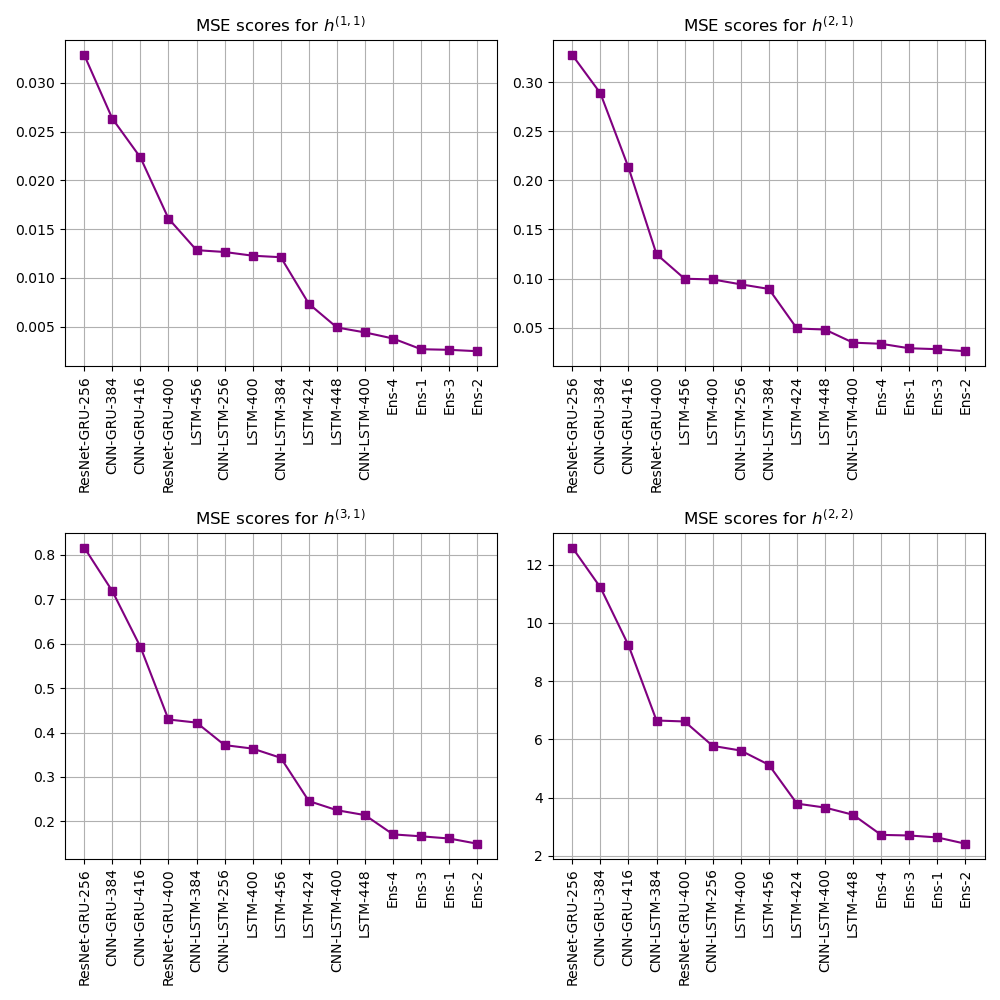}
\caption{Model performances ranked in terms of the MSE scores for the 4 Hodge numbers (with worst to best models arranged from the left to the right in each diagram) .}\label{mse-i-f}
\end{figure}
\begin{table}[H]
\centering
\begin{tabular}{clcccc}
\hline\hline
 & Model & $h^{1,1}$ & $h^{2,1}$ & $h^{3,1}$ & $h^{2,2}$ \\
\hline  \hline     
& CNN-GRU-384 & 0.020 & 0.184 & 0.387 & 1.745\\
& CNN-GRU-416 & 0.019 & 0.136 & 0.295 &1.336\\
&ResNet-GRU-256 &0.025 & 0.200 & 0.400 & 1.718\\
&ResNet-GRU-400  &  0.010& 0.088 & 0.182& 0.886\\
& CNN-LSTM-256  &0.009 & 0.065&  0.153 & 0.777\\
& CNN-LSTM-384  &0.010 &0.061& 0.164 &0.817\\
& CNN-LSTM-400  & 0.003 & 0.023& 0.068 &0.375\\
& LSTM-400  & 0.008 & 0.069 & 0.162 & 0.782 \\
& LSTM-424 & 0.005 & 0.034 & 0.091 & 0.494\\
& LSTM-448 & 0.003& 0.030 & 0.081  &0.436\\
& LSTM-456  & 0.008 & 0.070& 0.162 & 0.771\\
& Ens-1 & 0.002 & 0.020 & 0.058 & 0.335 \\
& Ens-2 &$\mathbf{0.002}$ & $\mathbf{0.019}$ &$\mathbf{0.055}$ &$\mathbf{0.324}$\\
& Ens-3 & 0.002& 0.020 & 0.062 &0.375\\
& Ens-4 & 0.003 & 0.024 & 0.067& 0.383 \\  
\hline
\end{tabular}
\caption{MAE scores of all models for each of the Hodge numbers. The best results are noted in bold. Taking the mean of the four MAE scores of each model gives us the MAE score (in Table \ref{mse-mae-r2-all}) for the same model.}\label{mae-i}
\end{table}
 
\begin{figure}[H]
\centering
\includegraphics[width = 0.8\textwidth]{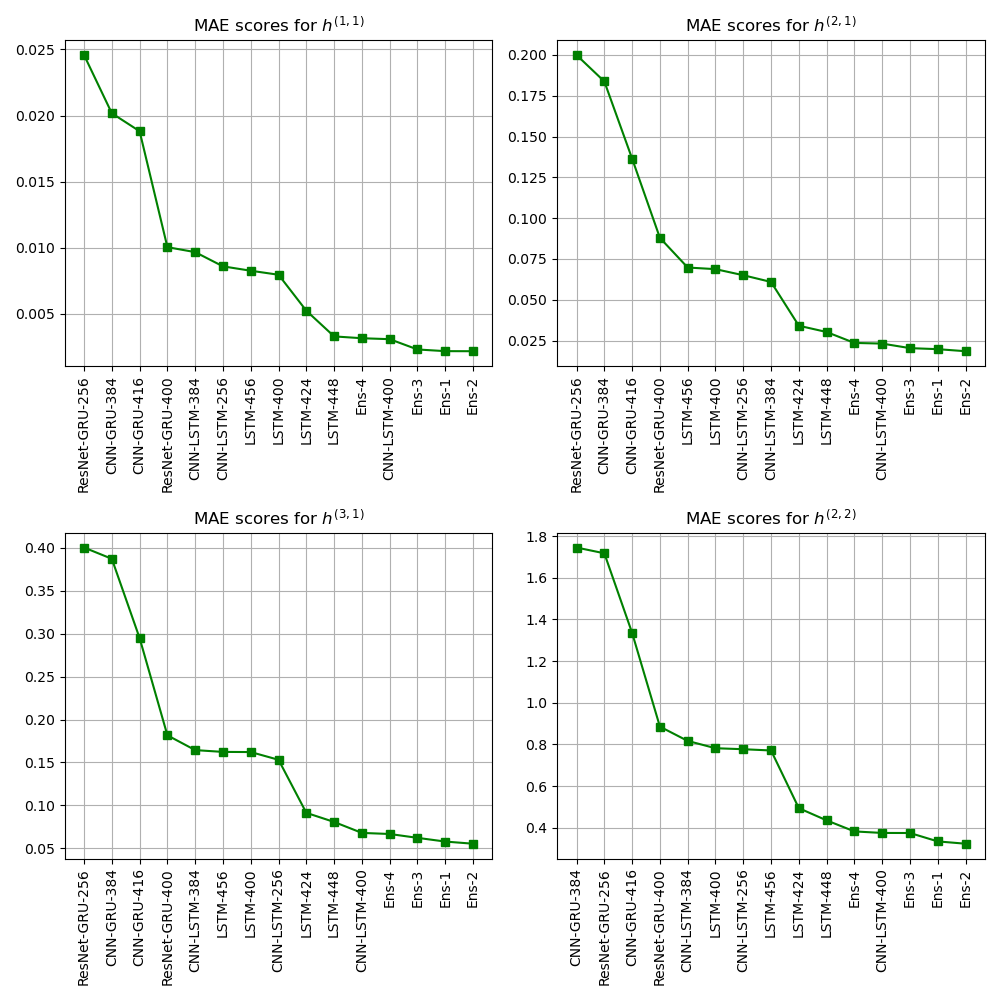}
\caption{Model performances ranked in terms of the MAE scores for the 4 Hodge numbers (with worst to best models arranged from the left to the right in each diagram). }\label{mae-i-f}
\end{figure}
\begin{table}[H]
\centering
\begin{tabular}{clcccc}
\hline\hline
 & Model & $h^{1,1}$ & $h^{2,1}$ & $h^{3,1}$ & $h^{2,2}$ \\
\hline  \hline     
& CNN-GRU-384 &   0.996 & 0.919 & 0.996 & 0.995\\
& CNN-GRU-416 & 0.996& 0.940 &0.997& 0.996\\
&ResNet-GRU-256 & 0.994 & 0.908 & 0.996 & 0.995\\
&ResNet-GRU-400  & 0.997& 0.965 &0.998 &0.997 \\
& CNN-LSTM-256  & 0.998 & 0.973 & 0.998 &0.998\\
& CNN-LSTM-384  & 0.998 &0.975 &0.998 &0.997\\
& CNN-LSTM-400  &  0.999& 0.990 & 0.999 & 0.999\\
& LSTM-400  &0.998 & 0.972& 0.998& 0.998  \\
& LSTM-424 & 0.999 & 0.986 & 0.999  &0.998\\
& LSTM-448 & 0.999 & 0.986 & 0.999 &0.999\\
& LSTM-456  & 0.998 & 0.972 & 0.998& 0.998\\
& Ens-1 & 1.000& 0.992 &0.999 &0.999\\
& Ens-2 &$\mathbf{1.000}$ &$\mathbf{0.993}$ & $\mathbf{0.999}$ & $\mathbf{0.999}$\\
& Ens-3 & 1.000& 0.992& 0.999 &0.999\\
& Ens-4 & 0.999 &0.990& 0.999 &0.999\\  
\hline
\end{tabular}
\caption{$R^2$ scores of all models for each of the Hodge numbers. The best results are noted in bold. Taking the mean of the four $R^2$ scores for each model gives us the $R^2$ score (in Table \ref{mse-mae-r2-all}) for the same model.}\label{r2-i}
\end{table}

\begin{figure}[H]
\centering
\includegraphics[width = 0.8\textwidth]{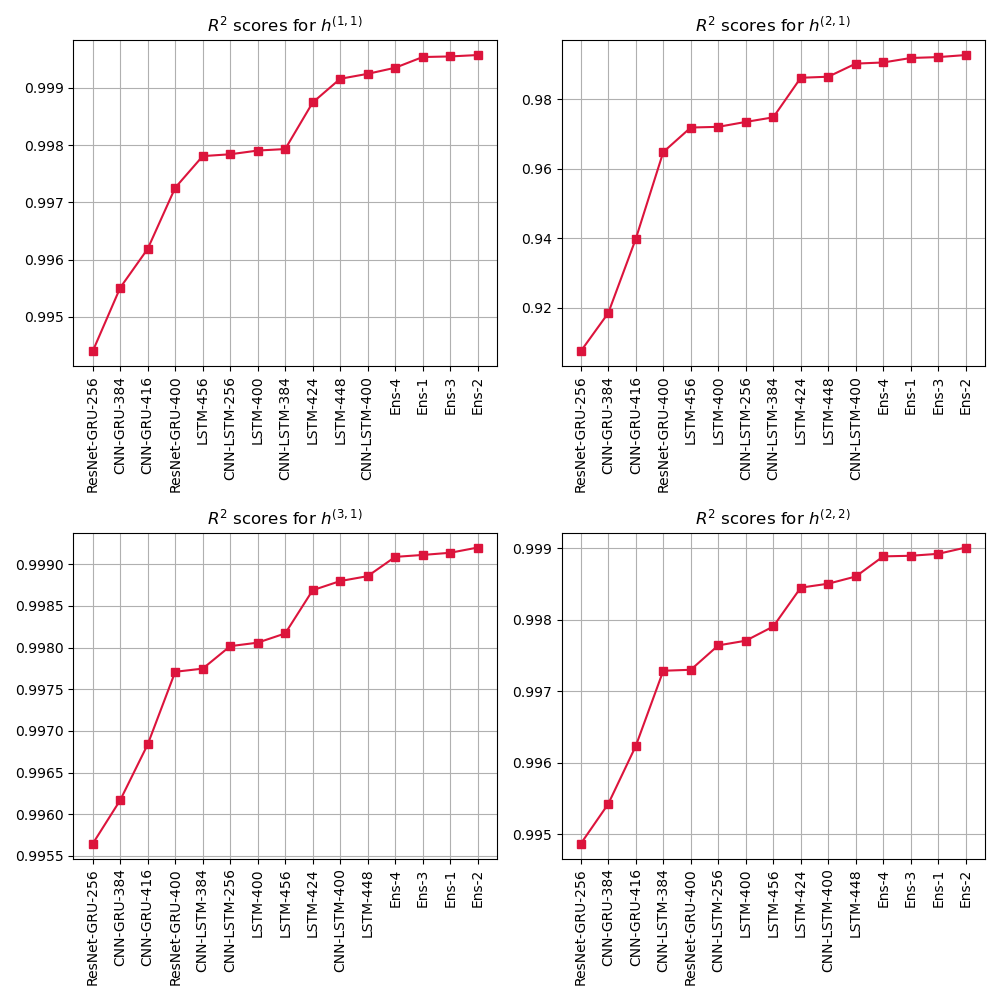}
\caption{Model performances ranked in terms of the $R^2$ scores for the 4 Hodge numbers (with worst to best models arranged from the left to the right in each diagram). }\label{r2-i-f}
\end{figure}
\newpage
\subsection{MSE, MAE, R-squared metrics (5-fold CV training)} \label{metrics-cv}
The MSE, MAE and $R^2$ scores of all models in Tables \ref{5f-sep} and \ref{5f-ens} are listed  in Table \ref
{mse-i-cv}, Table \ref{mae-i-cv} and Table \ref{r2-i-cv} below. The performance ranking of these models in terms of MSE, MAE and $R^2$ metrics are shown in Fig.\ref{mse-i-f-cv}, Fig. \ref{mae-i-f-cv} and Fig.\ref{r2-i-f-cv}, respectively. 

\begin{table}[H]
\centering
\begin{tabular}{clcccc}
\hline\hline
 & Model & $h^{1,1}$ & $h^{2,1}$ & $h^{3,1}$ & $h^{2,2}$ \\
 \hline\hline
&LSTM-448-f0 &0.006 &0.067 & 0.351 &5.587\\
&LSTM-448-f1 &0.003 &0.034 &0.308 &4.982\\
&LSTM-448-f2 &0.020 &0.156 &0.548 &8.368\\
&LSTM-448-f3 &0.009 &0.081 &0.420 &6.593\\
&LSTM-448-f4 &0.007 &0.044 &0.259 &4.040\\
&LSTM-448-5f &0.003 &0.033 &0.192 &3.085\\
&LSTM-448-f0f1f4& 0.003 & 0.027& 0.185& 2.977\\
&LSTM-448-f1f4 &0.003 &0.025 &0.189 &3.032\\
&Ens-f1f4-CL400 & 0.002& 0.021& 0.161& 2.617\\
&Ens-0f1f4-CL400 &$\mathbf{0.002}$ &$\mathbf{0.021}$& $\mathbf{0.143}$ & $\mathbf{2.324}$\\
\hline
\end{tabular}
\caption{MSE scores of all the models in Tables \ref{5f-sep}, \ref{5f-ens} for each of the Hodge numbers. The best results are noted in bold.}\label{mse-i-cv}
\end{table}
\begin{figure}[H]
\centering{}
\includegraphics[width = 0.8\textwidth]{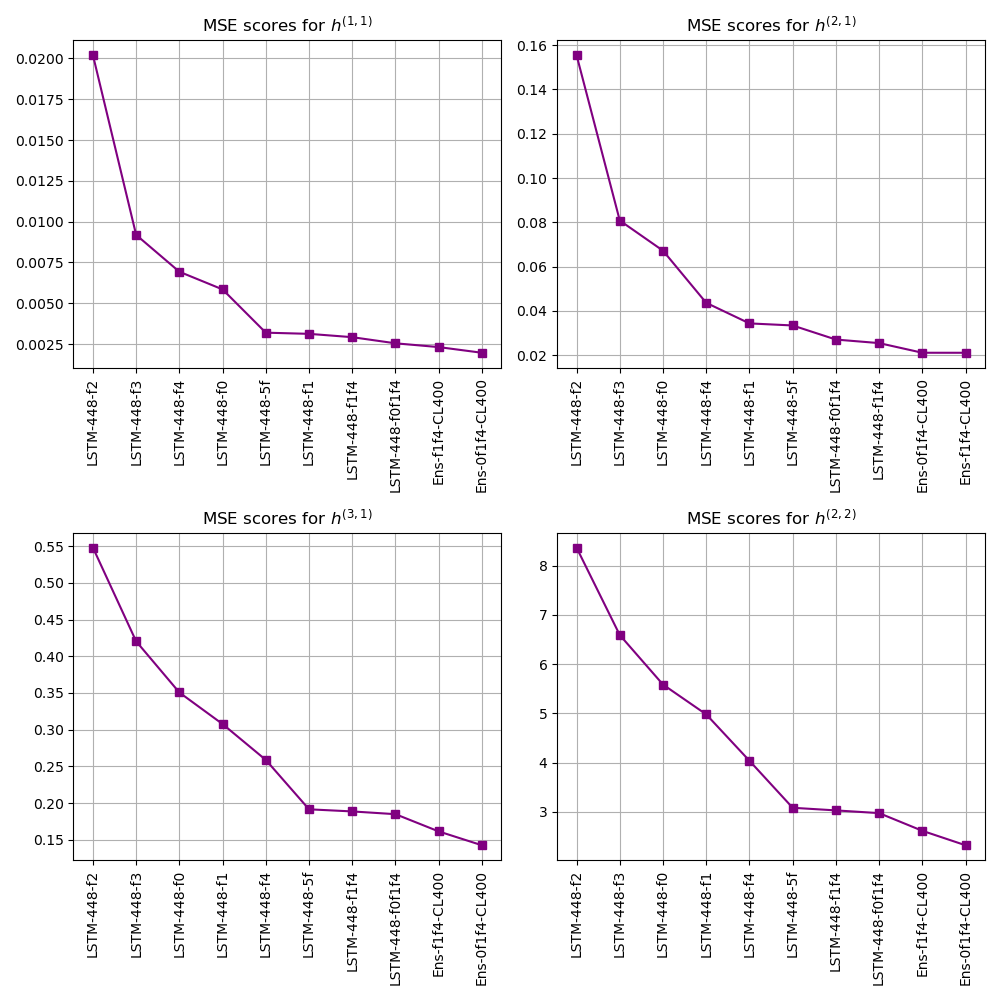}
\caption{Performances of the models in Tables \ref{5f-sep}, \ref{5f-ens} ranked in terms of the MSE scores for the 4 Hodge numbers (with worst to best models arranged from the left to the right in each diagram). }\label{mse-i-f-cv}
\end{figure}
\begin{table}[H]
\centering
\begin{tabular}{clcccc}
\hline\hline
 & Model & $h^{1,1}$ & $h^{2,1}$ & $h^{3,1}$ & $h^{2,2}$ \\
\hline  \hline
&LSTM-448-f0 & 0.004 &0.040 & 0.114 & 0.572\\
&LSTM-448-f1 & 0.002& 0.021 &0.069  &0.380\\
&LSTM-448-f2 &0.013 &0.103 &0.244 &1.068\\
&LSTM-448-f3& 0.006 &0.055 &0.134 &0.672\\
&LSTM-448-f4 &0.004 &0.029 &0.088 &0.453\\
&LSTM-448-5f &0.003 &0.025 &0.076 &0.447\\
&LSTM-448-f0f1f4 & 0.002& 0.020 &0.064 &0.361\\
&LSTM-448-f1f4 &0.002 &0.018 &0.062 &0.346\\
&Ens-f1f4-CL400 & 0.002& 0.015 &0.053 &0.305\\
&Ens-0f1f4-CL400 &$\mathbf{0.002}$  &$\mathbf{0.015}$ &$\mathbf{0.050}$ &$\mathbf{0.295}$\\
\hline
\end{tabular}
\caption{MAE scores of all the models in Tables \ref{5f-sep}, \ref{5f-ens} for each of the Hodge numbers. The best results are noted in bold.}\label{mae-i-cv}
\end{table}
 
\begin{figure}[H]
\centering
\includegraphics[width = 0.8\textwidth]{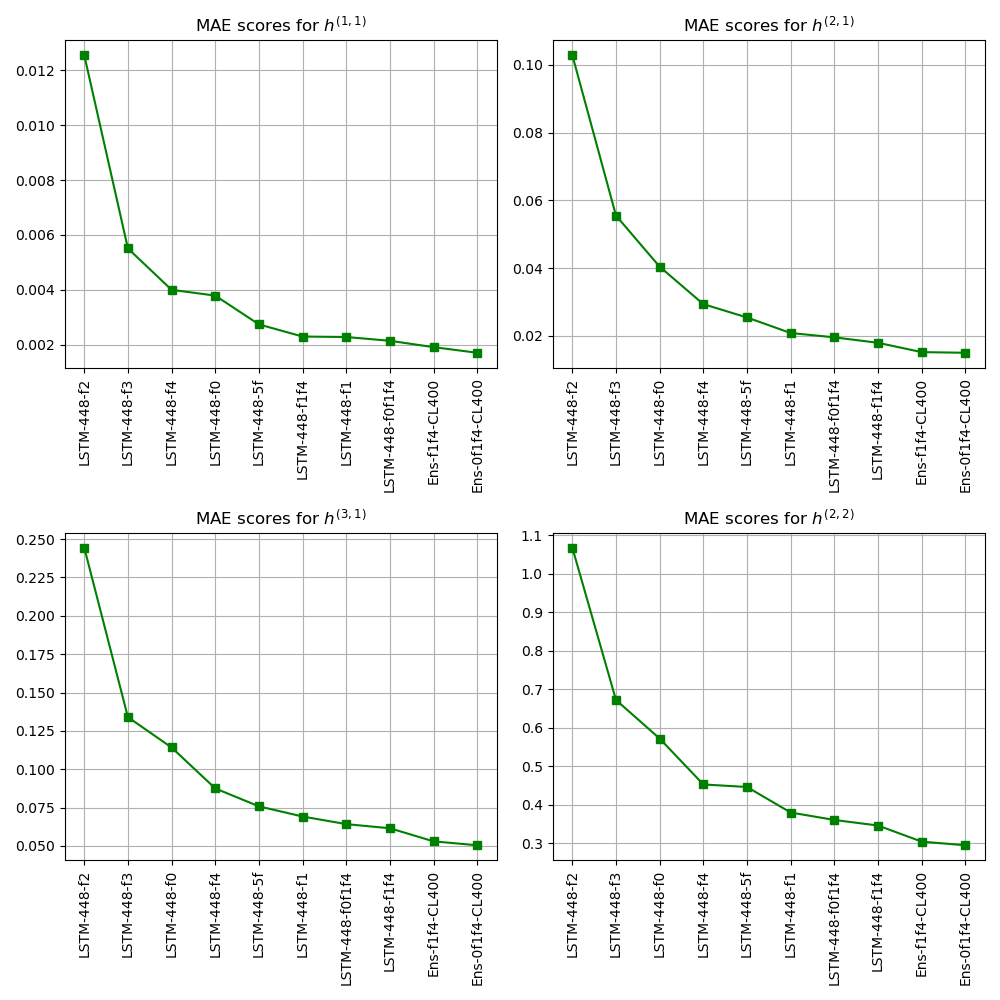}
\caption{Performances of the models in Tables \ref{5f-sep}, \ref{5f-ens} ranked in terms of the MAE scores for the 4 Hodge numbers (with worst to best models arranged from the left to the right in each diagram). }\label{mae-i-f-cv}
\end{figure}
\begin{table}[H]
\centering
\begin{tabular}{clcccc}
\hline\hline
 & Model & $h^{1,1}$ & $h^{2,1}$ & $h^{3,1}$ & $h^{2,2}$ \\
\hline  \hline
&LSTM-448-f0 & 0.999& 0.981 &0.998 &0.998\\
&LSTM-448-f1 &0.999 &0.990 &0.998 &0.998\\
&LSTM-448-f2 &0.997 &0.956 &0.997 &0.997\\
&LSTM-448-f3 &0.998 &0.977 &0.998 &0.997\\
&LSTM-448-f4 &0.999 &0.988 &0.999 &0.998\\
&LSTM-448-5f &0.999 &0.991 &0.999 &0.999\\
&LSTM-448-f0f1f4 &1.000 & 0.992& 0.999 &0.999\\
&LSTM-448-f1f4 & 1.000 &0.993 &0.999 &0.999\\
&Ens-f1f4-CL400 & 1.000 &0.994 &0.999 &0.999\\
&Ens-0f1f4-CL400 &$\mathbf{1.000}$ &$\mathbf{0.994}$ & $\mathbf{0.999}$& $\mathbf{0.999}$\\
\hline
\end{tabular}
\caption{$R^2$ scores of all models in Tables \ref{5f-sep}, \ref{5f-ens} for each of the Hodge numbers. The best results are noted in bold.}\label{r2-i-cv}
\end{table}

\begin{figure}[H]
\centering
\includegraphics[width = 0.8\textwidth]{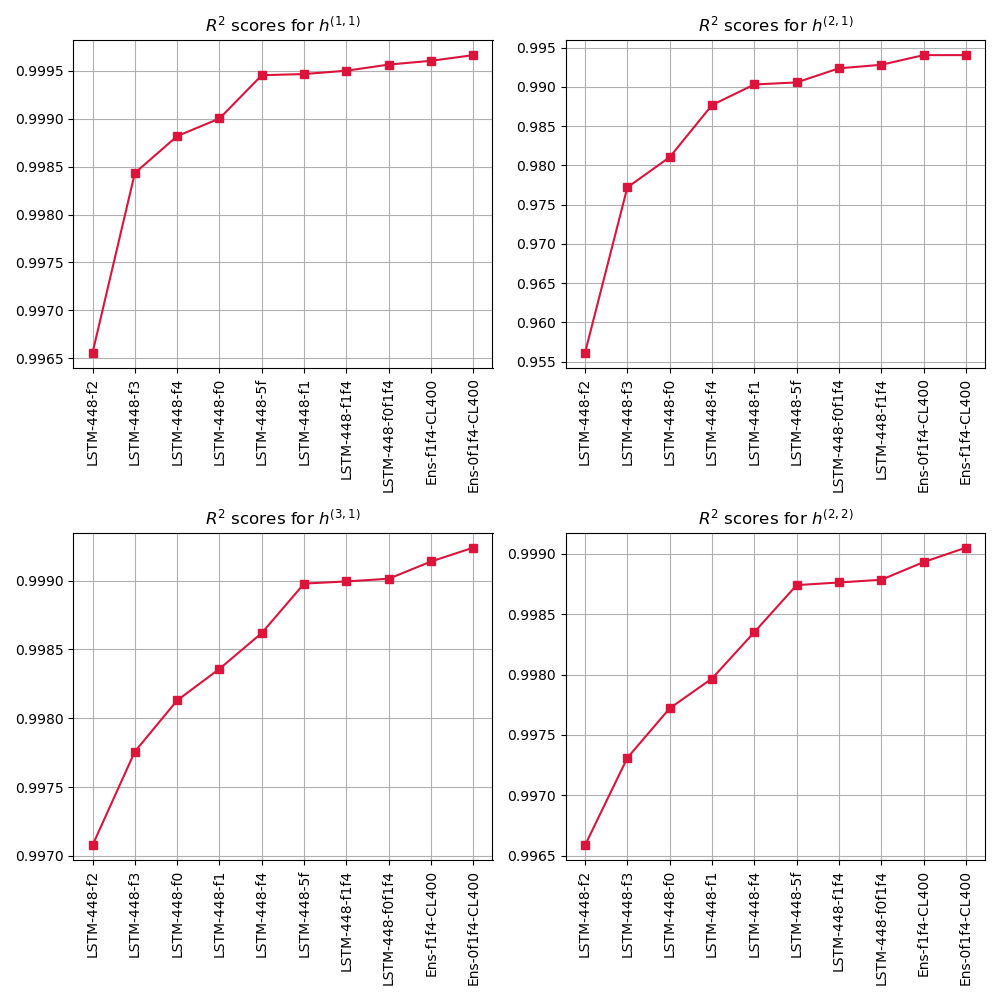}
\caption{Performances of the models in Tables \ref{5f-sep}, \ref{5f-ens} ranked in terms of the $R^2$ scores for the 4 Hodge numbers (with worst to best models arranged from the left to the right in each diagram). }\label{r2-i-f-cv}
\end{figure}
Taking the mean of the four MSE (MAE/$R^2$) scores of each model corresponding to the four Hodge numbers gives the overall MSE score for that particular model, and the results for MSE (MAE/$R^2$) scores of the 11 models of Tables \ref{5f-sep}, \ref{5f-ens} are listed in Table \ref{scores-all-CV}. The corresponding model ranking is shown in Fig. \ref{scores-all-CV-f}.
In terms of the overall MSE/MAE/$R^2$ metric, Ens-0f1f4-CL400 has the best performance, as is consistent with its accuracy ranking shown in Fig.\ref{acc-CV-4x}.
\begin{table}[H]
\centering
\begin{tabular}{clccc}
\hline\hline
 & Model & MSE & MAE & $R^2$  \\
\hline  \hline
&LSTM-448-f0 &1.503& 0.183 &0.994\\
&LSTM-448-f1 &1.332 &0.118 &0.997\\
&LSTM-448-f2 &2.273 &0.357 &0.987\\
&LSTM-448-f3 &1.776 &0.217 &0.993\\
&LSTM-448-f4 &1.087 &0.144 &0.996\\
&LSTM-448-5f &0.828 &0.138 &0.997\\
&LSTM-448-f0f1f4& 0.798 &0.112& 0.997\\
&LSTM-448-f1f4 &0.812 &0.107& 0.998\\
&Ens-f1f4-CL400 & 0.701& 0.094 &0.998\\
&Ens-0f1f4-CL400 &$\mathbf{0.622}$ & $\mathbf{0.091}$&  $\mathbf{0.998}$\\
\hline
\end{tabular}
\caption{MSE, MAE and $R^2$ scores of all models in Tables \ref{5f-sep}, \ref{5f-ens}. The best results are noted in bold.}\label{scores-all-CV}
\end{table}

\begin{figure}[H]
\centering
\includegraphics[width = 0.8\textwidth]{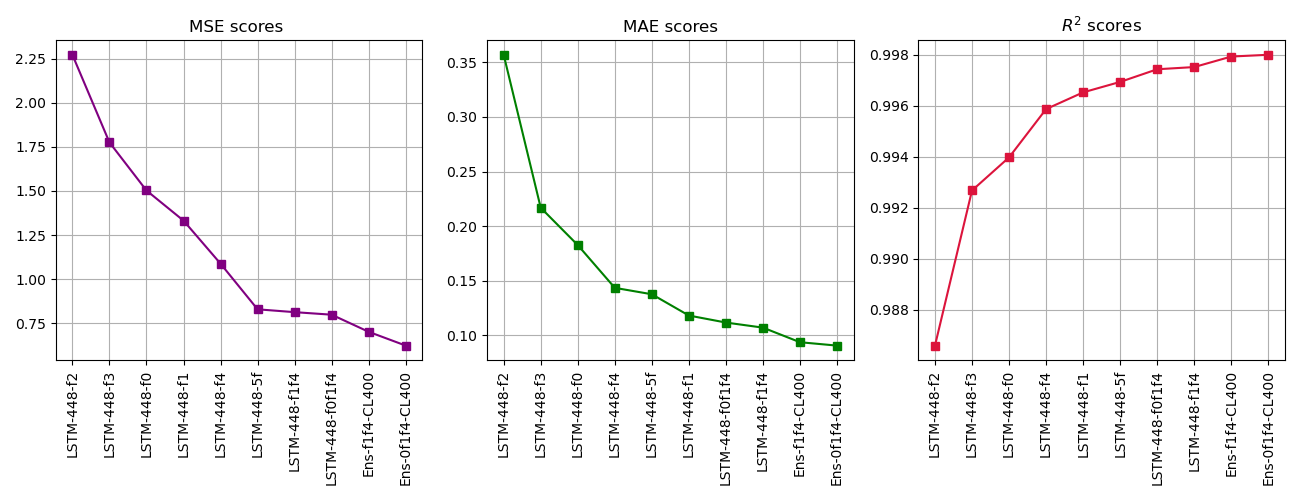}
\caption{Performances of the models in Tables \ref{5f-sep}, \ref{5f-ens} (evaluated on the test set of the 72\% dataset) ranked in terms of the MSE/MAE/$R^2$ scores (with worst to best models arranged from the left to the right in each diagram).}\label{scores-all-CV-f}
\end{figure}
\subsection{MSE, MAE, R-squared metrics (80\% dataset)} \label{metrics-80}
The individual MSE, MAE and $R^2$ scores (for each of the Hodge numbers) of all models in Tables \ref{res-80}, \ref{ens-res-80} evaluated on the test set of the 80\% dataset are recorded in Tables \ref{mse-80-i}, \ref{mae-80-i}, \ref{r2-80-i}.
 In each table, the best results are noted in bold. The visualization of the model rankings in terms of each of these metrics for each of the Hodge numbers can be found in Figs.\ref{mse-80-i-f}, \ref{mae-80-i-f}, \ref{r2-80-i-f}.
\begin{table}[H]
\centering
\begin{tabular}{clcccc}
\hline\hline
 & Model & $h^{1,1}$ & $h^{2,1}$ & $h^{3,1}$ & $h^{2,2}$ \\
\hline  \hline 
&CNN-LSTM-400&  0.005 &0.036 &0.238 &3.863\\
&LSTM-424  &0.002 & 0.034 &0.178& 2.820\\
&LSTM-448  &0.003 &0.030 &0.157 &2.517\\
& LSTM-448-f1& 0.003 &0.033 & 0.370 & 5.948 \\
&Ens-80-1  &0.002 &0.021 &0.122 &1.972\\
&Ens-80-2  &0.002 &0.022 &0.131 &2.165\\
& Ens-80-3 & 0.001 &0.018 &$\mathbf{0.113}$ &$\mathbf{1.864}$\\
&Ens-80-4  &$\mathbf{0.001}$ &$\mathbf{0.017}$ & 0.133 & 2.144 \\
& Ens-80-5  &0.001 & 0.018  &0.140 &2.262\\
\hline
\end{tabular}
\caption{MSE scores of all models in Tables \ref{res-80}, \ref{ens-res-80} for each of the Hodge numbers.}\label{mse-80-i}
\end{table}

\begin{figure}[H]
\centering
\includegraphics[width = 0.8\textwidth]{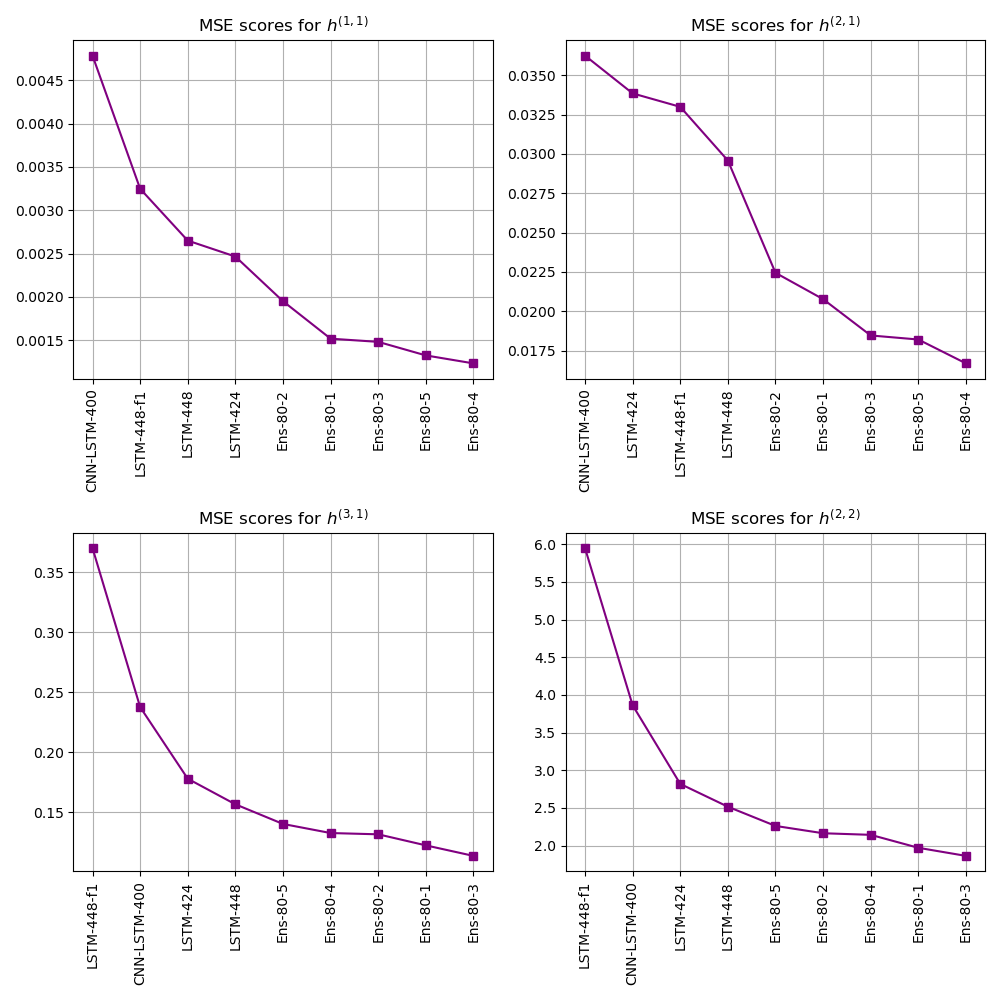}
\caption{Performances of the models in Tables \ref{res-80}, \ref{ens-res-80} ranked in terms of the MSE scores for the 4 Hodge numbers (with worst to best models arranged from the left to the right in each diagram). }\label{mse-80-i-f}
\end{figure}
\begin{table}[H]
\centering
\begin{tabular}{clcccc}
\hline\hline
 & Model & $h^{1,1}$ & $h^{2,1}$ & $h^{3,1}$ & $h^{2,2}$ \\
\hline  \hline 
&CNN-LSTM-400 & 0.003 &0.024 & 0.070 &0.384\\
&LSTM-424  &0.002 &0.022 &0.059 &0.349\\
&LSTM-448  &0.002 &0.017 &0.053 &0.304\\
& LSTM-448-f1 & 0.002 & 0.021 & 0.071 & 0.388\\
&Ens-80-1  &0.001 &0.014 &0.044 &0.264\\
&Ens-80-2  &0.002 &0.015 &0.047 &0.278\\
& Ens-80-3 & 0.001 &0.013 &0.041& 0.253\\
&Ens-80-4 &  $\mathbf{0.001}$ & $\mathbf{0.011}$ &$\mathbf{0.041}$ & $\mathbf{0.247}$\\
& Ens-80-5 &  0.001& 0.013 & 0.043 &0.255 \\
\hline
\end{tabular}
\caption{MAE scores of all models in Tables \ref{res-80}, \ref{ens-res-80} for each of the Hodge numbers.}\label{mae-80-i}
\end{table}

\begin{figure}[H]
\centering
\includegraphics[width = 0.8\textwidth]{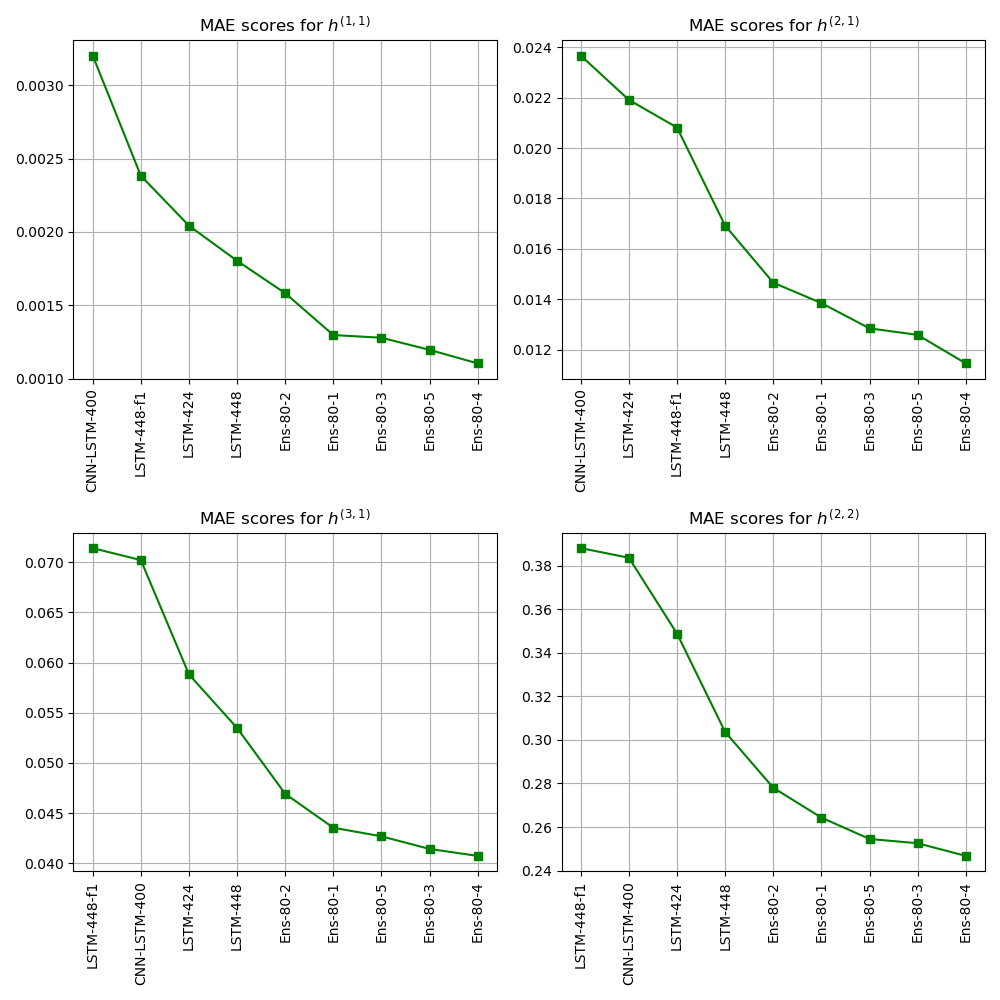}
\caption{Performances of the models in Tables \ref{res-80}, \ref{ens-res-80} ranked in terms of the MAE scores for the 4 Hodge numbers (with worst to best models arranged from the left to the right in each diagram). }\label{mae-80-i-f}
\end{figure}
\newpage
\begin{table}[H]
\centering
\begin{tabular}{clcccc}
\hline\hline
 & Model & $h^{1,1}$ & $h^{2,1}$ & $h^{3,1}$ & $h^{2,2}$ \\
\hline  \hline     
& CNN-LSTM-400& 0.999 & 0.990 & 0.999 &0.998 \\
& LSTM-424 & 1.000 & 0.990 & 0.999 & 0.999\\
& LSTM-448 & 1.000 & 0.992 & 0.999 &0.999\\
&LSTM-448-f1 &0.999& 0.991& 0.998 &0.998 \\
&Ens-80-1 & 1.000 & 0.994 &0.999 & 0.999 \\
&Ens-80-2  &1.000 &0.994 &0.999 &0.999\\
&Ens-80-3  &1.000  & 0.995 &$\mathbf{0.999}$ &$\mathbf{0.999}$\\
&Ens-80-4 & $\mathbf{1.000}$& $\mathbf{0.995}$ & 0.999 &0.999 \\
&Ens-80-5  &1.000 &0.995 &0.999 &0.999\\
\hline
\end{tabular}
\caption{$R^2$ scores of all models in Tables \ref{res-80}, \ref{ens-res-80} for each of the Hodge numbers.}\label{r2-80-i}
\end{table}

\begin{figure}[H]
\centering
\includegraphics[width = 0.8\textwidth]{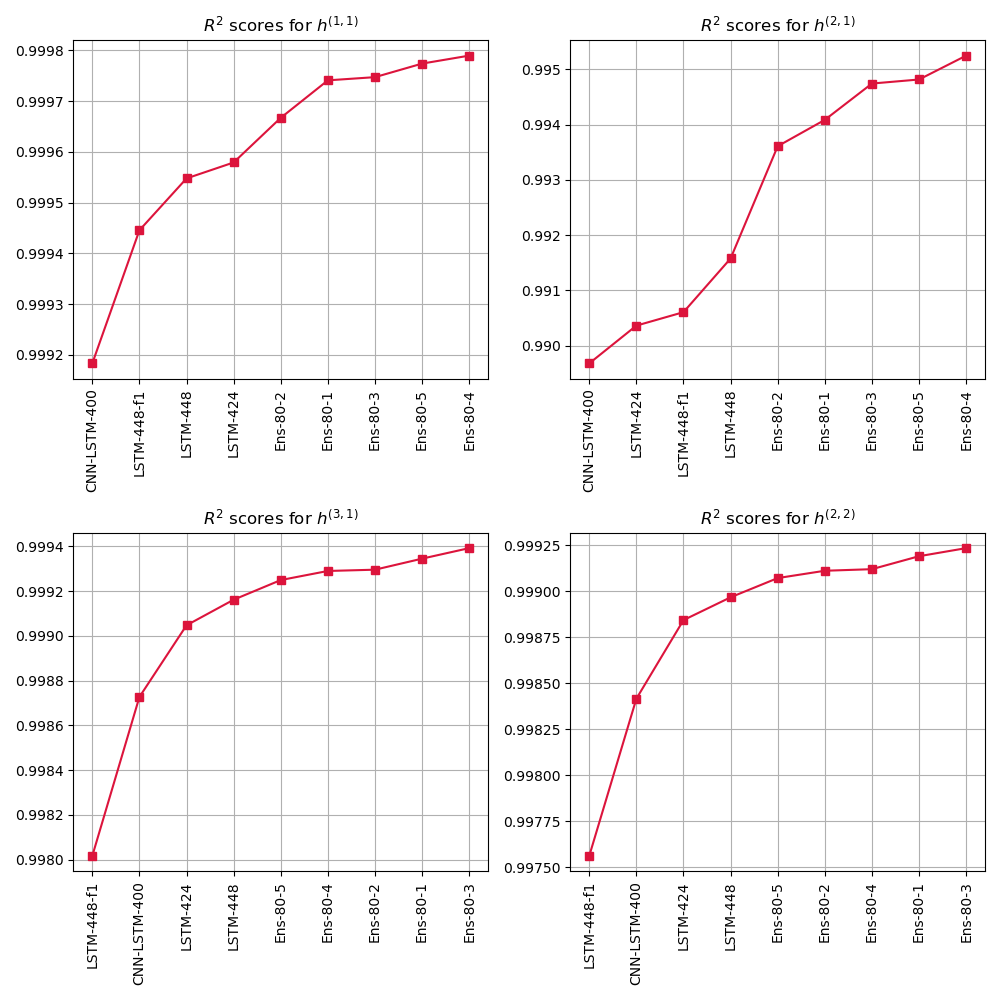}
\caption{Performances of the models in Tables \ref{res-80}, \ref{ens-res-80} ranked in terms of the $R^2$ scores for the 4 Hodge numbers (with worst to best models arranged from the left to the right in each diagram). }\label{r2-80-i-f}
\end{figure}
\newpage

The overall MSE/MAE/$R^2$ scores (which are the means of the individual MSE/MAE/$R^2$ scores for the four Hodge numbers from Tables \ref{mse-80-i}, \ref{mae-80-i}, \ref{r2-80-i}) are recorded in Table \ref{scores-80-all}.
In terms of the overall MSE and $R^2$ metrics, Ens-80-4 is ranked first (consistent with the accuracy results shown in Fig.\ref{acc-80-4x}), while in terms of the MSE metric, Ens-80-3 is ranked first. 

\begin{table}[H]
\centering
\begin{tabular}{clcccc}
\hline\hline
& Model & MSE & MAE & $R^2$ &\\
\hline \hline
& CNN-LSTM-400 &  1.035 & 0.120 & 0.997 &\\
& LSTM-424        &  0.758 & 0.108 & 0.997&\\
& LSTM-448        &  0.676 & 0.094 & 0.997 &\\
&LSTM-448-f1 & 1.589 & 0.121& 0.996 &\\
& Ens-80-1 & 0.529 & 0.081 & 0.998&\\
& Ens-80-2  &0.580 &0.085 &0.998&\\
& Ens-80-3  & $\mathbf{0.499}$ &0.077 &0.998 &\\
&Ens-80-4 &  0.574 &$\mathbf{0.075}$ & $\mathbf{0.998}$ &\\
&Ens-80-5  &0.605 &0.078 &0.998 &\\
\hline\hline
\end{tabular}
\caption{MSE, MAE, R2 scores of the models in Table \ref{res-80} plus the ensembles in Table \ref{ens-res-80}.} \label{scores-80-all}
\end{table}
A visualization of the ranked model performances in each of the three metrics (MSE, MAE, $R^2$) can be found in Fig \ref{scores-all-80}.
\begin{figure}[H]
\centering
\includegraphics[width = \textwidth]{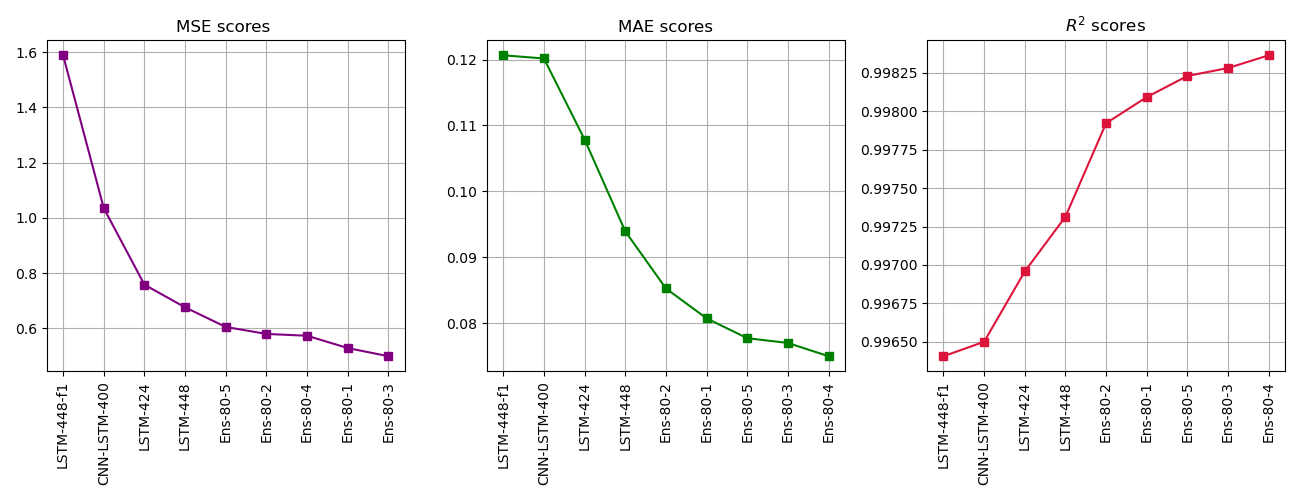}
\caption{Performances of the models in Tables \ref{res-80}, \ref{ens-res-80}  (evaluated on the test set of the 80\% dataset) ranked in terms of the MSE/MAE/$R^2$ scores (with worst to best models arranged from the left to the right in each diagram).}\label{scores-all-80}
\end{figure}
\subsection{Distributions of the Hodge numbers in 5-fold cross validation training} \label{5fold_cv_hodge_dist}
\begin{figure}[H]
\centering
\includegraphics[width = 0.6\textwidth]{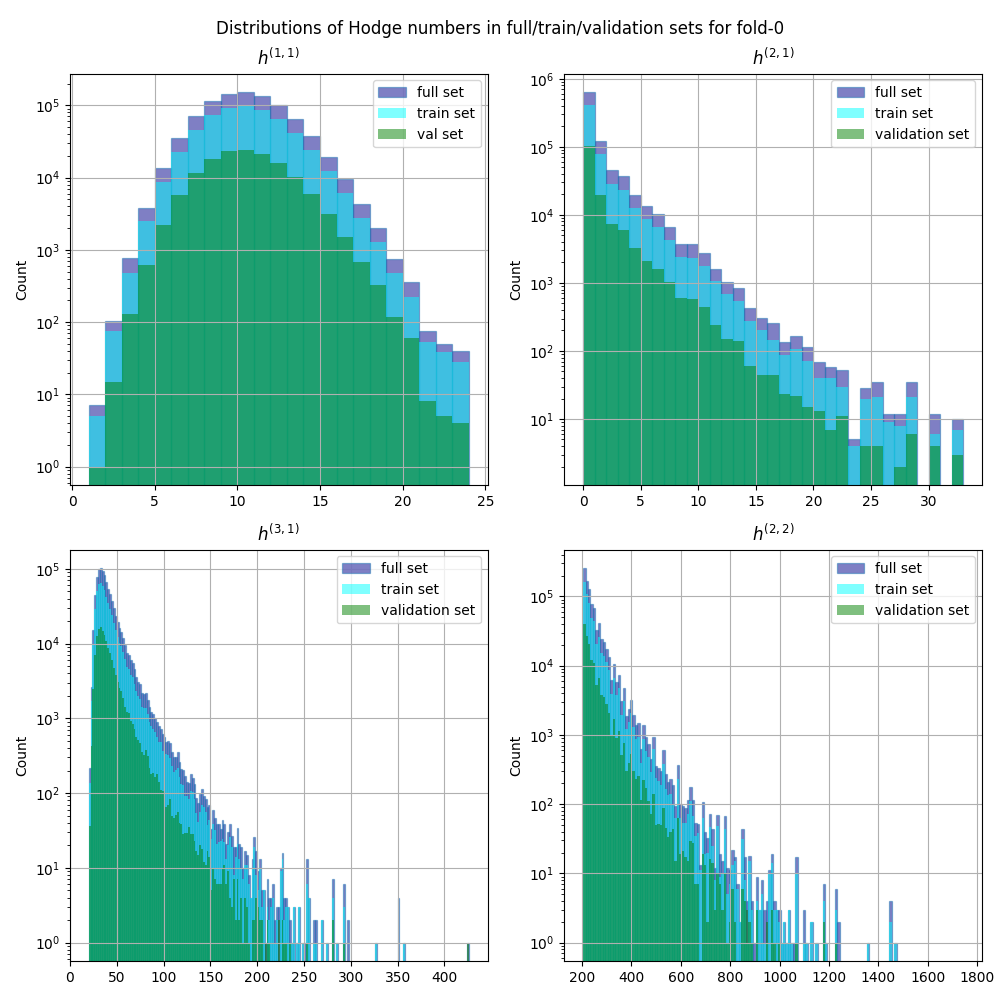}
\caption{Distributions of the Hodge numbers in the first fold of 5-fold CV training}
\label{hodge_fold_1}
\end{figure}

\begin{figure}[H]
\centering
\includegraphics[width = 0.6\textwidth]{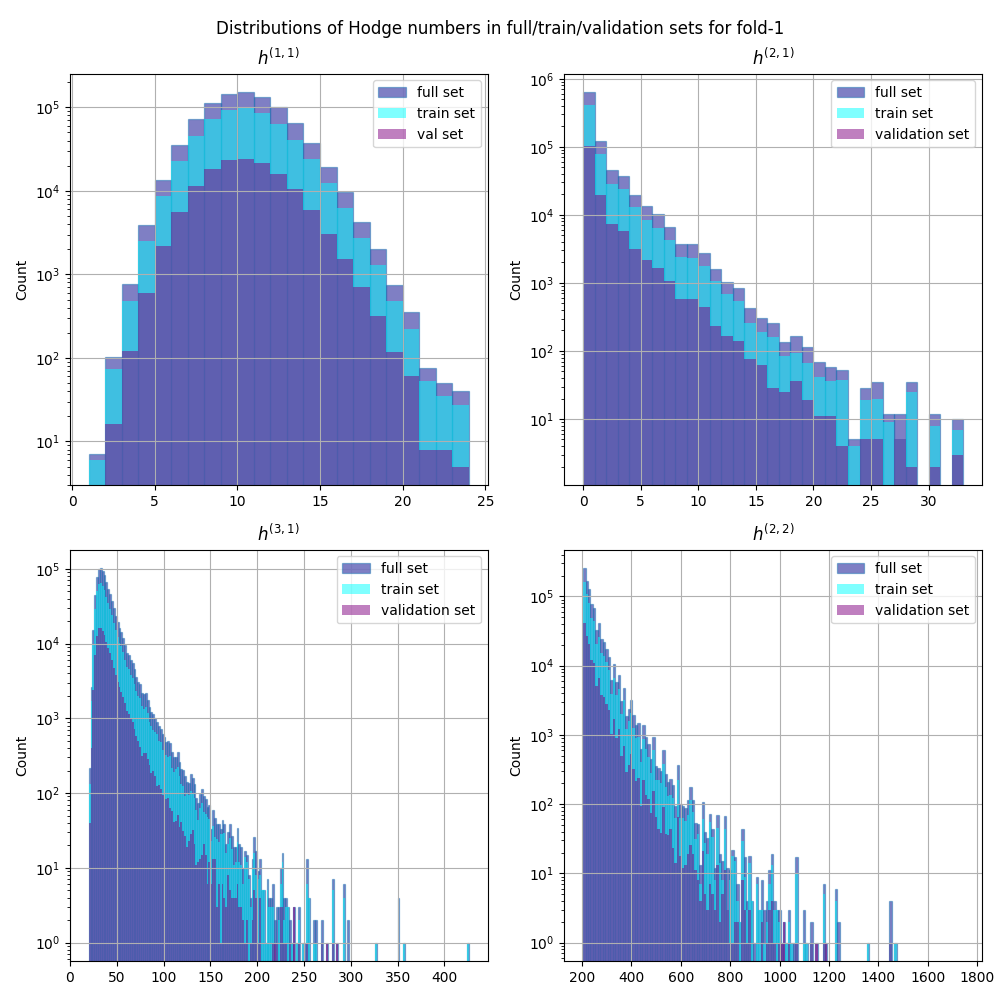}
\caption{Distributions of the Hodge numbers in the second fold of 5-fold CV training}
\label{hodge_fold_2}
\end{figure}

\begin{figure}[H]
\centering
\includegraphics[width = 0.6\textwidth]{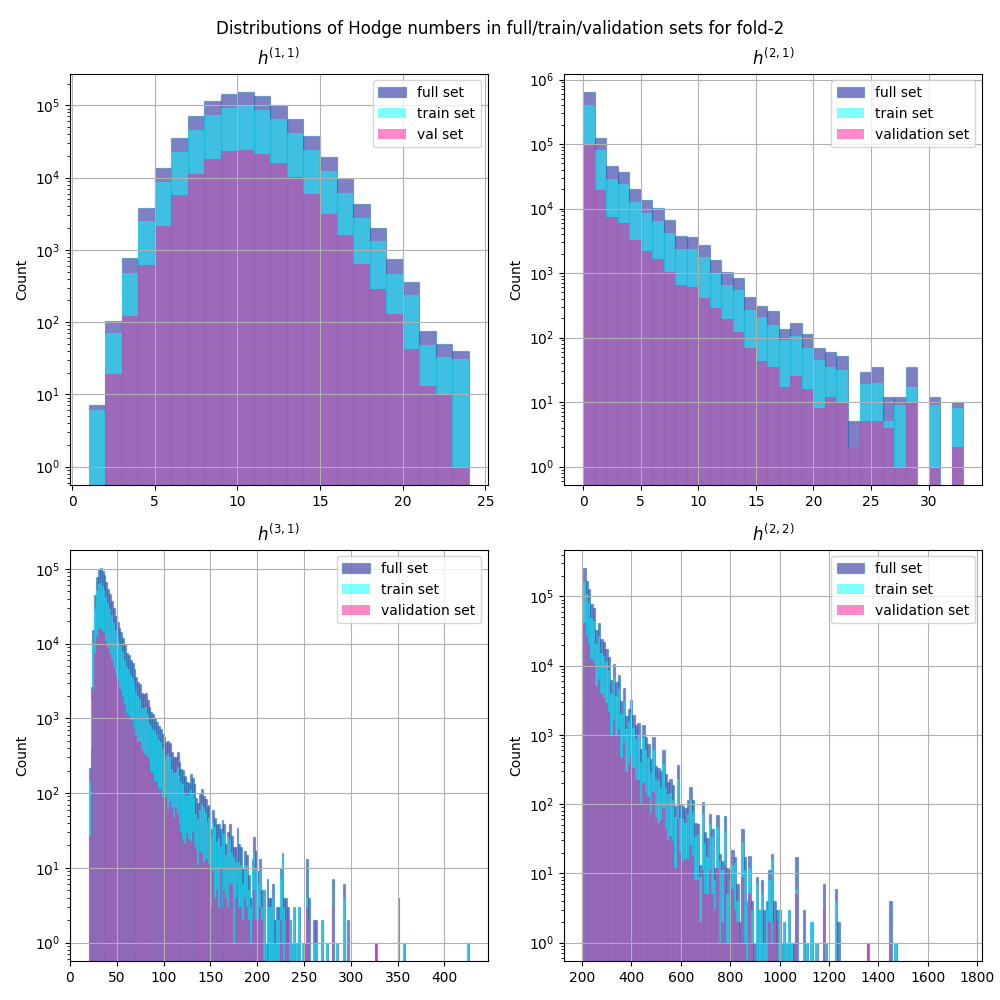}
\caption{Distributions of the Hodge numbers in the third fold of 5-fold CV training}
\label{hodge_fold_3}
\end{figure}

\begin{figure}[H]
\centering
\includegraphics[width = 0.6\textwidth]{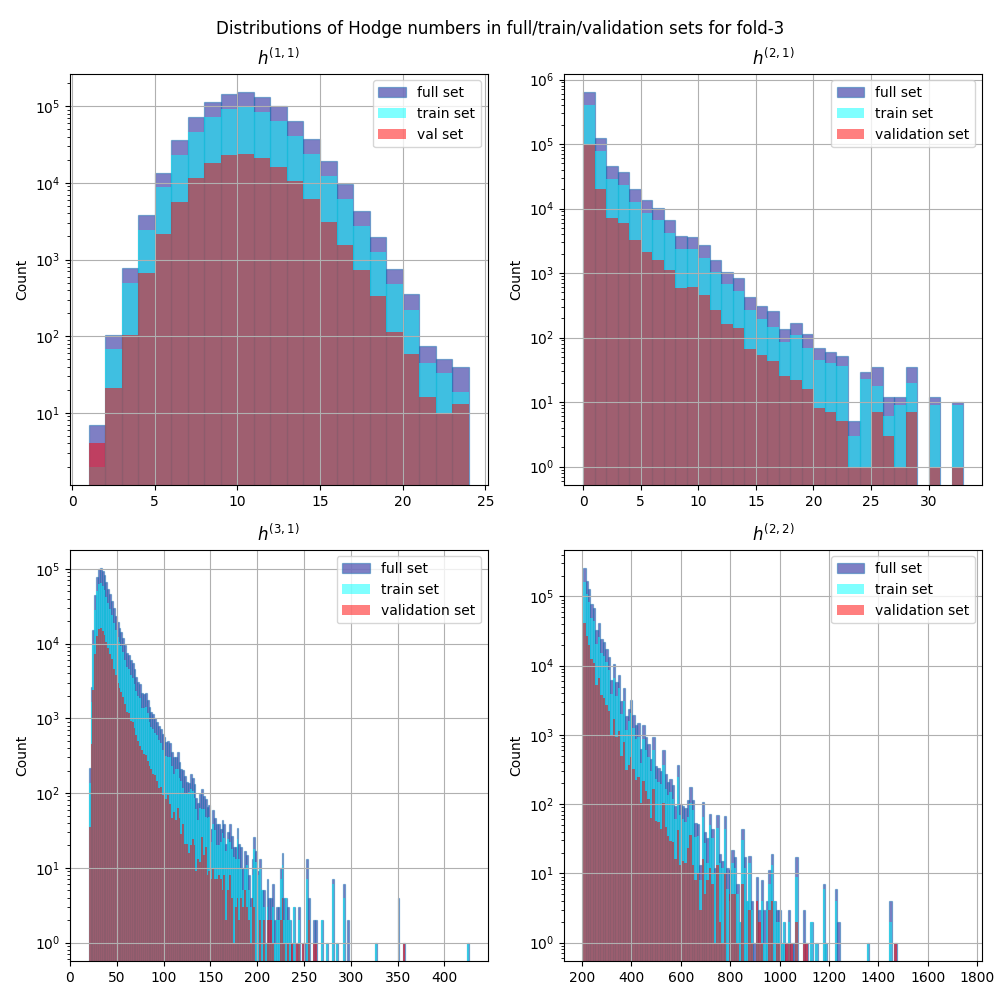}
\caption{Distributions of the Hodge numbers in the fourth fold of 5-fold CV training}
\label{hodge_fold_4}
\end{figure}

\begin{figure}[H]
\centering
\includegraphics[width = 0.6\textwidth]{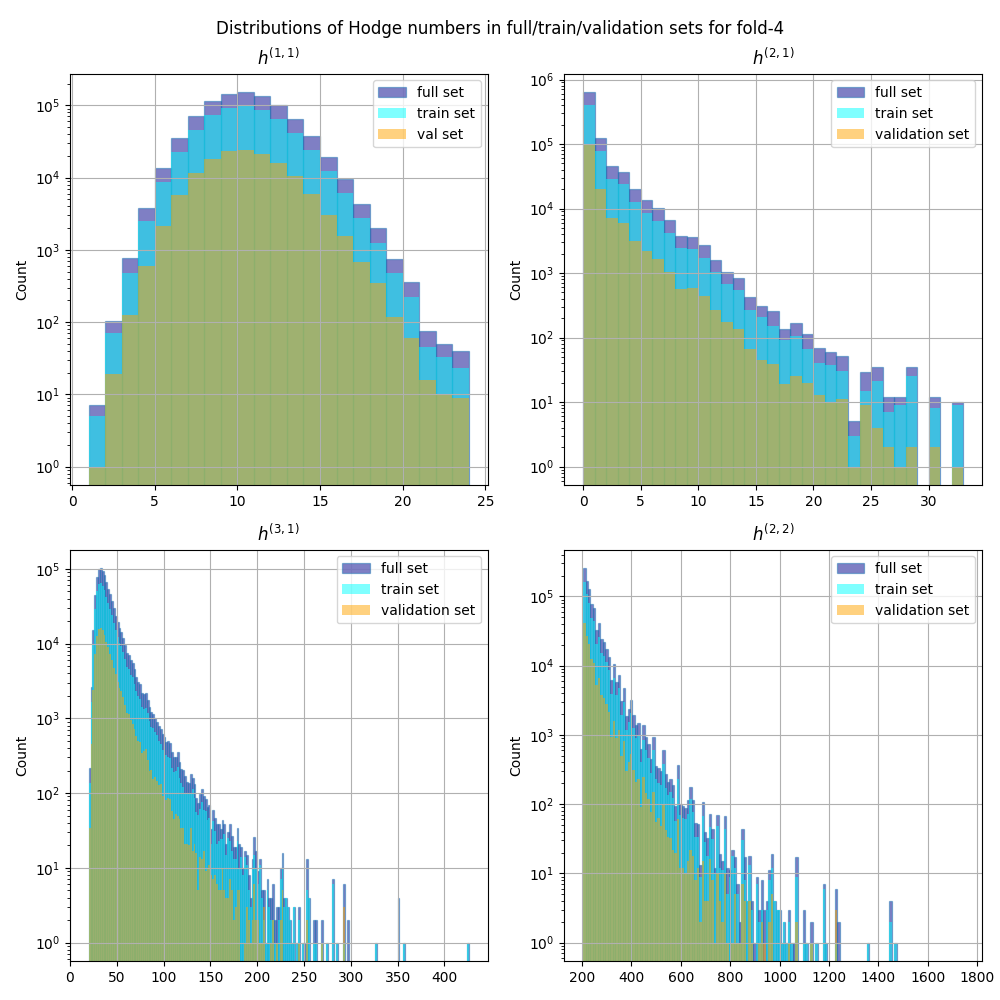}
\caption{Distributions of the Hodge numbers in the fifth fold of 5-fold CV training}
\label{hodge_fold_5}
\end{figure}
\subsection{Training curves of all neural networks} \label{curves}
\begin{figure}[H]
\centering
\includegraphics[width = 0.7\textwidth]{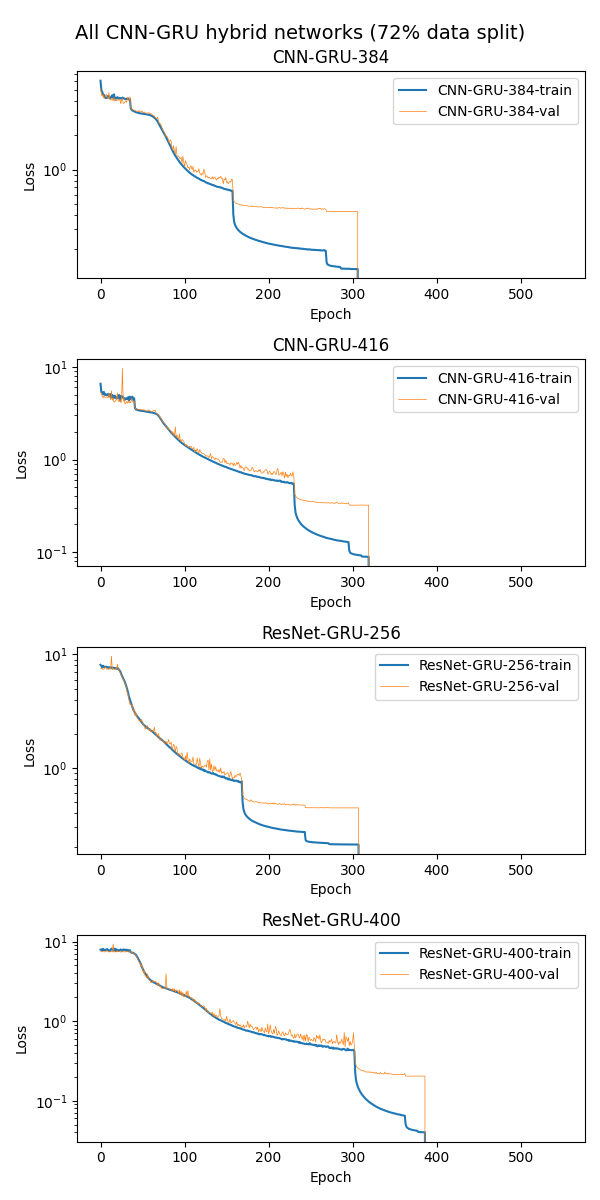}
\caption{Train and validation losses of CNN-GRU and ResNet-GRU models (listed in Table \ref{cnn-gru-acc}). CNN-GRU-384, CNN-GRU-416, and ResNet-GRU-256 achieved convergence after around 300 epochs of training, while ResNet-GRU-400 achieved convergence after nearly 400 epochs.}\label{cnn-gru-results}
\end{figure}
\begin{figure}[H]
\centering
\includegraphics[width = 0.7\textwidth]{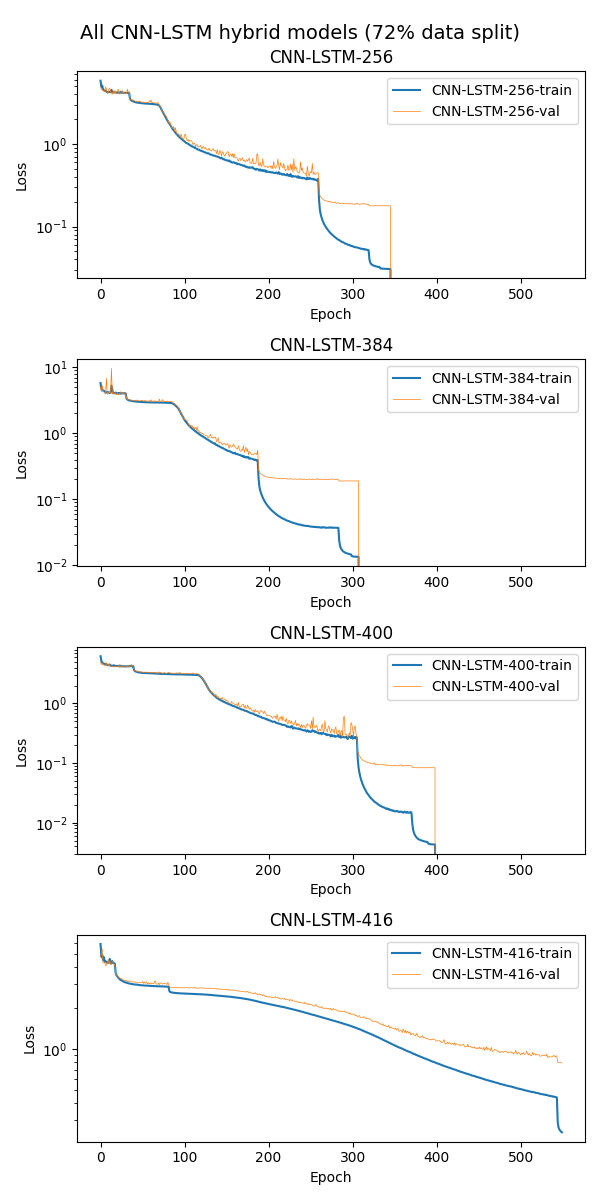}
\caption{Train and validation losses of all CNN-LSTM models (listed in Table \ref{cnn-lstm-acc}). CNN-LSTM-256 converged at around epoch 350, CNN-LSTM-384 converged at around epoch 300, CNN-LSTM-400 converged at around epoch 400. CNN-LSTM-416 fails to converge at the maximal epoch 550. }\label{cnn-lstm-all}
\end{figure}

\begin{figure}[H]
\centering
\includegraphics[width = 0.7\textwidth]{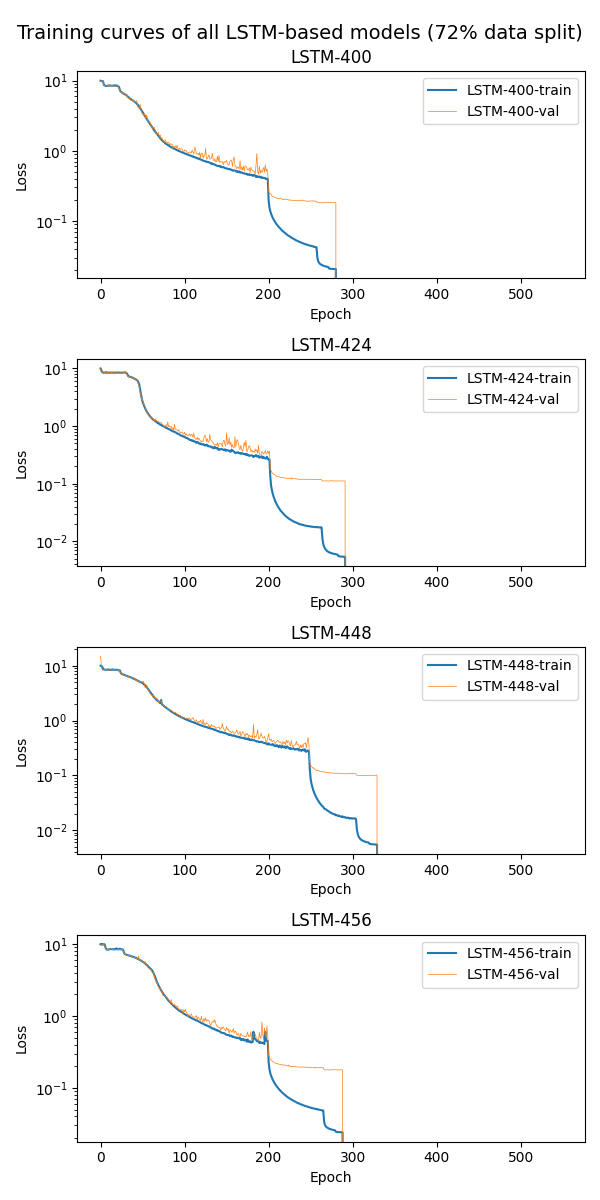}
\caption{Train and validation losses of all LSTM models listed in Table \ref{lstm-acc}. LSTM-400 converged at around epoch 280, LSTM-424 converged at around epoch 300, LSTM-448 converged at around epoch 330, LSTM-456 converged at around epoch 290.}\label{lstm-all-models}
\end{figure}

\begin{figure}[H]
\centering
\includegraphics[width = 0.5\textwidth]{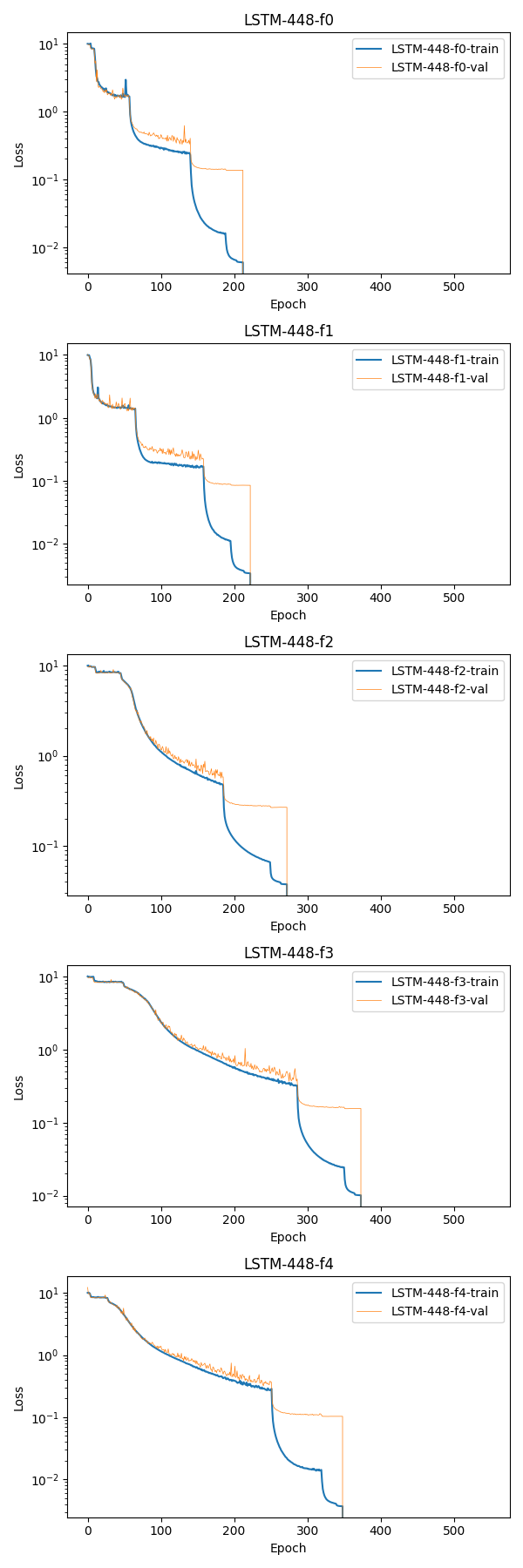}
\caption{Train and validation losses of the five LSTM models in Table \ref{5f-sep} in 5-fold cross validation training.}\label{lstm-5fold-models}
\end{figure}
\begin{figure}[H]
\centering
\includegraphics[width = 0.7\textwidth]{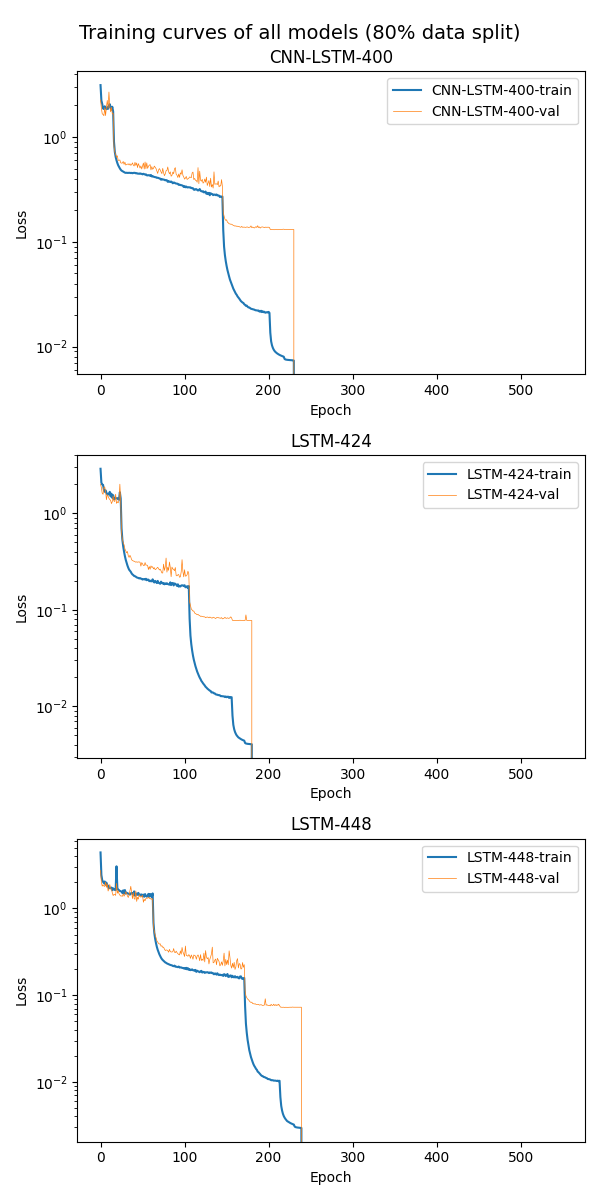}
\caption{Train and validation losses of the 3 models in Table \ref{res-80} trained from scratch on the enlarged 80\% dataset. The models were trained with the starting point being the saved checkpoints of the same models after being trained on the 72\% dataset. }\label{2nd-round-results}
\end{figure}
\newpage

\end{document}